\documentclass[aps,preprint,amsmath,amssymb,11pt]{revtex4}
\usepackage[T1]{fontenc}
\usepackage{mathrsfs}
\usepackage{epsfig}
\usepackage{slashed}
\usepackage{graphicx}
\usepackage{epstopdf}
\usepackage{graphics}
\usepackage{subfigure,psfrag}
\usepackage{amssymb}
\usepackage{amsmath}
\usepackage{amsthm}
\usepackage{multirow}
\usepackage{color}
\usepackage{makecell}

\begin{document}

\title{Exploring the Anomalous Top-Higgs FCNC Couplings at the electron proton colliders}
\author{Hao Sun\footnote{Corresponding author: haosun@mail.ustc.edu.cn \hspace{0.2cm} haosun@dlut.edu.cn}}
\author{Xuan Luo}
\author{Jing Li}
\affiliation{
Institute of Theoretical Physics, School of Physics,
Dalian University of Technology,
No.2 Linggong Road, Dalian, Liaoning, 116024, P.R.China}

\begin{abstract}

We perform an updated analysis on the searches for the anomalous FCNC
Yukawa interactions between the top quark, the Higgs boson, and either an up or charm quark ($\rm tqh,\ q=u,\ c$).
We probe the observability of the FCNC top-Higgs couplings through the processes
$\rm e^- p\rightarrow \nu_e \bar{t} \rightarrow \nu_e h \bar{q}$ (signal.I)
and $\rm \ e^- p \to \nu_e h b$ (singal.II) at the proposed electron proton (ep) colliders,
where the Higgs boson decays to a $\rm b\bar{b}$ pair.
We find that at the high luminosity (1 $\rm ab^{-1}$) ep colliders
where the electrons have a polarisation of $\rm 80\%$ and electron energy is typical 60 GeV,
the 2$\sigma$ upper limit on $\rm Br(t\to uh)$ are $0.15\times 10^{-2}$($2.9\times 10^{-4}$)
at the 7TeV@LHeC(50TeV@FCC-eh) for signal.I
and $0.15\times 10^{-2}$($2.2\times 10^{-4}$) for signal.II.
We also give an estimate on how the sensitivity (take signal.I as examples) would change
when we reduce the electron beam energy from 60 GeV to 50 GeV or even 40 GeV due to the cost
reason. The conclusion is that the discovery potential reduce $8.7\%$($29.4\%$) if the electron beam
change from 60GeV to 50(40) GeV at the 7TeV LHeC, and $16.8\%$($19.8\%$) at the 50 TeV FCC-eh.

\end{abstract}

\maketitle

\section{Introduction}

The discovery of the Higgs boson at the Large Hadron
Collider (LHC)\cite{SMHiggs_ATLAS}\cite{SMHiggs_CMS}
is a major step towards understanding
the electroweak symmetry breaking (EWSB)
mechanism and marks a new era in particle physics.
The precise measurement of the Higgs boson and the top quark
properties would provide the possibility of searching for
the anomalous flavor changing neutral current (FCNC)
Yukawa interactions between them and either an up or charm quark ($\rm tqh,\ q=u,\ c$).
According to the Standard Model (SM),
FCNC processes are forbidden at tree level
and very much suppressed at higher orders due to
the Glashow-Iliopoulos-Maiani (GIM) mechanism\cite{GIM_mechanism}.
For instance, the $\rm t\rightarrow qh\ (q=u,c)$
branching ratio is of the order of $\sim10^{-10}$ or even below.
In models beyond the SM (BSM), the GIM suppression can be relaxed,
yielding effective tqh couplings orders of magnitude much larger than those of the SM
and therefore being detectable using current experimental data.
Observations of such anomalous top-Higgs couplings
would provide a clear signal of new physics.
Examples of such model extensions \cite{tqh_review} are, for instance,
the Minimal Supersymmetric Model (MSSM) with/without R-parity Violating
\cite{MSSM_tqh_1, MSSM_tqh_2, MSSM_tqh_3, MSSM_tqh_4, MSSM_tqh_5, MSSM_tqh_6, MSSM_tqh_7,
MSSM_tqh_8, MSSM_tqh_9, MSSM_tqh_10, MSSM_tqh_11, MSSM_tqh_12, MSSM_tqh_13, MSSM_tqh_14, MSSM:DiazCruz:2001gf},
the two Higgs Doublet Model (2HDM) \cite{2HDM_tqh_1, 2HDM_tqh_2, 2HDM_tqh_3, 2HDM_tqh_4, 2HDM_tqh_5, 2HDM_tqh_6,
2HDM_tqh_7, 2HDM_tqh_8, 2HDM_tqh_9, 2HDM_tqh_10, 2HDM:He:2002fd, 2HDM:He:2002ak, 2HDM_tqH_Altunkaynak, 2HDM_tqH:He:1998ie},
the Warped Extra Dimensions Model \cite{WED_tqH_1,WED_tqH_2},
the Alternative Left-Right symmetric Model (ALRM)\cite{ALRM_tqH},
the Little Higgs with T parity Model (LHT)\cite{LHT_tqH_1},
the Quark Singlet model (QS)\cite{QS_tqh_1, QS_tqh_2, QS_tqh_3}, etc.

The searches for the anomalous FCNC top-Higgs couplings have been investigated at the LHC,
and the direct limits on the branching ratio are set from the collider experiments.
The most stringent constraint through direct measurements was reported by the CMS and ATLAS collaborations.
They have set upper limits on the FCNC couplings in the top sector
through the top pair production, with one top decaying to wb
and the other assumed to decay to hq. The w boson is considered decaying leptonically
and the Higgs decaying either to two photons\cite{tqh_rr_ATLAS, tqh_rr_ATLAS_2017, tqh_rr_CMS}
or to $\rm b\bar{b}$\cite{tqh_bb_ATLAS,tqh_bb_CMS}.
Combining the analysis of the different Higgs decay channels,
corresponding to 20.3 (19.7) $\rm fb^{-1}$ data at the center-of-mass energy of 8 TeV for ATLAS (CMS),
the 95$\%$ confidence level (C.L.) upper limits has found to be
$\rm Br(t\to uh) \leq 4.5(5.5)\times 10^{-3}$ \cite{tqh_rr_ATLAS}
and $\rm Br(t\to ch) \leq 4.6(4.0)\times 10^{-3}$ \cite{CMS_tqh_limit_comb}.
In addition to the direct collider measurements,
indirect constrains on the anomalous tqh vertex can be obtained from
the low energy measurements in flavor mixing processes, like, for example,
neutral meson oscillations ($\rm K^0-\bar{K}^0$, $\rm B^0-\bar{B}^0$
and $\rm D^0-\bar{D}^0$) \cite{Indirect_KK,Indirect_BB,Indirect_DD}.
Typically, at one-loop level, the $\rm D^0-\bar{D^0}$
mixing observable can receive sizeable contributions
with such an unvanishing flavor violating tqh coupling\cite{Indirect_DD}.
Use data observed on $\rm D^0-\bar{D^0}$ mixing,
the upper limit of $\rm Br(t\to qh)\leq 5\times 10^{-3}$ can be obtained.
The tqh coupling also affects the $\rm Z\to c\bar{c}$ decay at the loop level
and is therefore constrained by the electroweak precision observables of the Z boson\cite{tqh_Ztocc}.
On the phenomenological side the sensitivity to these non-standard flavor violating
couplings in the top sector has been explored in great details.
A lot of works have been done at the LHC, through top pair production
\cite{tqh_LHC_tt_Aguilar, tqh_LHC_tt_Atwood, SM_tqH_multileptons, tqh_LHC_tt_Kobakhidze, tqh_LHC_tt_Wu},
single top plus Higgs production\cite{tqh_review, tqh_LHC_th_Greljo, SM_tqH_ChargeRatio},
and also single top plus W production\cite{tqh_LHC_Whj_Liu}.
Some have been done at the $\rm e^+e^-$ colliders
\cite{tqh_ee_Han, tqh_ee_Behnke, tqh_ee_Aicheler, tqh_ee_Hesari, tqh_ee_Monalisa},
and several at the ep colliders\cite{tqh_LHeC_haosun1, tqh_LHeC_haosun2}.
Some other related studies include, for example, Ref.\cite{DipoleM_tqH}, which derives model-independent constraints
on the tqh couplings that arise from the bounds on hadronic electric dipole moments.

In our present paper, we perform an updated study on the anomalous FCNC Yukawa interactions
at the ep colliders. A former study was performed in Ref.\cite{tqh_LHeC_haosun1}.
There we briefly reviewed the search of this anomalous couplings at the basic parton level.
A comparison between different charge current(CC) and neutral current(NC) production channels
were provided. There comes the conclusion that the CC induced
$\rm e^- p\rightarrow \nu_e \bar{t} \rightarrow \nu_e h \bar{q}$ (signal.I)
production with $\gamma\gamma$, $\rm b\bar{b}$ pair and $\tau^+\tau^-$ decays are the favored candidate channels.
$\rm H\to \gamma\gamma$ channel was chosen because of its demonstrated high importance for
inclusive Higgs boson studies, with a rather clean signature at the normal LHC.
However, for a Higgs boson mass around 125 GeV, $\rm e^- p\rightarrow \nu_e \bar{t} \rightarrow \nu_e h \bar{q}$
production with $\rm h\to \gamma\gamma$ decay at the ep collider,
suffers from its small branching ratio (0.23$\%$), thus is not the most favored one.
For $\rm h\to \tau^+\tau^-$ channel, the $\tau$ event reconstruction is not easy,
thus not been concentrated on here at this moment. In this paper we choose the $\rm h\to b\bar{b}$ mode
which is more interesting than the other channels.
In addition to signal.I, we consider a second production $\rm e^- p \to \nu_e h b$ (singal.II).
Different from signal.I that the tqh couplings mainly come from the single top decays,
in signal.II, the couplings are induced through light quarks that directly emitting from the protons.
We present the discovery potentials from both channels and compare them with each other.

Our paper is organized as follows:
Section 2 present a short description of the anomalous top-Higgs FCNC couplings.
Section 3 is arranged to present the analysis and numerical results in detail.
There comes the subsections include signal and background analysis,
simulation and the discovery potential, etc.
The discovery potentials are compared with the LHC limits and the other studies.
Typically, its dependence on the electron beam energy is also presented due to the cost reason.
Finally we summarize our conclusion in the last section.

\section{The Anomalous Top-Higgs FCNC Couplings}

Considering the FCNC Yukawa interactions in the effective field theory framework,
the SM Lagrangian can be extended simply by allowing the following terms,
\begin{eqnarray}\label{lagrangian}
\rm {\cal L} = \kappa_{tuh} \bar{t}uh + \kappa_{tch} \bar{t}ch + h.c.,
\end{eqnarray}
where $\rm \kappa_{tuh}$ and $\rm \kappa_{tch}$ are the real parameters
and denote the flavor changing couplings of Higgs to up-type quarks.
Now we have $\rm m_t$ minus $\rm m_h$ larger than $\rm m_c$, $\rm m_u$ and $\rm m_b$,
therefore, in addition to the usual decay mode $\rm t\rightarrow w^\pm b$,
the top quark can also decay into a charm or up quark associated with a Higgs boson.
Similarly, the new tqh interactions can also affect the width of the Higgs boson,
through its additional decay into an off-shell top, that subsequently leads to a single w,
namely, $\rm h\rightarrow u(c) (t^*\rightarrow wb)$ where $\rm t^*$ denotes off-shell top quark.
The total decay width of the top quark ($\rm \Gamma_t$) is
\begin{eqnarray}
\rm \Gamma_t &=&\rm  \Gamma^{SM}_{t\rightarrow w^-b}+\Gamma_{t\rightarrow ch}+\Gamma_{t\rightarrow uh}.
\end{eqnarray}
Here $\rm \Gamma_{t\to w^-b}^{SM}$ is the normal top decay width in the SM.
Its analytical formula up to next-to-leading order(NLO) can be found in Ref.\cite{twb_NLO}.
The $\rm t\rightarrow u(c)h$ partial decay width is given as\cite{decay_tqH}
\begin{eqnarray}
\rm \Gamma_{t\rightarrow u(c)h}=\frac{\kappa^2_{tu(c)h}}{16\pi} m_t  [(\tau_{u(c)}+1)^2-\tau^2_h]
\sqrt{1-(\tau_h-\tau_{u(c)})^2}\sqrt{1-(\tau_h+\tau_{u(c)})^2}
\end{eqnarray}
where $\rm \tau_h=\frac{m_h}{m_t}$, $\rm \tau_{u(c)}=\frac{m_{u(c)}}{m_t}$.
The total decay width of the Higgs boson($\rm \Gamma_h$) is given by
\begin{eqnarray}
\rm \Gamma_h&=&\rm  \Gamma_{h}^{SM} +
\Gamma_{h\rightarrow u(\bar{t}^*\rightarrow \bar{b}w^-)} +
\Gamma_{h\rightarrow \bar{u}(t^*\rightarrow bw^+)}+
\Gamma_{h\rightarrow c(\bar{t}^*\rightarrow \bar{b}w^-)} +
\Gamma_{h\rightarrow \bar{c}(t^*\rightarrow bw^+)}.
\end{eqnarray}
Here $\rm \Gamma_{h}^{SM}$ is the normal two body Higgs decay width in the SM.
The terms related to the Higgs boson three-body decays are numerically estimated
with FFL-package\cite{FFL}. Thus we have 
$\rm \Gamma_h \simeq \Gamma^{SM}_h + 
\sum^{\bar{t}^*\rightarrow \bar{b}w^-}_{q=u,c} 0.28 \kappa^2_{tq} + 
\sum^{t^*\rightarrow bw^+}_{q=u,c} 0.28 \kappa^2_{tq}
\simeq \Gamma^{SM}_h + 0.56(\kappa^2_{tu} + \kappa^2_{tc})$ in unit of MeV.
After assuming the top quark decay width is dominated by the SM and
neglecting the light quark mass, the branching ratio for $\rm t \rightarrow qh$ is then approximately given by
\begin{eqnarray}
\rm B(t\rightarrow u(c)h) =
\frac{\kappa_{tu(c)h}^2}{\sqrt{2} G_F m_t^2 } \frac{(1-\tau_h^2)^2}{(1-\tau_w^2)^2(1+2 \tau_w^2)} K_{QCD} \simeq 0.58\kappa^2_{tu(c)h}
\end{eqnarray}
where $\rm \tau_w=\frac{m_w}{m_t}$ and $\rm G_F$ is the fermi constant.
The factor $\rm K_{QCD}$ is the NLO QCD correction to $\rm Br(t\to qh)$, which is calculated
to be $\rm K_{QCD}=1+0.97\alpha_{s}\simeq 1.1$ by the results of high order corrections to
$\rm t\to wb$\cite{twb_NLO} and $\rm t\to qh$\cite{tqh_NLO}.

\section{PROCESS ANALYSIS AND NUMERICAL CALCULATIONS}

\subsection{The signal and background analysis}

Here we start to present our study on the anomalous tqh couplings at the ep colliders.
Ep colliders are hybrids between the $\rm e^+e^-$ and pp colliders, which consist of a hadron beam with an electron beam.
They provide a cleaner environment compared to the pp colliders and higher center-of-mass (c.m.) energies to the $\rm e^+e^-$ ones.
Currently, the proposed ep collider is the Large Hadron Electron Collider (LHeC) \cite{LHeC_1, LHeC_2, LHeC_3, LHeC_4},
which is a combination of 60 GeV electron beam and 7 TeV proton beam of the LHC.
It can deliver up to 100 $\rm fb^{-1}$ integrated luminosity per year at a c.m. energy of
around 1 TeV and 1 $\rm ab^{-1}$ over its lifetime.
This may later be extended to the future circular electron-hadron collider (FCC-eh)\cite{FCC-eh_1},
which features a 60 GeV (or maybe higher or maybe lower) electron beam
with the 50 TeV proton beam from the future circular hadron-hadron collider (FCC-hh).
This would result in a c.m. energy up to 3.5 TeV with comparable luminosities to the LHeC\cite{FCC-eh_2}.
There are a lot of works, for example,
\cite{LHeC_BSM_Lindner, LHeC_BSM_Fischer, LHeC_BSM_Subhadeep, LHeC_BSM_ShouHua, LHeC_BSM_Kumar, LHeC_BSM_haosun,LHeC_BSM_ohan}, etc,
that have been done in the content of new physics searches,
based on such proposed colliders, in order trying hard to enrich the physics motivations.

\begin{figure}[hbtp]
\centering
\includegraphics[scale=0.25]{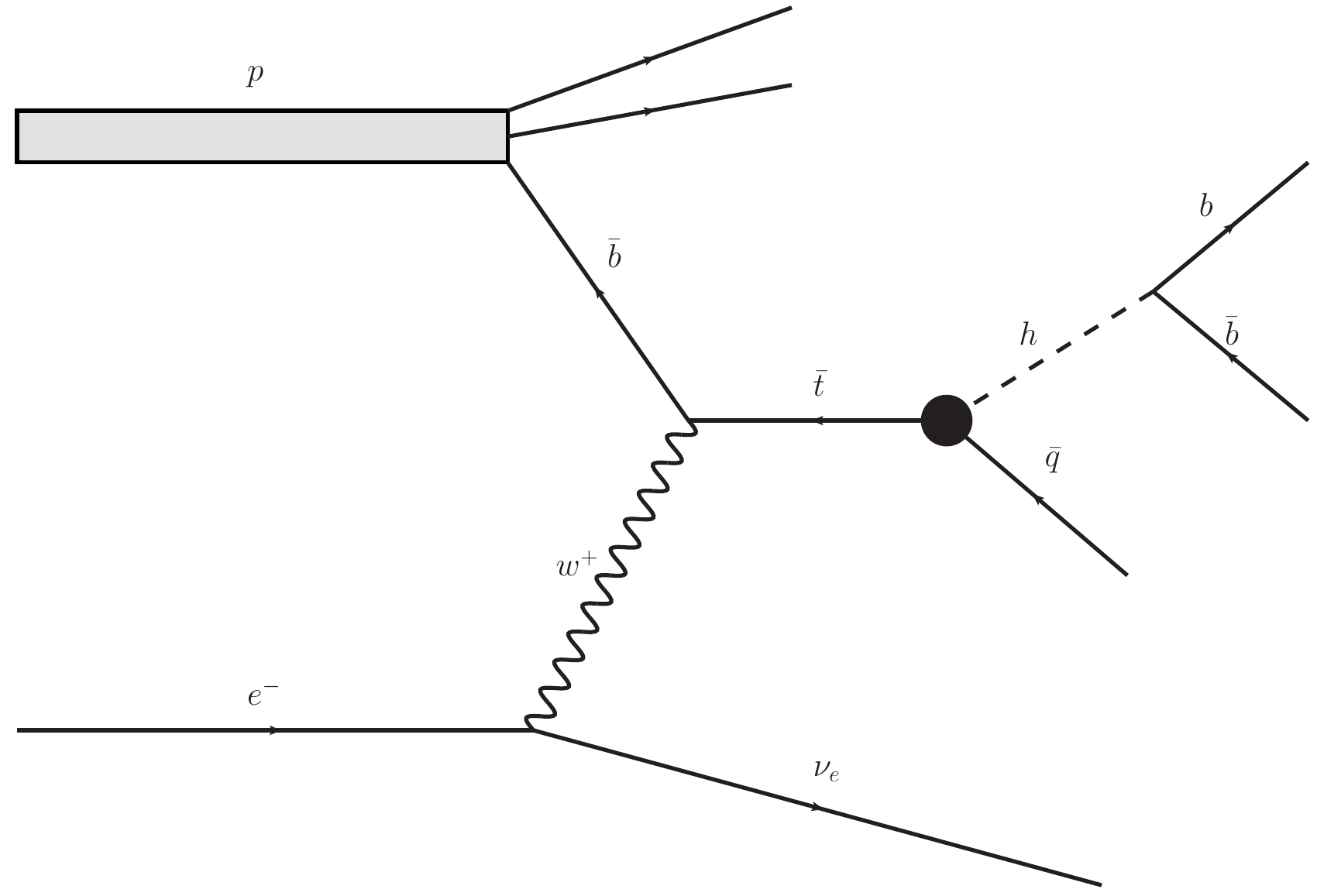}
\includegraphics[scale=0.25]{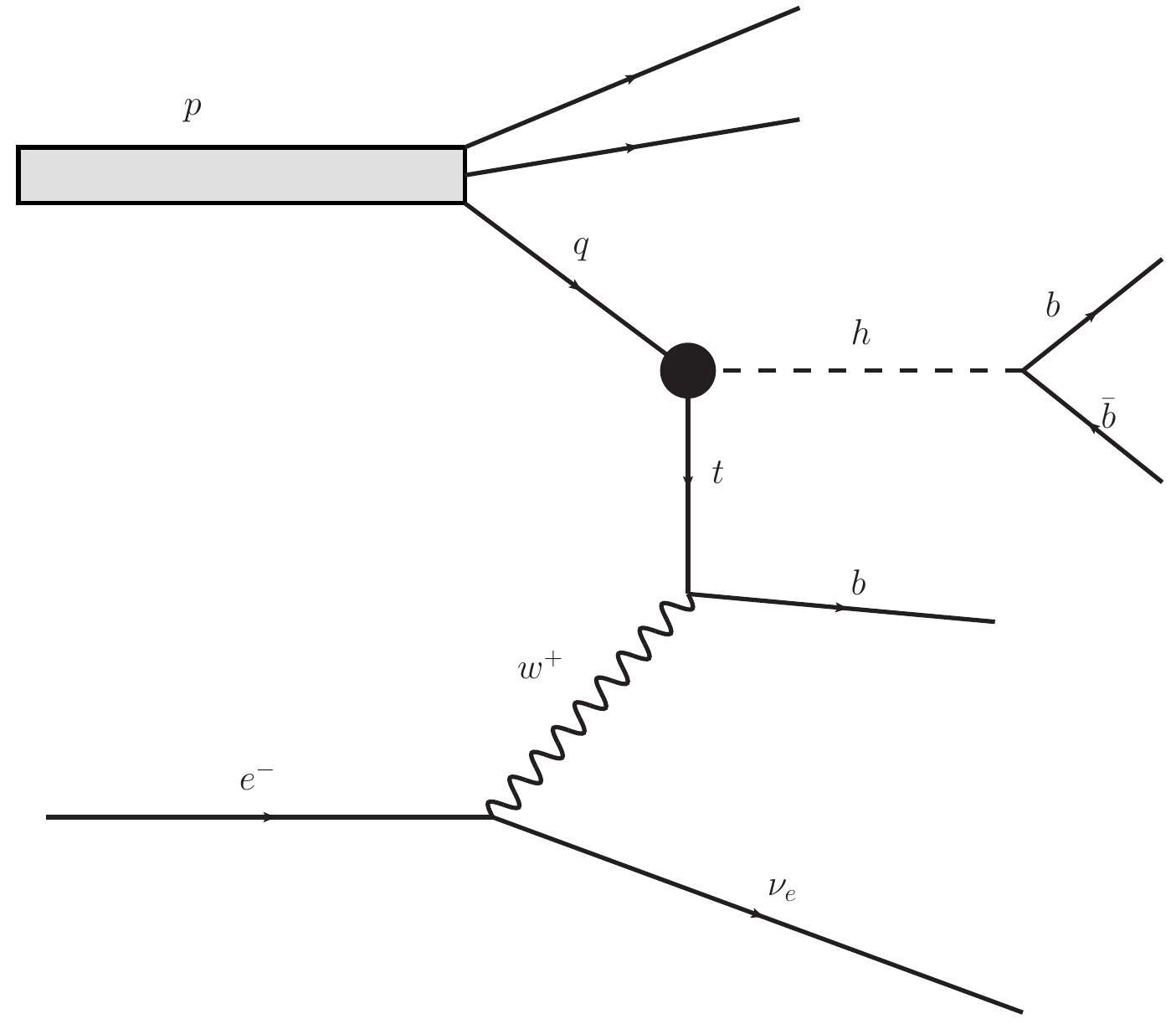}
\caption{\label{fig1_sig_Feynman}
Illustrated Feynman diagrams for the processes $\rm e^{-} p \to \nu_e \bar{t} \to \nu_e h \bar{q} \to \nu_e b\bar{b}\bar{q}$ (signal.I)
and $\rm e^{-} p \to \nu_e hb \to \nu_e b\bar{b} b$ (signal.II) at the ep colliders that contain flavor changing top-Higgs interactions.}
\end{figure}

At the ep colliders, the first signal production which contains the top-Higgs FCNC couplings,
that we considered, can be written as
\begin{eqnarray}
\rm signal.I:\ e^- p \to \nu_e \bar{t} \to \nu_e h \bar{q} \to \nu_e b\bar{b}\bar{q},
\end{eqnarray}
where q=u or c, which is the largest channel compare to the other productions\cite{tqh_LHeC_haosun1}.
In this case, the five flavor scheme should be applied
and an initial state bottom quark will collide with a w boson to produce a single top,
which decay anomaly to a Higgs and a light quark.
The Feynman diagram is plotted in Fig.\ref{fig1_sig_Feynman}(left pannel for signal.I).
As can be seen, the studied topology gives rise to the $\rm E^{miss}_T + jets$ signature characterized by
three(or more than three) jets and a missing transverse momentum($\rm E^{miss}_T$) from the undetected neutrino.
Two of the jets should be tagged as B-jets.
The combination of the two B-jets should appear as a narrow resonance
centered around the SM Higgs boson mass. Together with the remaining light jet(s),
they should be able to reconstruct a resonant top quark.

The second channel we considered is
\begin{eqnarray}
\rm singal.II:\ e^- p \to \nu_e h b \to \nu_e b\bar{b} b.
\end{eqnarray}
In this case, the FCNC tqh couplings are induced through light quarks that directly emitting from the proton,
which is different from signal.I that the tqh couplings are coming from the single top decays.
At a first glance, this contribution may not small,
because of the larger parton distribution functions (pdfs) of light quarks that inside the proton.
However, its cross section is found to be much smaller than the first one,
due to the suppression of the three body phase space integration (before Higgs decay).
There is another thing that may also be interesting and worth to be noticed.
Usually, the analysis between the $\rm t\to ch$ and $\rm t\to uh$
final states have similar acceptances. This is true for our signal.I, but not for signal.II.
For signal.II, the charm quark pdfs are much suppressed than that of the up-quark, thus the
analysis between $\rm t\to ch$ and $\rm t\to uh$ are quite different.
In our analysis, we only concentrate on $\rm t\to uh$ mode as reference throughout this work.
Even for signal.I, we should comment that if the $\rm t\to ch$ mode is considered,
the charm mis-tagging rate would also affect the signal acceptance.
If so, one can use the technique based on studies in, for example, Ref.\cite{seperate_tuh_tch},
in order to differentiate the $\rm t\to ch$ and $\rm t\to uh$ decays.
Considering the studied topology, we require that there should be three tagged B-jets for our signal.II.
This is indeed a critical selection, by applying which
the backgrounds can be strongly suppressed, thus providing a much clean channel.
This is the reason that signal.II is also included in our study, though its production rate is small.
Further more, we can find that in some case the discovery potential
through this channel can be even better than the former one.
The related Feynman diagram is plotted in Fig.\ref{fig1_sig_Feynman}(right panel for signal.II).
Both the signal channels are belonging to the charge current productions at the ep colliders
and their backgrounds are also quite similar, as which will be discussed in the following.

\begin{figure}[hbtp]
\centering
\includegraphics[scale=0.1]{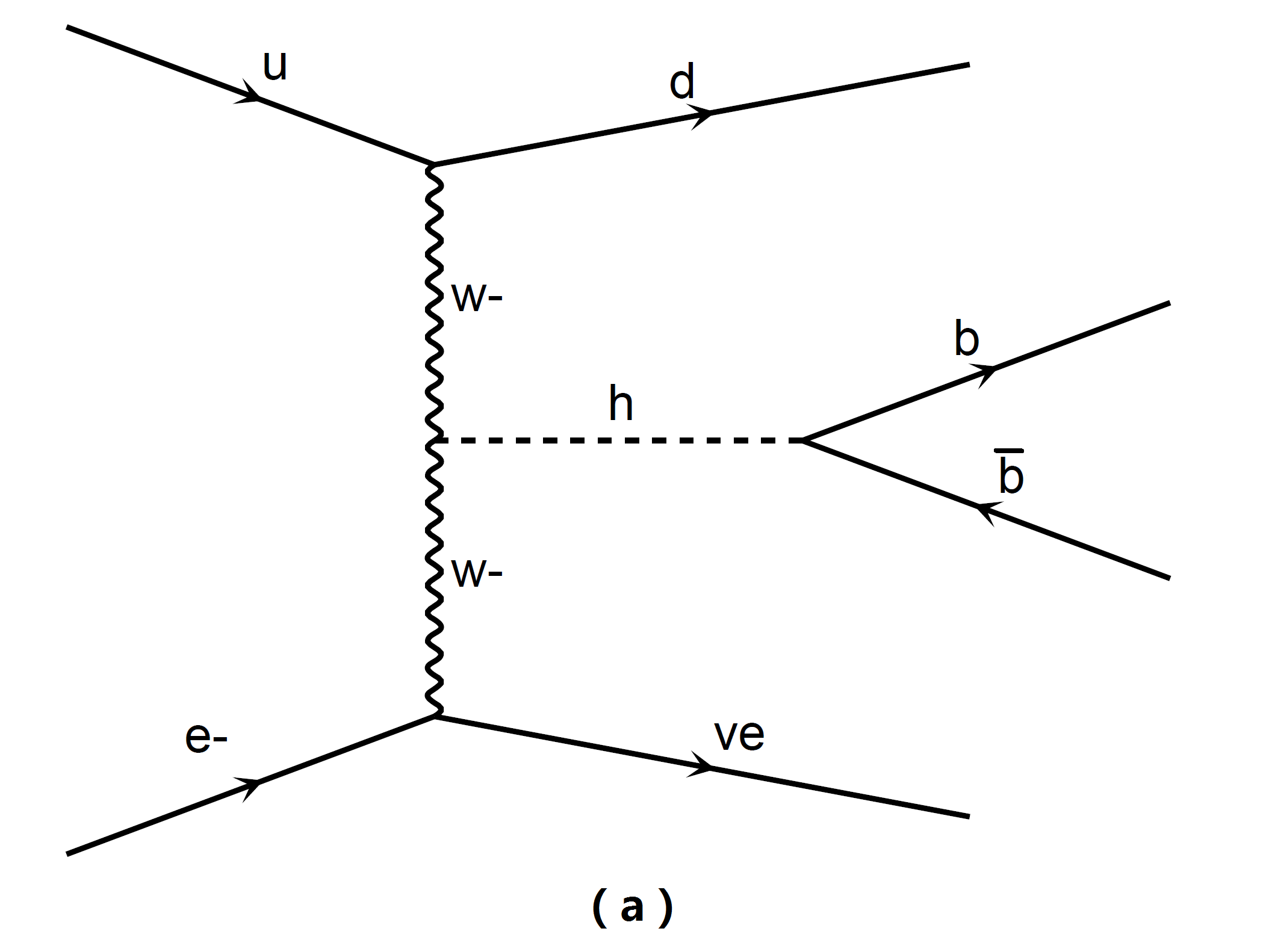}
\includegraphics[scale=0.1]{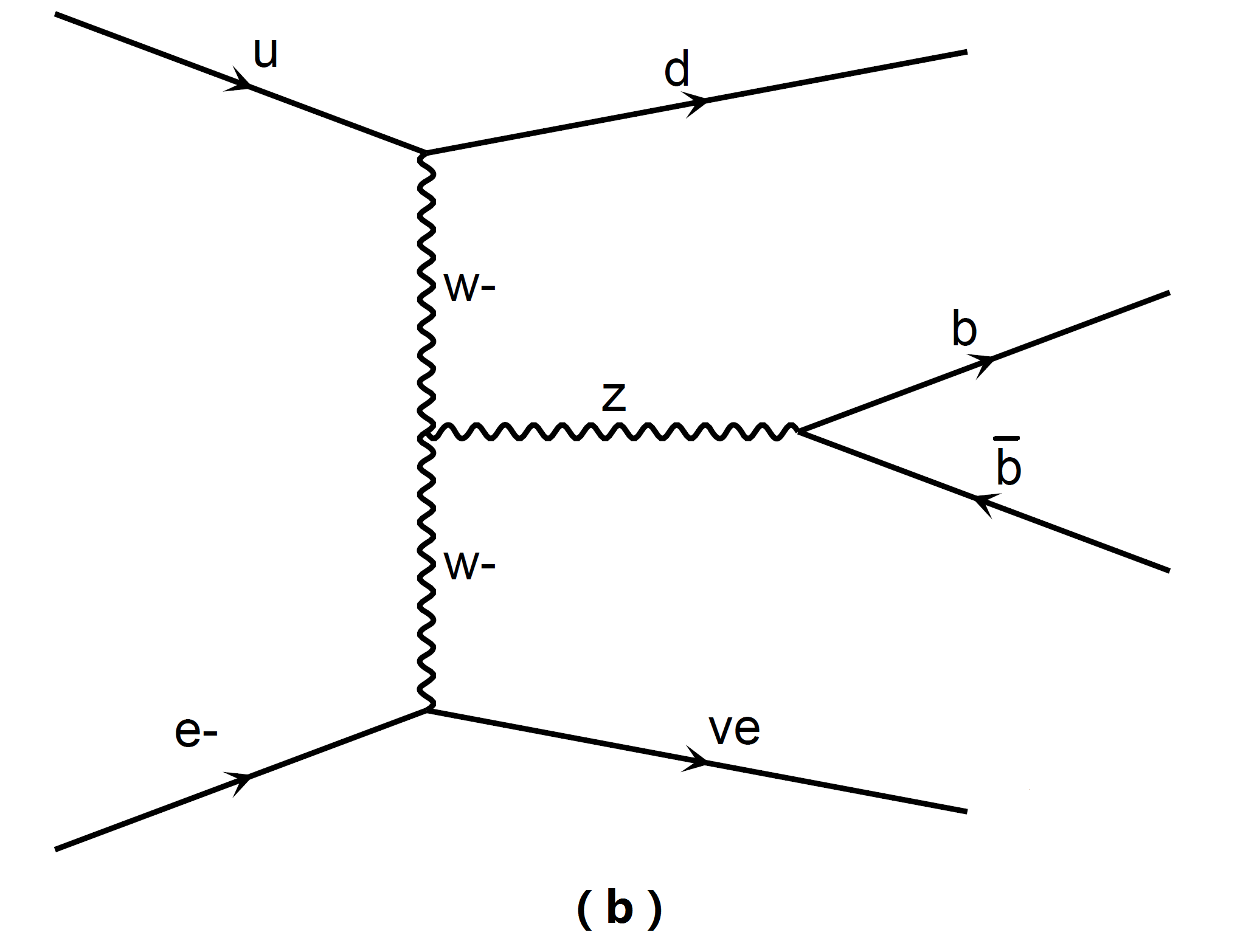}
\includegraphics[scale=0.1]{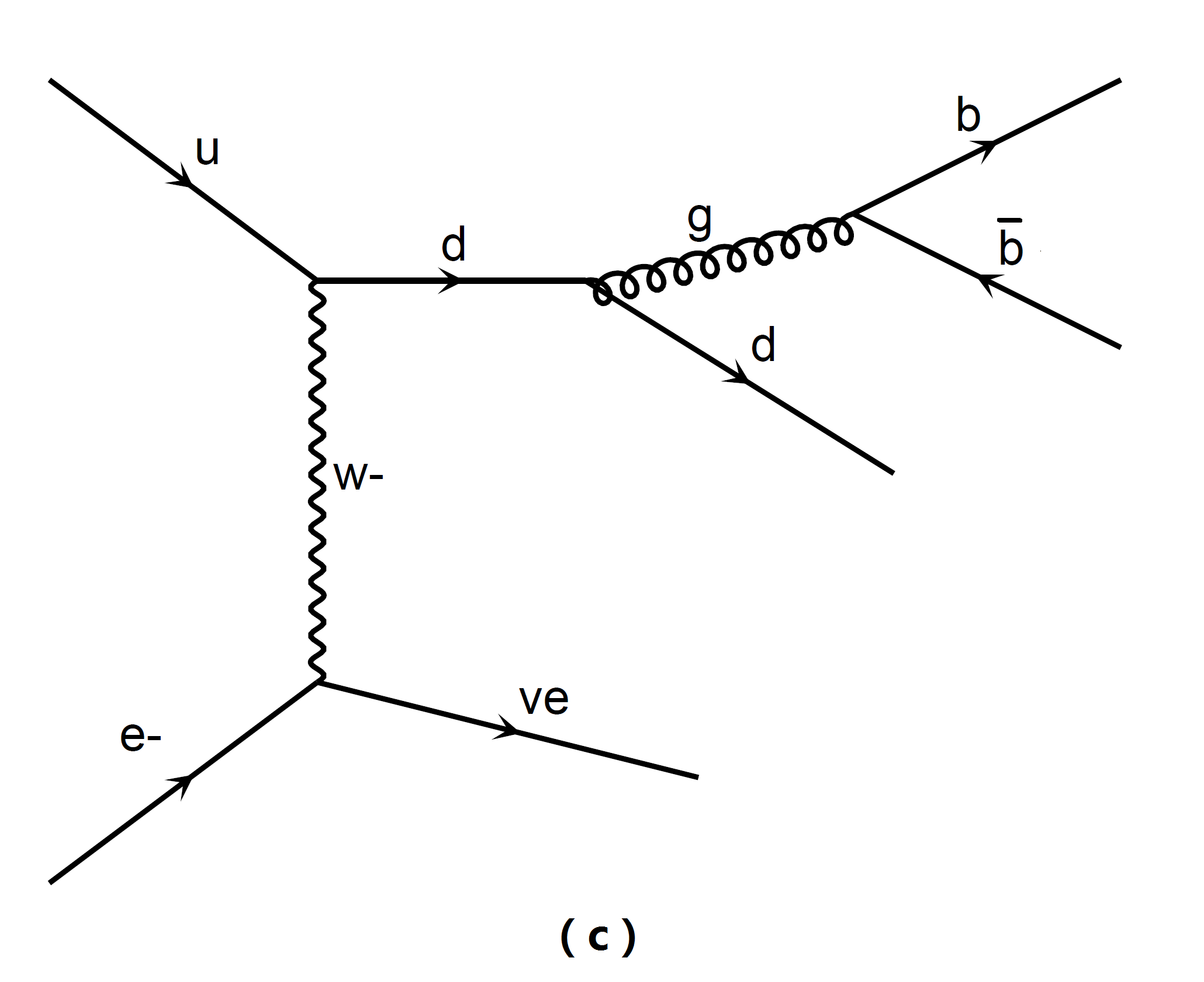}
\includegraphics[scale=0.1]{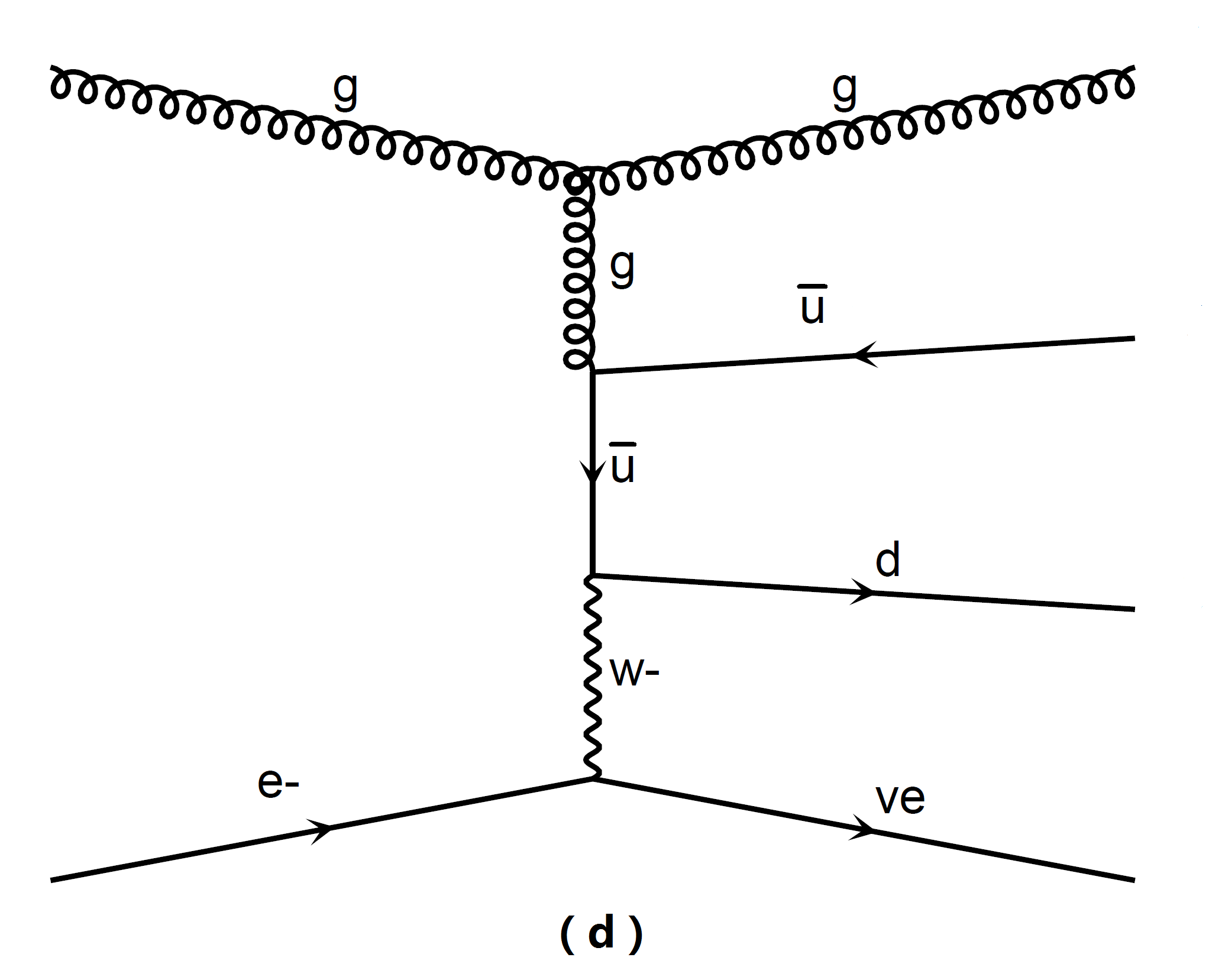}
\includegraphics[scale=0.1]{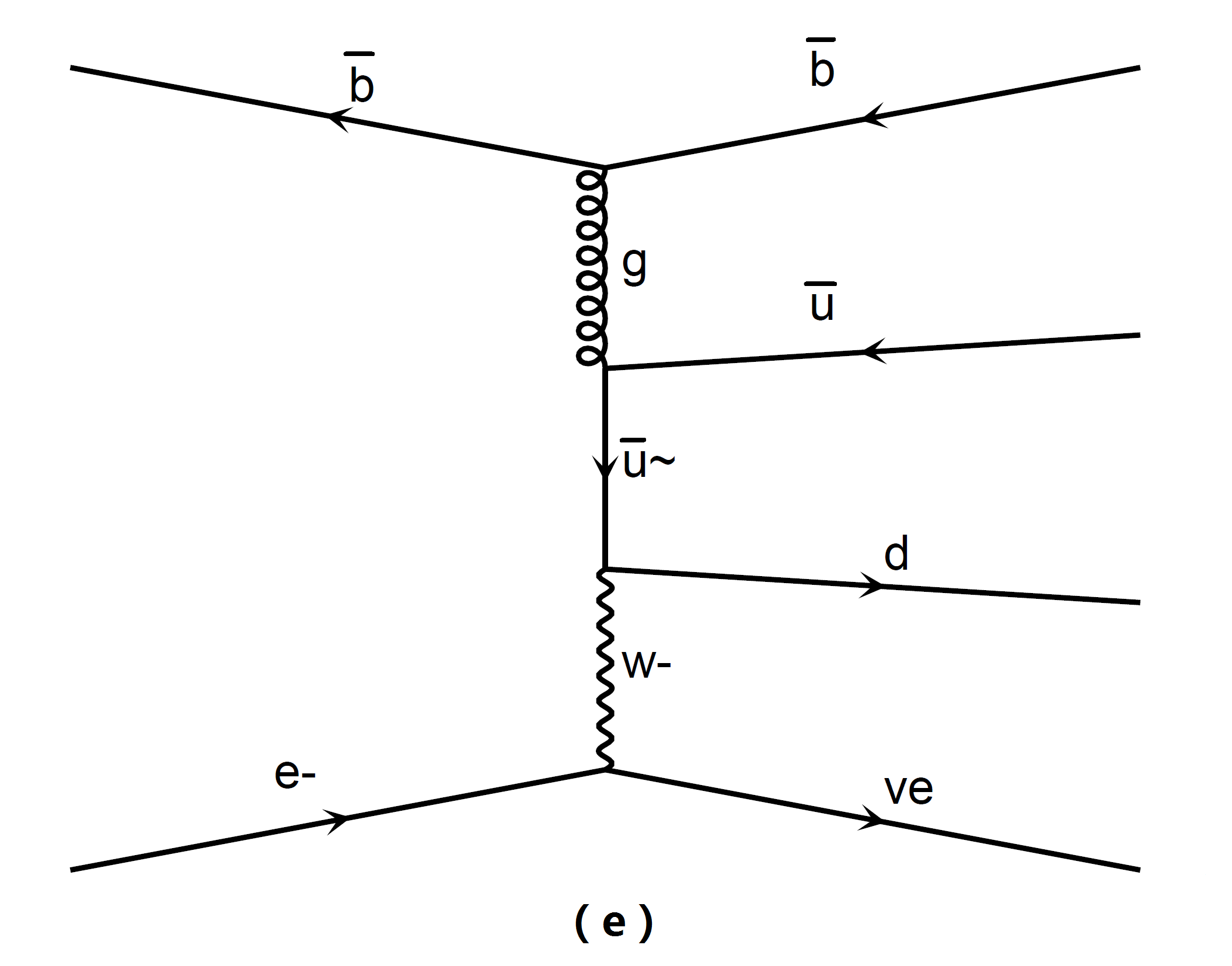}
\includegraphics[scale=0.1]{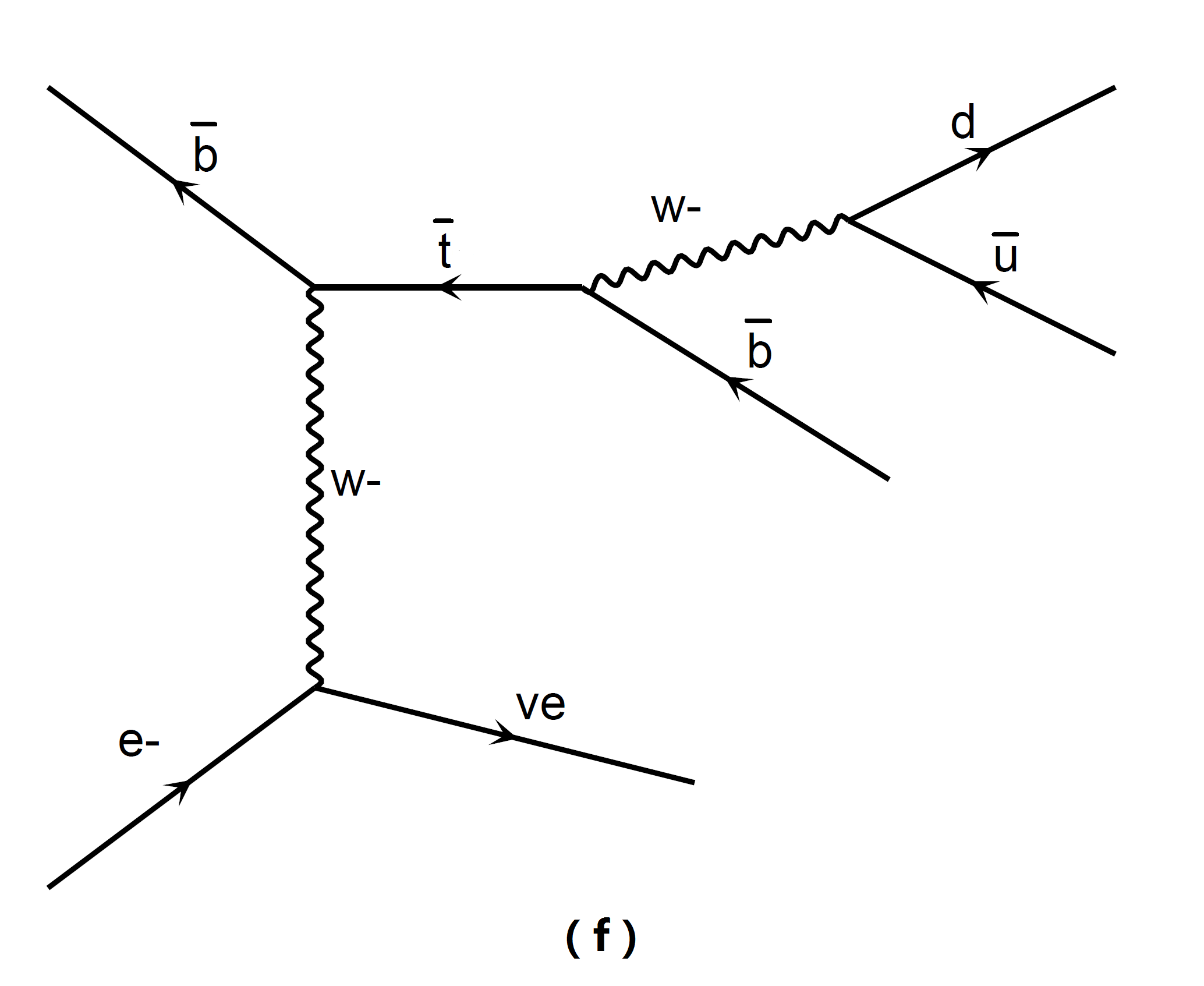}
\includegraphics[scale=0.1]{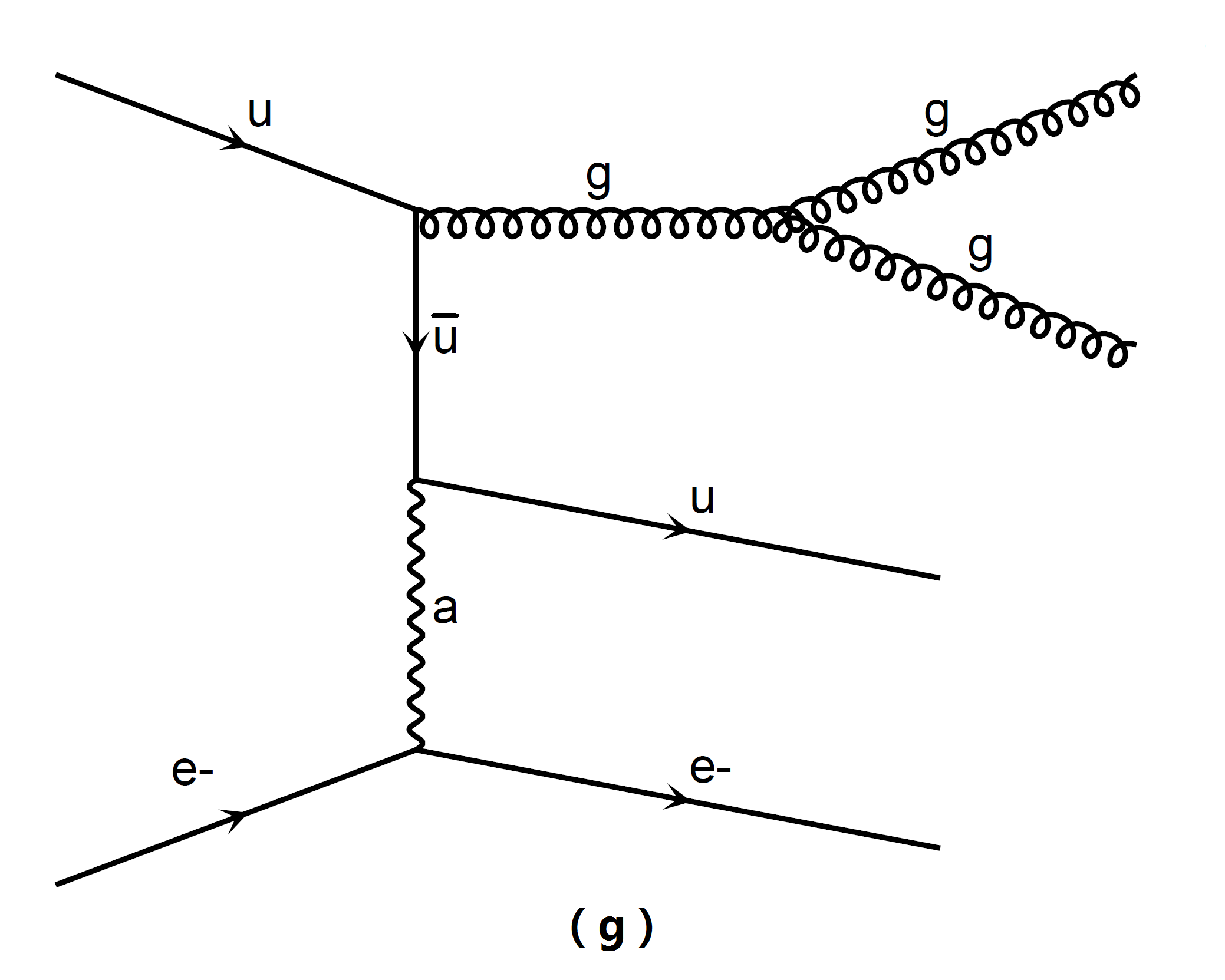}
\includegraphics[scale=0.1]{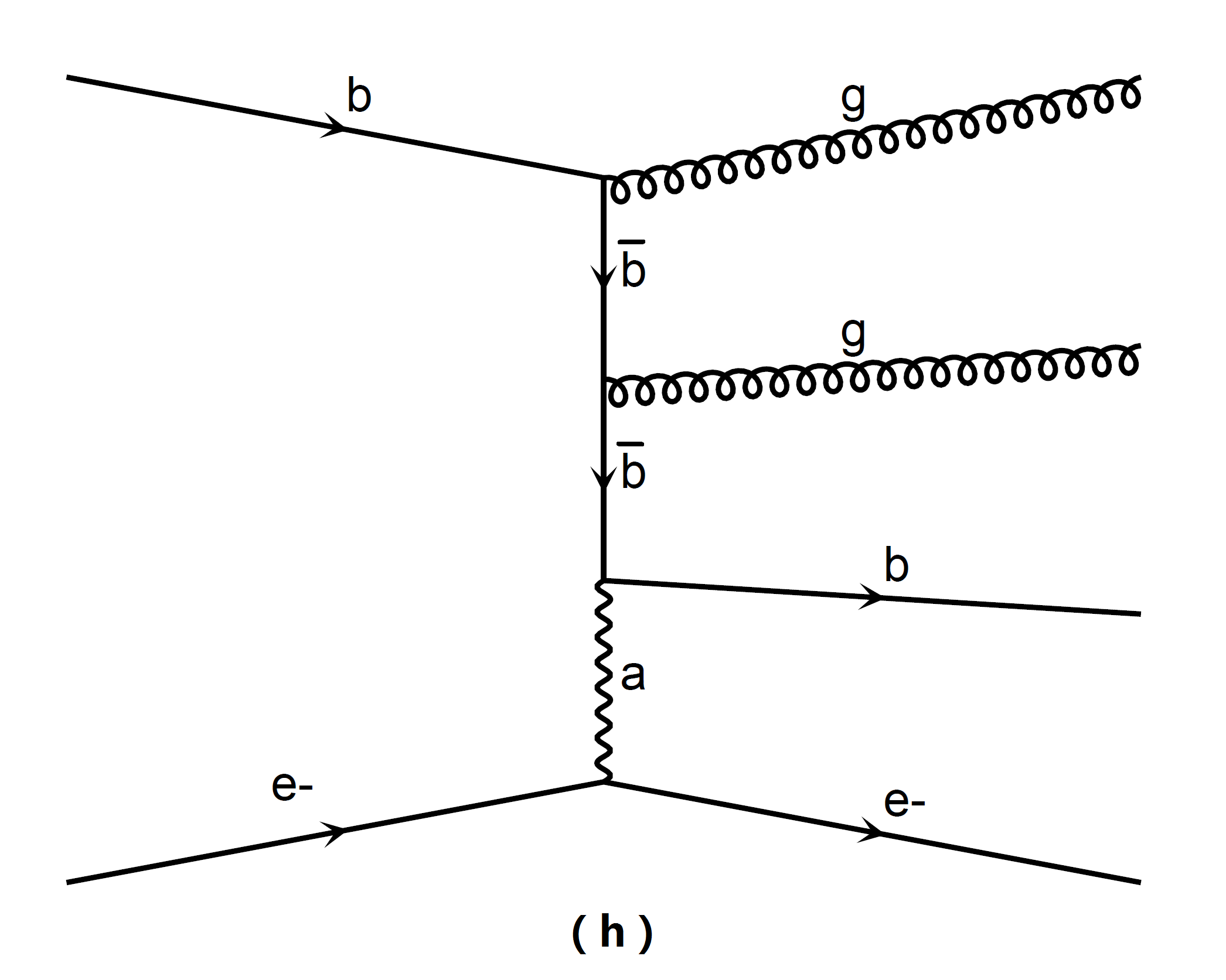}
\caption{\label{fig2_bak_Feynman}
Some examples of partonic Feynman diagrams for the reducible and irreducible backgrounds correspond to the signal.}
\end{figure}

The main backgrounds come from both the reducible and irreducible ones.
The crucial irreducible backgrounds
which yield exactly the same final states to signal.I are listed bellow. See,
\begin{eqnarray}
&&\rm e^-p \rightarrow \nu_e (h\rightarrow b \bar{b} ) j  \\
&&\rm e^-p \rightarrow \nu_e (z\rightarrow b \bar{b} ) j
\end{eqnarray}
which contain three QED couplings, are noted as ``bakh'' and ``bakz'' respectively,
\begin{eqnarray}
&&\rm e^-p \rightarrow \nu_e (g\rightarrow b \bar{b}) j
\end{eqnarray}
which contains two QED couplings and two QCD couplings, is noted as ``bakg''.
Notice here and bellow, j = g, u, $\rm \bar{u}$, d, $\rm \bar{d}$, c,
$\rm  \bar{c}$, s, $\rm \bar{s}$.
One source of the most important potentially reducible backgrounds is
\begin{eqnarray}\nonumber
&&\rm e^-p \rightarrow \nu_e jjj \\
&&\rm e^-p \rightarrow \nu_e jjb/\bar{b}
\end{eqnarray}
due to a mis-identification of one or more of the final state light jets to B-jets.
These processes contain two QED couplings and two QCD couplings as well.
We refer them as ``bakjjj'' (including ``bakg'' backgrounds).
Another source of reducible background is single top production.
As can be seen, the signal process studied in our paper
is essentially single top production at the ep collider,
followed by a particular decay chain. This means that SM single top production
and decay is an important background to our signal production under consideration.
We refer these backgrounds as ``bakt''. The production is
\begin{eqnarray}
&&\rm e^-p \rightarrow \nu_e (\bar{t} \rightarrow ( w^- \rightarrow jj) \bar{b} ).
\end{eqnarray}
The produced top quark will decay to a w boson and a B-jet.
The hadronic decay of the w boson to non-B-jets final states,
which might mis-tagged as a B-jet, make this background a dangerous one.
We have also looked into some neutral current (NC) production backgrounds:
\begin{eqnarray}\nonumber
&&\rm e^-p \rightarrow e^- jjj  \\\nonumber
&&\rm e^-p \rightarrow e^- jjb/\bar{b}   \\
&&\rm e^-p \rightarrow e^- (g\to b\bar{b}) j.
\end{eqnarray}
These are NC multi-jet backgrounds ("bakejjj") and belong to reducible ones. Applying a no-lepton selection,
they can be strongly reduced and safely ignored, thus not considered.
To be clear, we present some Feynman diagrams for the backgrounds in Fig.\ref{fig2_bak_Feynman}.
Typically, Fig.\ref{fig2_bak_Feynman} (a),(b),(c),(d, e),(f) and (g, h)
correspond to bakh, bakz, bakg, bakjjj, bakt and bakejjj respectively.
All the backgrounds listed above are also belonging to the backgrounds of signal.II,
but all are the irreducible ones, due to a mis-identification of one or more of the final state light jets to B-jets.

\subsection{The simulation}

For the simulation of the collider phenomenology,
we use FeynRules\cite{FeynRules2.0} to extract the Feynman Rules from the Lagrangian.
The model is generated into Universal FeynRules Output (UFO) files\cite{UFO}
and then fed to the Monte Carlo event generator MadGraph@NLO\cite{MadGraph5} for the generation of event samples.
We pass the generated parton level events on to PYTHIA6.4\cite{Pythia}
which handles the initial and final state parton shower, hadronization, heavy hadron decays, etc.
Then, we pass the events on to Delphes\cite{Delphes} which handles the detector effect.
The detector is assumed to have a cylindrical geometry comprising a central tracker
followed by an electromagnetic and a hadronic calorimeter.
The forward and backward regions are also covered by a tracker,
an electromagnetic and a hadronic calorimeter.
The angular acceptance for charged tracks in the pseudorapidity
range $-4.3 < \eta < 4.9$ and the detector performance,
in terms of momentum and energy resolution of electrons, muons and jets,
is based on the LHeC detector design~\cite{LHeC_2,LHeC_4}.
We use FASTJET\cite{Fastjet} for jet clustering. Jets are anti-kt clusted\cite{antiKT}
with a cone of radius $\rm \Delta R(j)=\sqrt{\Delta \eta^2+\Delta \phi^2}=0.7(0.4)$ at the LHeC(FCC-eh).
The B-jet tagging technique is applied and the C(light)-quarks mis-tagging rates as B-jets are included.
We use NN23LO1\cite{Ball:2012cx}\cite{Deans:2013mha} parton distribution functions for all event generations.
The factorisation and renormalisation scales for both the signal and the background simulation
are done with the default MadGraph5 dynamic scales.
The electron polarization is assumed to be unit for the unpolarized case,
and the results may increase by a factor of $\rm 1+P_e$ if the polarized electron beam is considered,
where $\rm P_e$ is the degree of the longitudinal polarization of the beam.
Notice this is only true for CC productions at the ep colliders,
while for NC production the results should be calculated correspondingly.
We take $\rm P_e = 0.8$ as the default value.
We take all the low flavored quarks, gluon and also the b-quark fluxes inside proton.
In our numerical calculation, the SM inputs are
$\rm \alpha_{M_Z}=1/127.9$, $\rm G_f=1.1663787\times 10^{-5}GeV^{-2}$,
$\rm \alpha_s=0.1182$, $\rm M_Z=91.1876GeV$, $\rm M_w=79.82\ GeV$, $\rm M_{top}=173.2\ GeV$ and $\rm M_h=125.09GeV$.
Typical fixed value of $\rm \kappa_{tuh}=0.1$ is chosen as the benchmark point if there is no other statement.
To estimate the event rate at parton level for the signal,
we apply the following basic pre-selections:
\begin{eqnarray}\label{generator-level-cuts}\nonumber
&&\rm   p_T^{k_0} \geq 20\ GeV,\\\nonumber
&&\rm  |\eta^{k_0}| < 10,  \ \ k_0=j, b, \ell,  \\
&&\rm  \Delta R(k_1k_2) >0.01, \ \ k_1k_2=jj, j\ell, jb, bb, b\ell,
\end{eqnarray}
where $\rm \Delta R = \sqrt{\Delta \Phi^2 + \Delta \eta^2}$ is the separation
in the rapidity($\rm \eta$)-azimuth($\rm \Phi$) plane, $\rm p_T^{j, b, \ell}$
are the transverse momentum of jets, B-jets and leptons. The cuts are defined in the lab frame.

\subsection{The cross sections and distributions}

\begin{figure}[hbtp]
\centering
\includegraphics[scale=0.27]{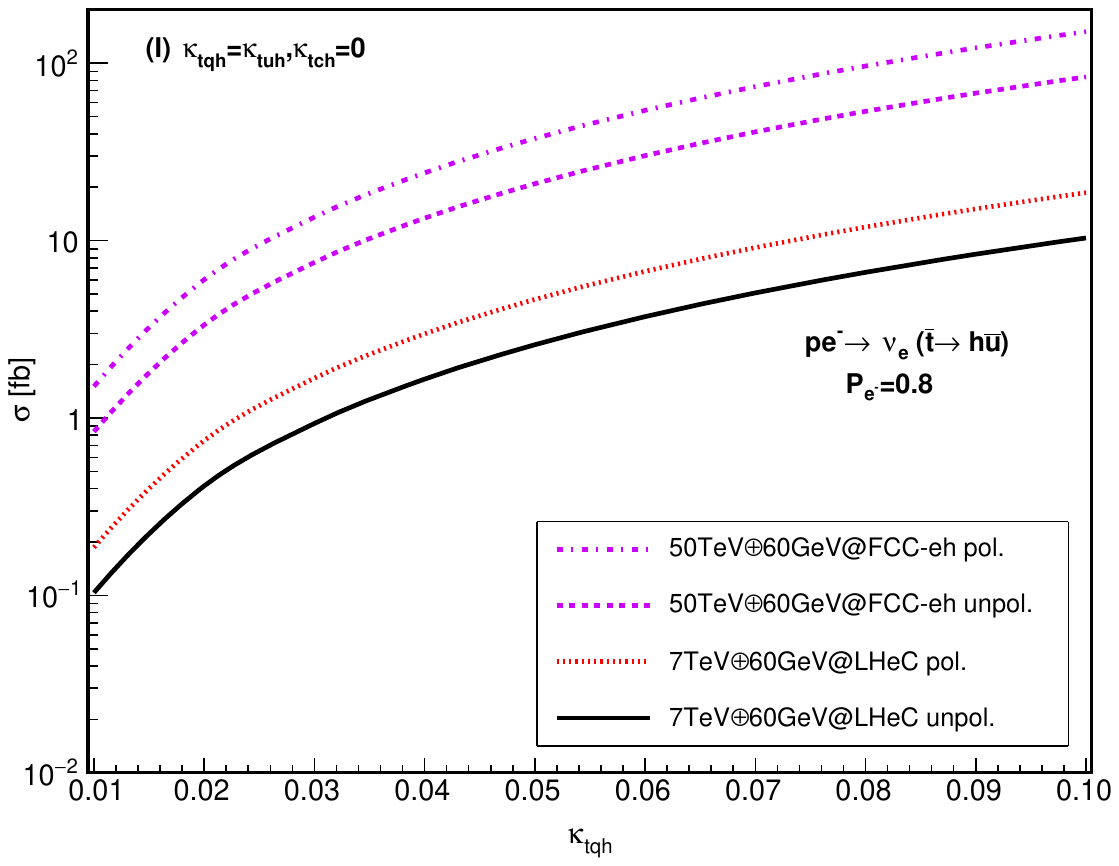}
\includegraphics[scale=0.27]{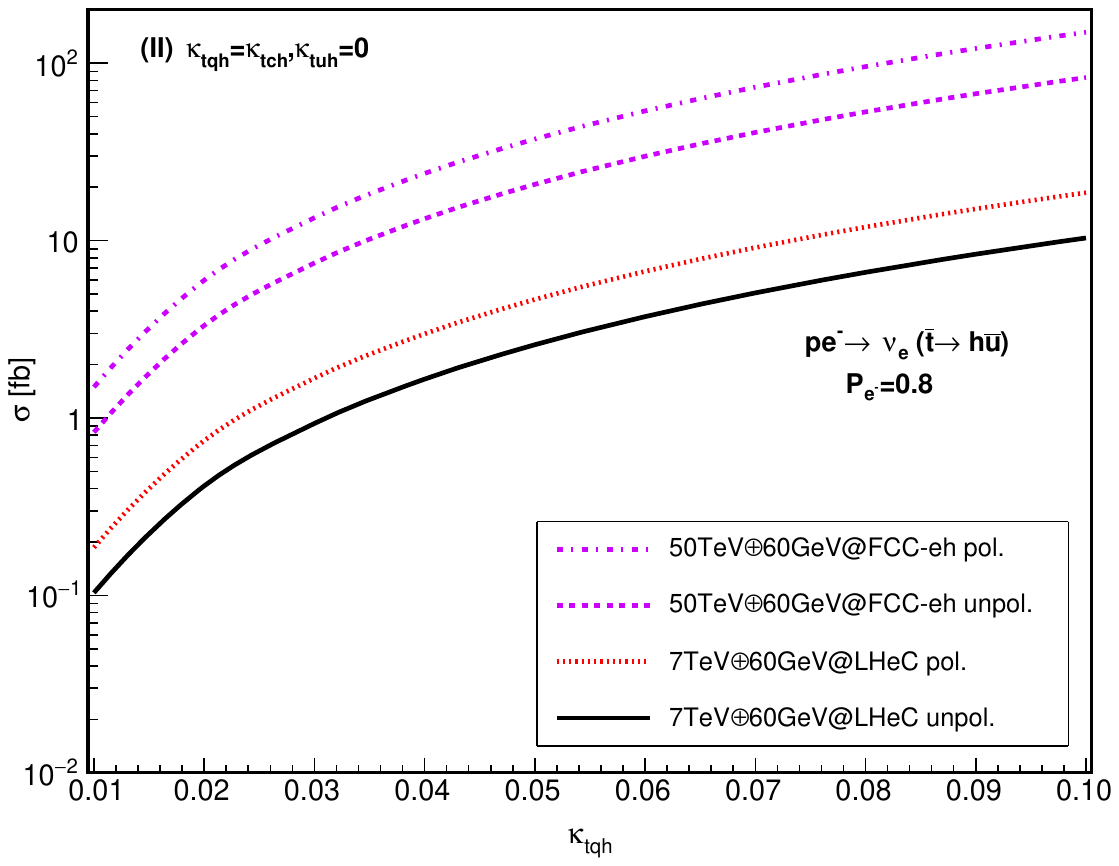}
\includegraphics[scale=0.27]{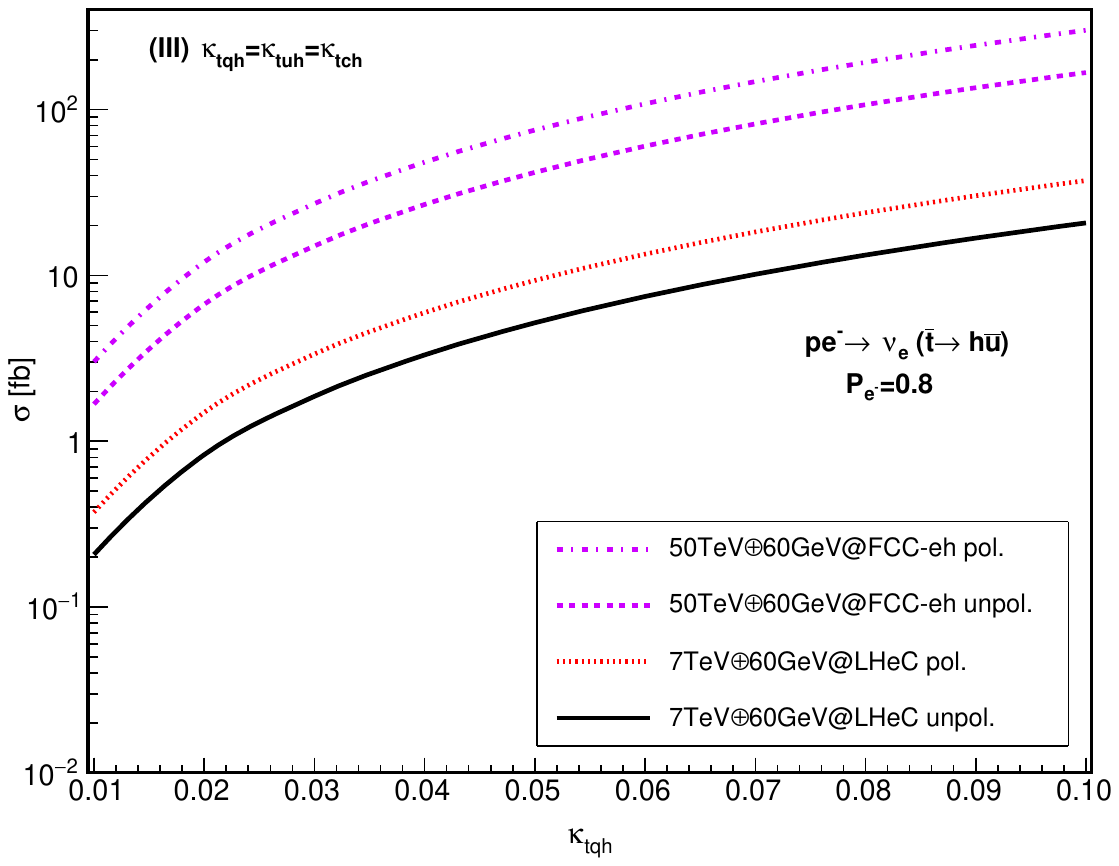}\\
\includegraphics[scale=0.27]{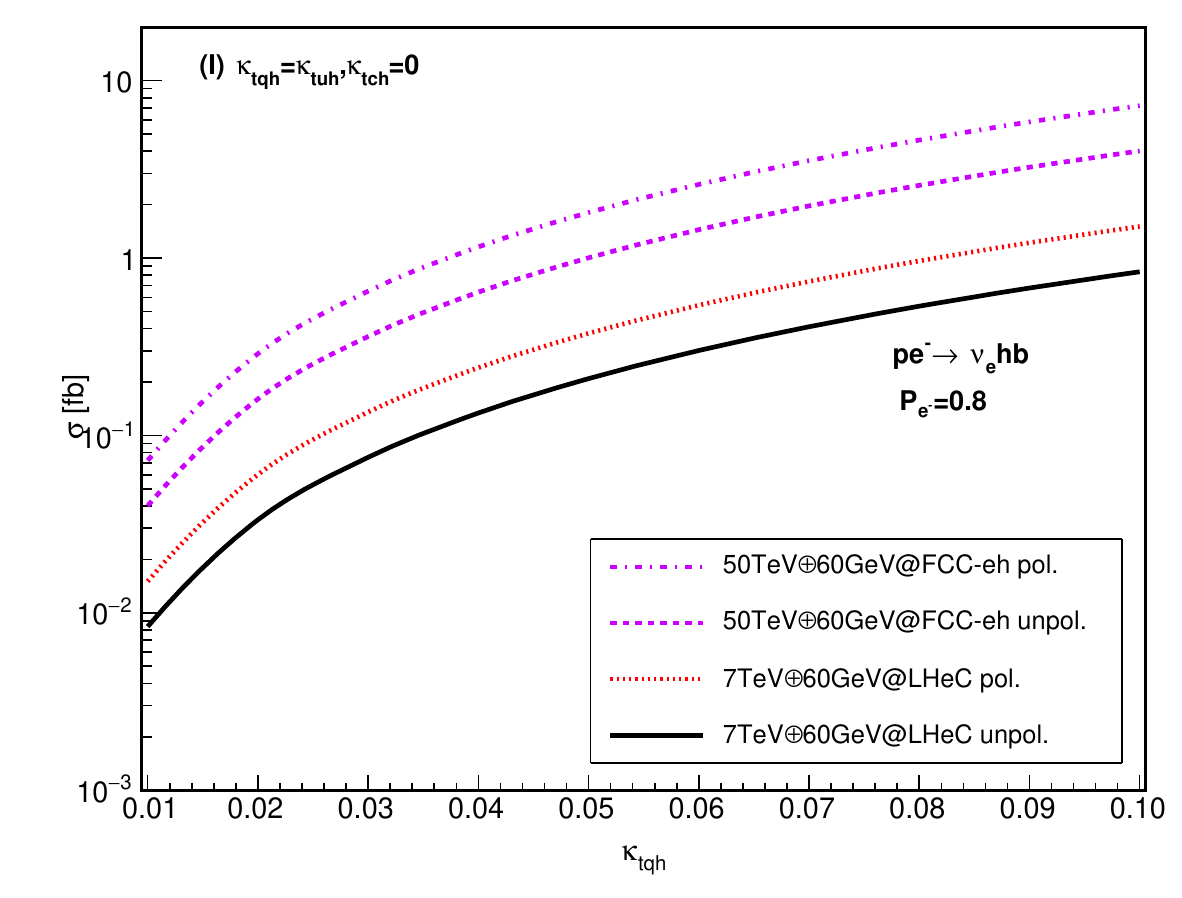}
\includegraphics[scale=0.27]{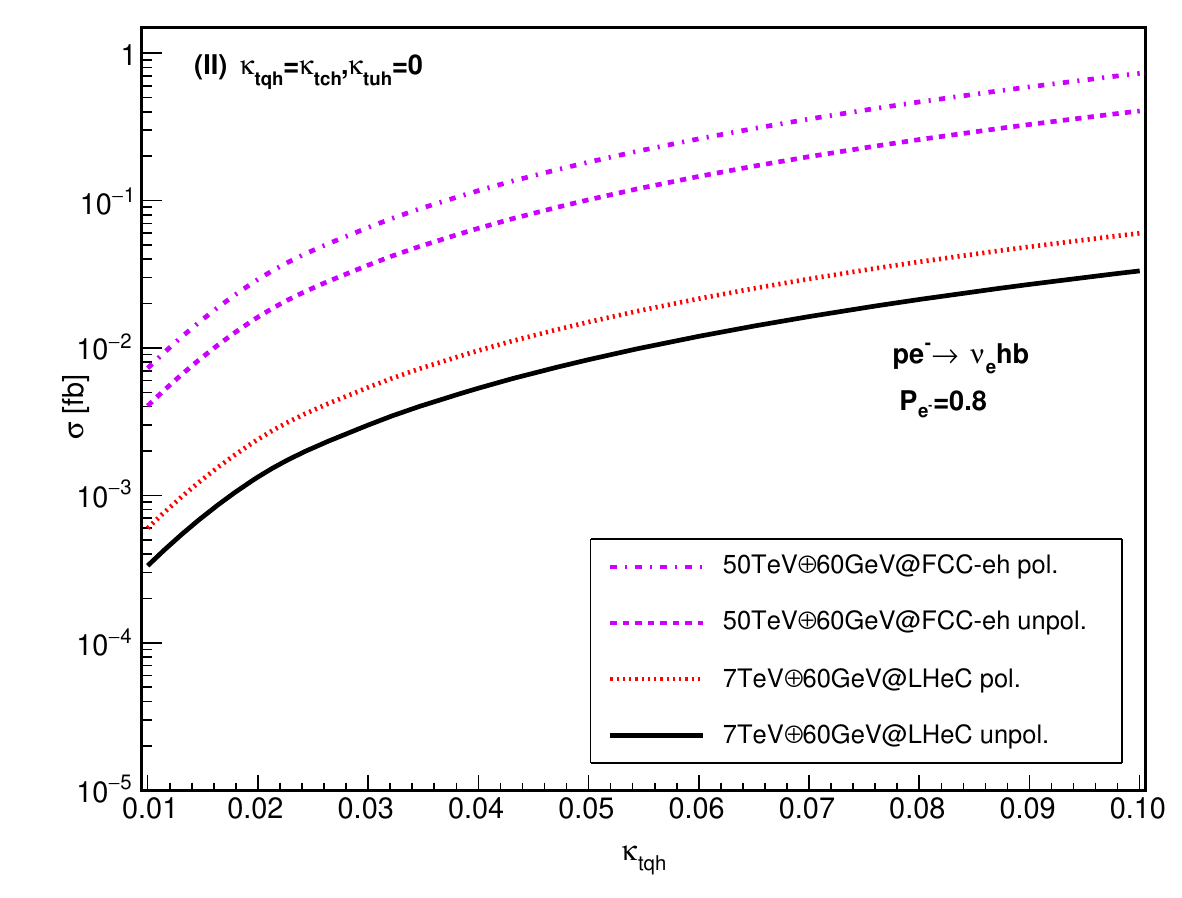}
\includegraphics[scale=0.27]{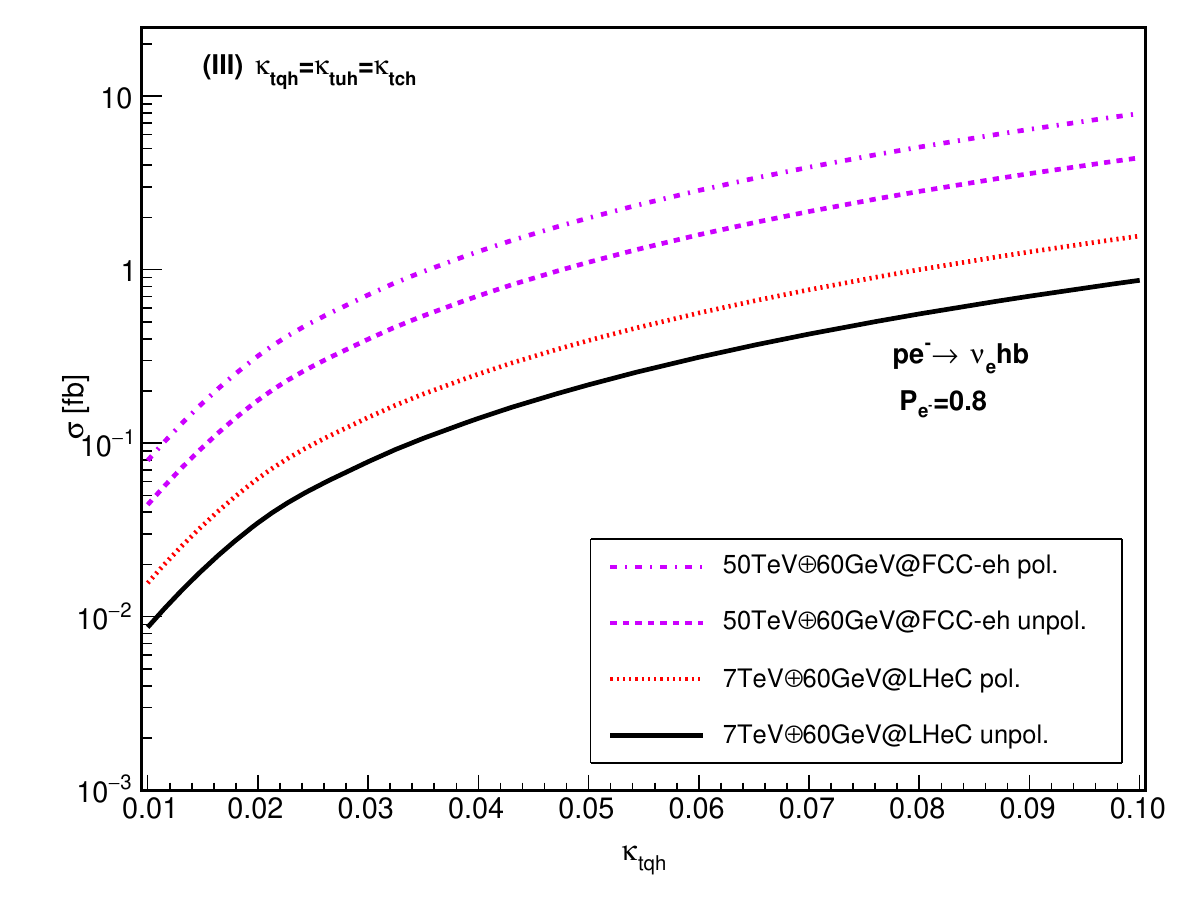}
\caption{\label{fig3_total_Xsection}
The cross sections for signal.I $\rm e^- p \to \nu_e \bar{t} \rightarrow \nu_e h \bar{q}$ and signal.II
$\rm e^- p \to \nu_e h b$ as a function of the top-Higgs FCNC couplings $\rm \kappa_{tqh}$
at the 7 TeV LHeC and 50 TeV FCC-eh. The electron beam is 60 GeV. }
\end{figure}

Before doing the full signal and background simulation,
we present the cross section of the signals (without Higgs to $\rm b\bar{b}$ decay)
in order to provide a basic idea of its production rate.
The proton beam energy is chosen to be 7(50) TeV and the electron beam is 60 GeV as proposed.
We show the dependence of the cross sections $\rm \sigma$ in units of fb
as a function of $\rm \kappa_{tqh}$ for three different cases:
\begin{itemize}
 \item (I)  $\rm \kappa_{tqh}= \kappa_{tuh}, \kappa_{tch} =0 $
 \item (II) $\rm \kappa_{tqh}= \kappa_{tch}, \kappa_{tuh} =0 $
 \item (III)$\rm \kappa_{tqh}= \kappa_{tuh} = \kappa_{tch}   $.
\end{itemize}
The results are plotted in Fig.\ref{fig3_total_Xsection}.
The first three figures present the cross sections for signal.I and the last three for signal.II.
The lower two curves are the results at the LHeC and the upper two ones are at the FCC-eh.
As shown in the figures, the production rate enhance obviously when $\rm \kappa_{tqh}$ is becoming larger.
We can see that the cross section at the FCC-eh is around ten times larger than that at the LHeC.
From the first two figures of signal.I, we can see that the $\rm t\to uh$ (case.(I)) and $\rm t\to ch$ (case.(II))
final states have similar production rates. This is not the same for signal.II, where the cross section in case.(II) is much
smaller than that in case.(I), due to the small values of the charm quark pdfs.

\begin{figure}[hbtp]
\centering
\includegraphics[scale=0.27]{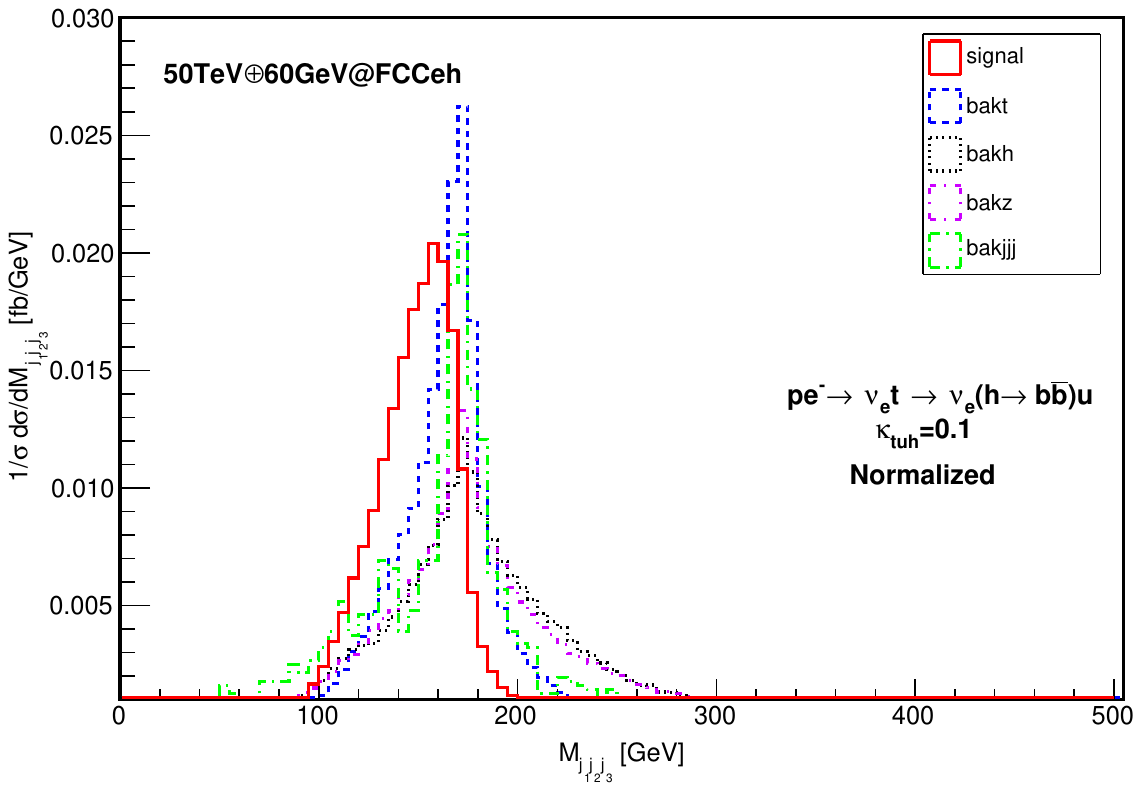}
\includegraphics[scale=0.27]{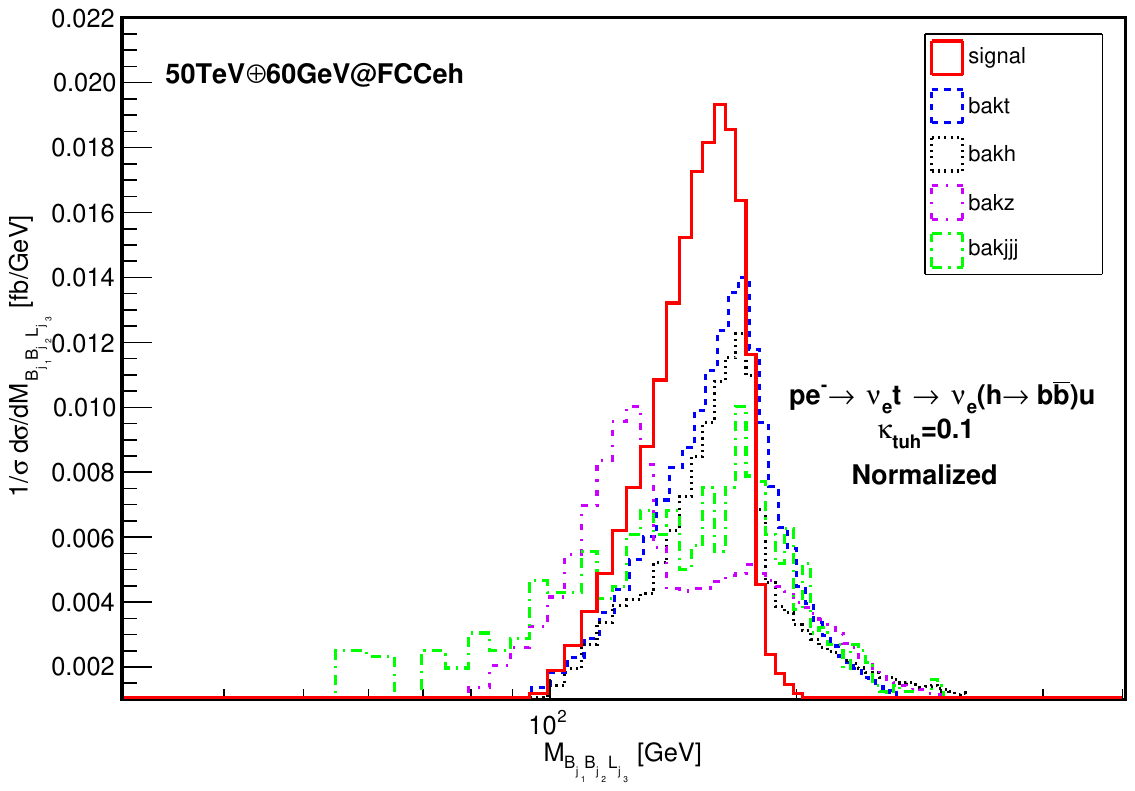}
\includegraphics[scale=0.27]{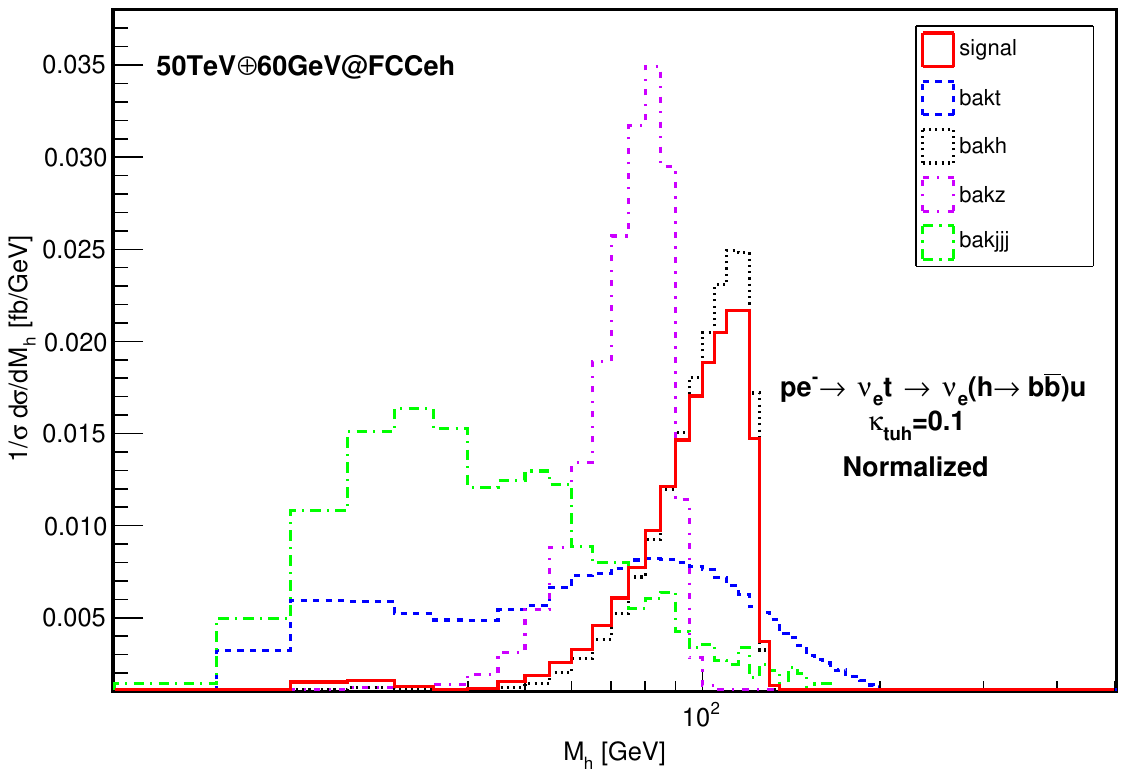} \\
\includegraphics[scale=0.27]{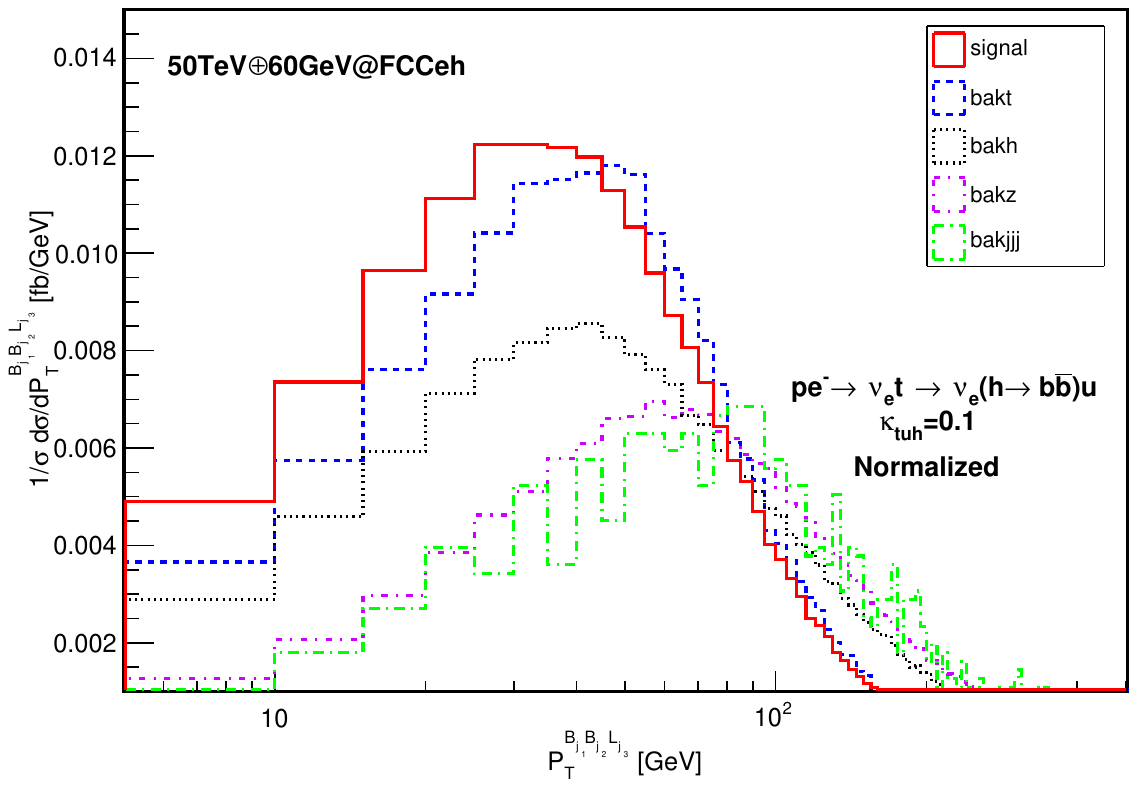}
\includegraphics[scale=0.27]{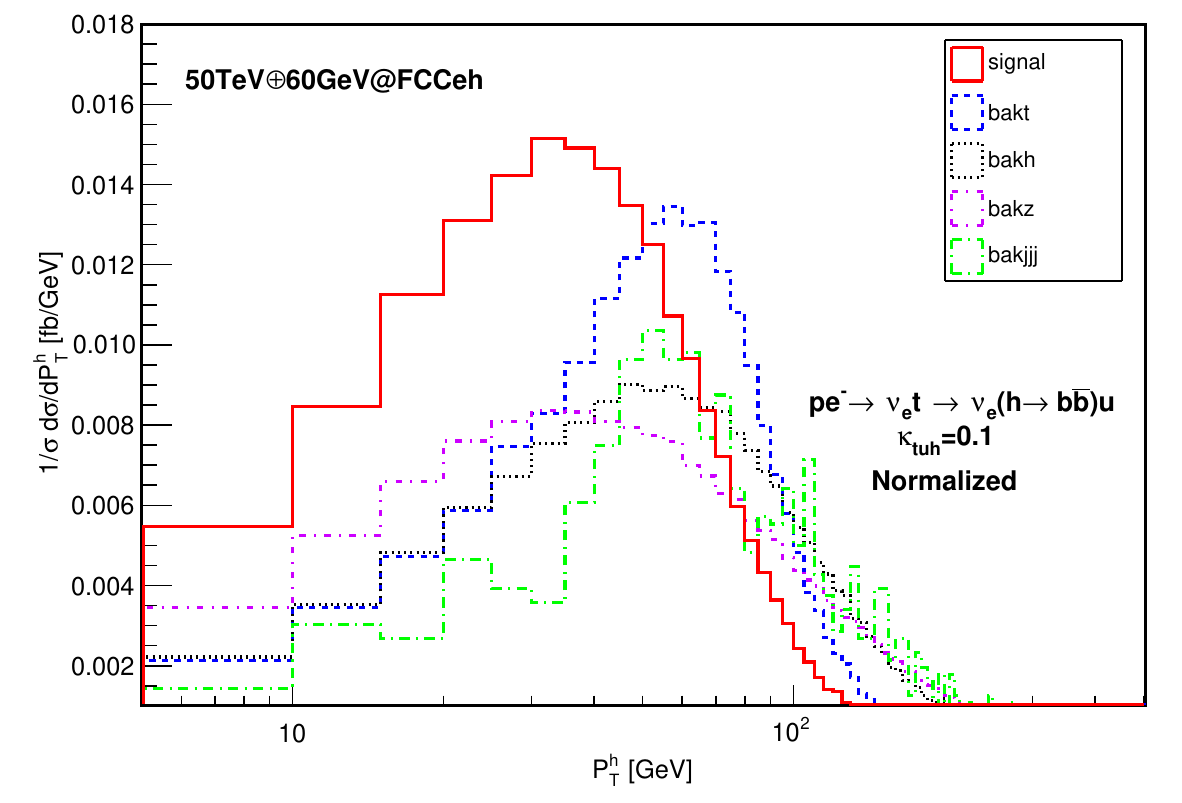}
\includegraphics[scale=0.27]{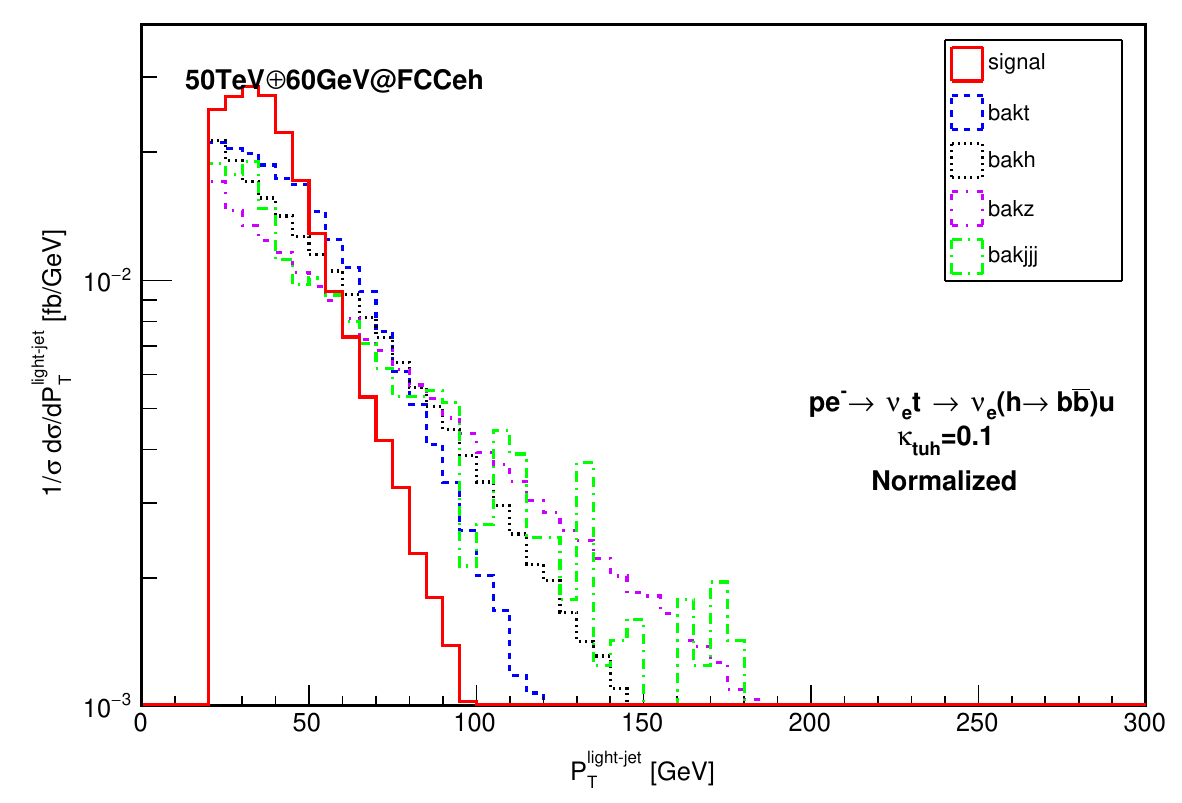}\\
\includegraphics[scale=0.27]{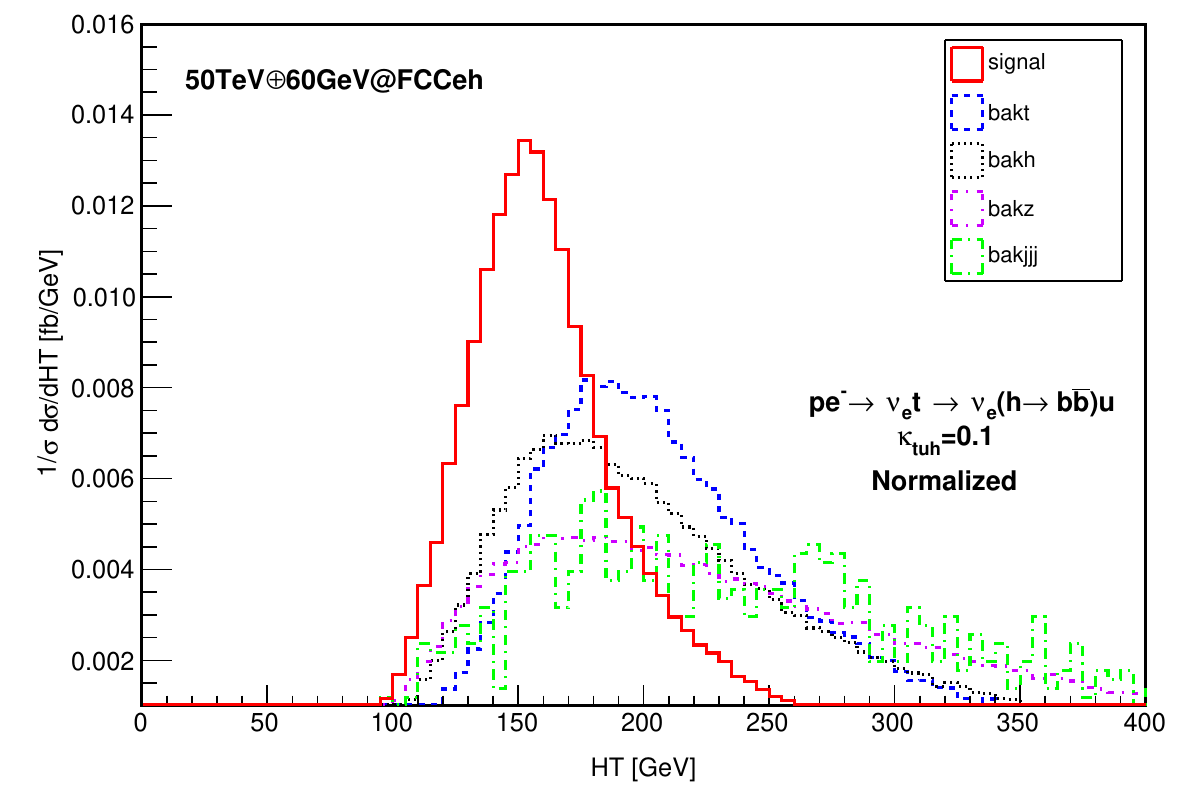}
\includegraphics[scale=0.27]{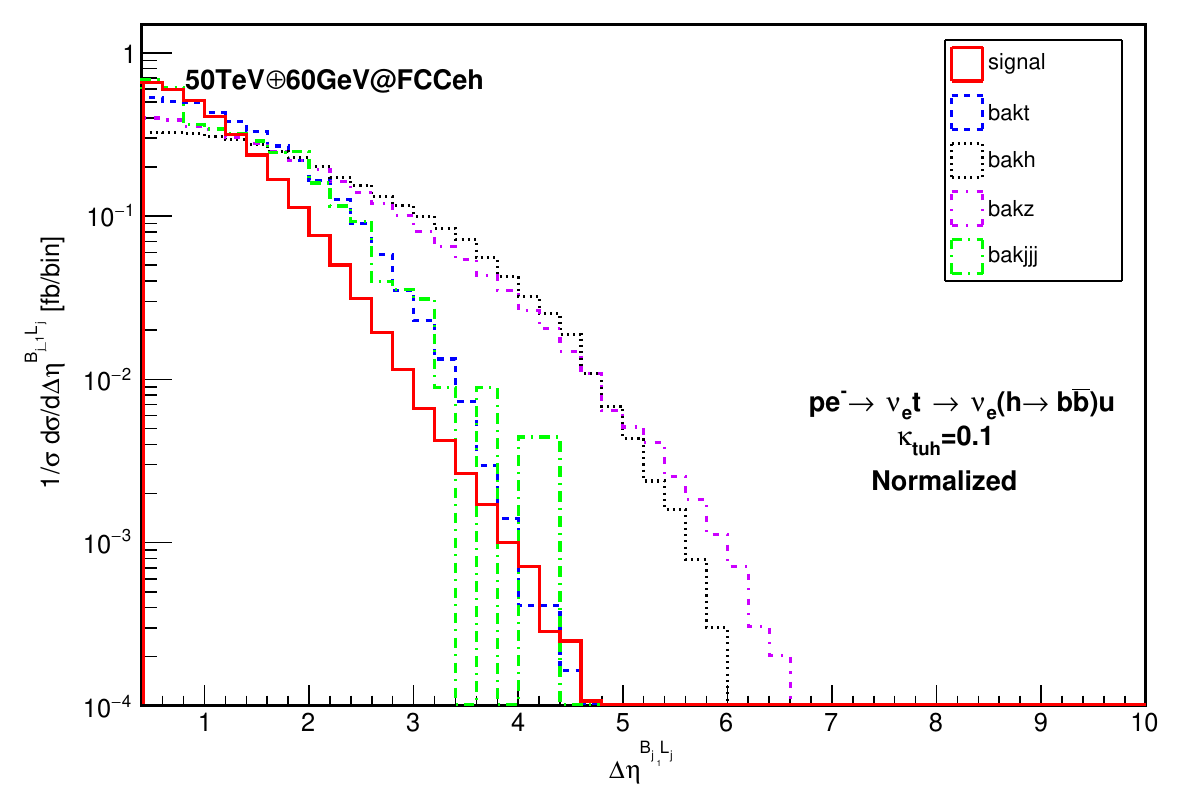}
\includegraphics[scale=0.27]{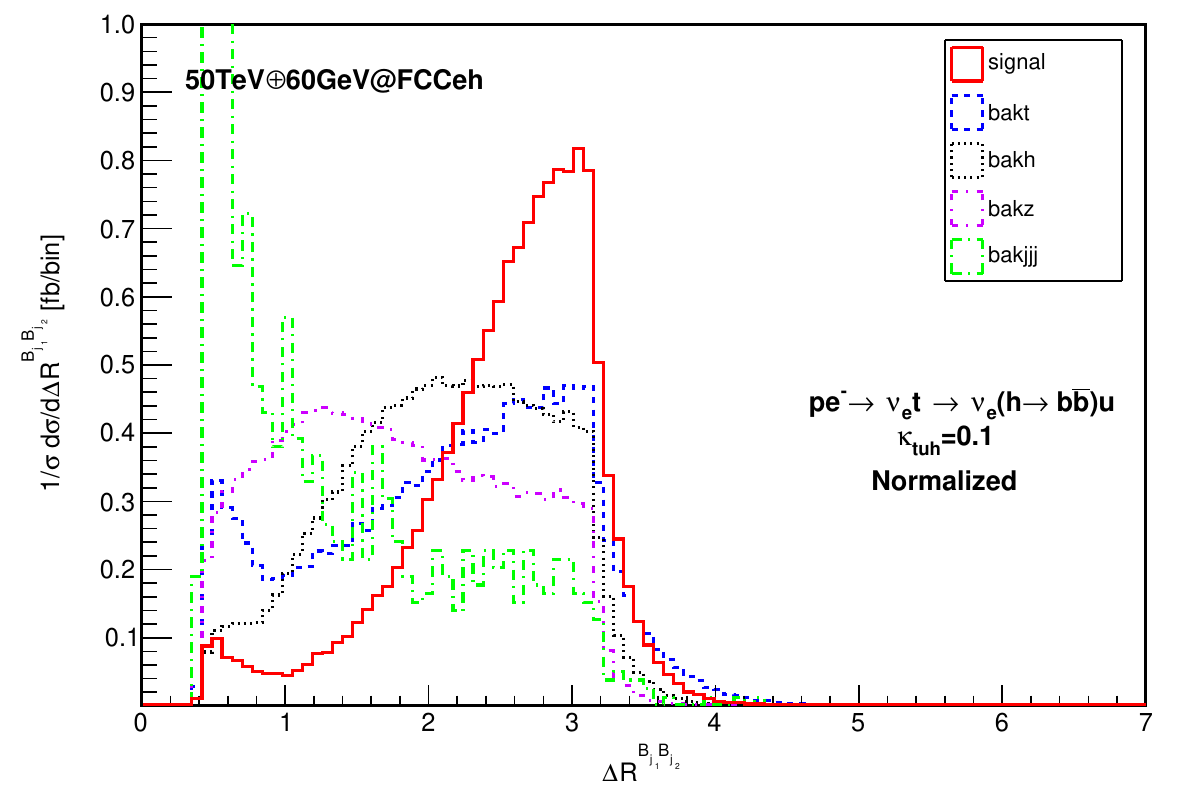}
\caption{\label{signalI_distributions}
Various kinematical distributions for signal.I and backgrounds at the FCCeh. The electron beam is 60 GeV.
Here $\rm \kappa_{tuh}=0.1$. Plots are unit normalized.}
\end{figure}

Now let's study the signal and the backgrounds at the distribution level.
After adopting the basic cuts in Eq.(\ref{generator-level-cuts}), our sample selection for signal.I is simply
\begin{eqnarray}
\rm  E^{miss}_T\ +\ 0\ \ell\ +\ \geq 3\ jets,\ \ (with\ 2\ tagged\ B-jets).
\end{eqnarray}
Taking the typical benchmark input for the signal ($\rm \kappa_{tuh}=0.1$),
the expected cross section before the selection is about 7.96(64.24) fb at the LHeC(FCC-eh),
and 1.05(18.06)fb after it.
In Fig.\ref{signalI_distributions}, we present some distributions, including
the reconstruction of the top mass ($\rm M_{top}$) and the Higgs mass ($\rm M_h$),
the transverse momentum distribution of the light jet ($\rm p_T^{light-jet}$),
the top system ($\rm p_T^{top}$) and the Higgs ($\rm p_T^{h}$),
the scalar sum of transverse momenta ($\rm HT$), as well as the rapidity separation
between the leading B-jet (one of which that reconstruct the Higgs boson) and the light jet ($\rm \Delta \eta^{B_{j_1}L_{j_3}}$),
and the rapidity-azimuth plane separation between the two B-jets that reconstruct the Higgs boson ($\rm \Delta R^{B_{j_1}B_{j_3}}$).
In order to reconstruct the top system, we use two ways. One is we choose three jets randomly,
find the three ones with their invariant mass close to the top mass ($\rm M_{j_{1}j_{2}j_{3}}$) and fill in the histogram.
The second one is we find the B jets that reconstruct the higgs boson first,
and then find the light jet from the top system similarly ($\rm M_{B_{j_1}B_{j_2}L_{j_{3}}}$).
The solid red curve is for the signal production.
We can find clear peaks around the Higgs mass and top mass,
unfortunately, both for signal and backgrounds.
The dashed blue curves, dotted black curves, short-dash-dotted violet curves and long-dash-dotted green ones
are for bakt, bakh, bakz and bakjjj, respectively. All the plots are unit normalized.
The results are for signal.I at the 50TeV$\oplus$60GeV@FCC-eh,
while for the results at the 7TeV$\oplus$60GeV@LHeC we can get similar ones thus not shown.

\begin{figure}[hbtp]
\centering
\includegraphics[scale=0.27]{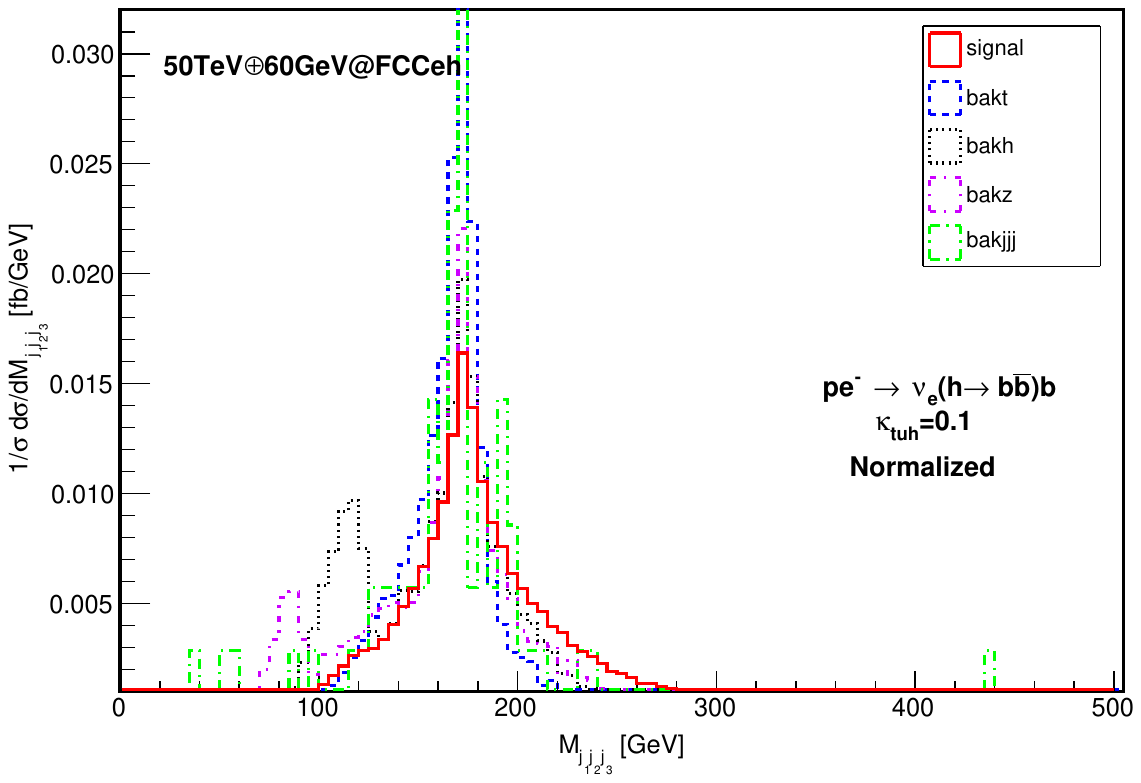}
\includegraphics[scale=0.27]{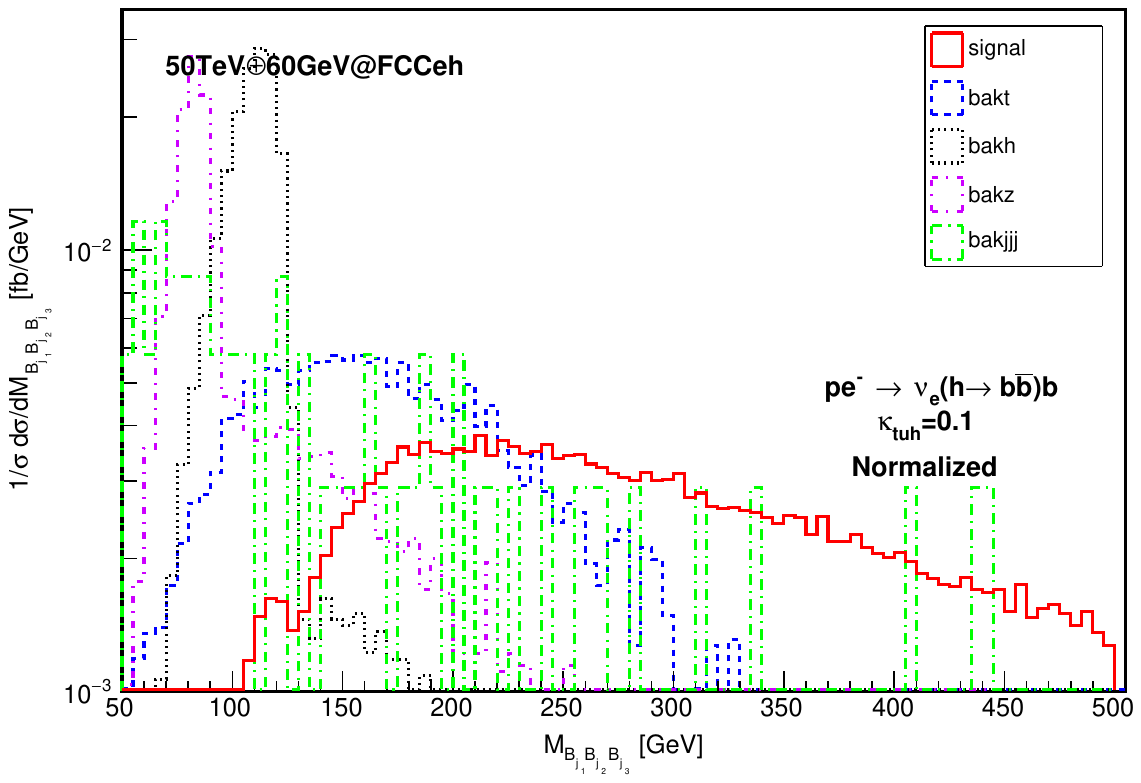}
\includegraphics[scale=0.27]{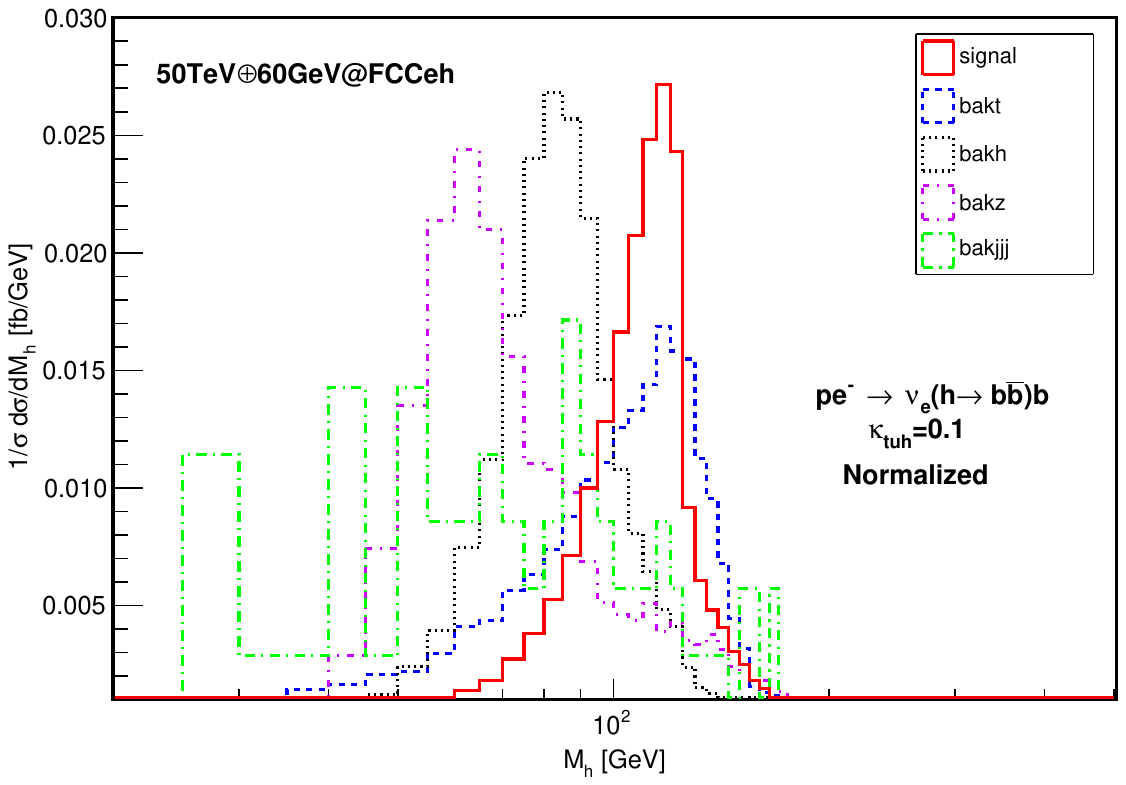} \\
\includegraphics[scale=0.27]{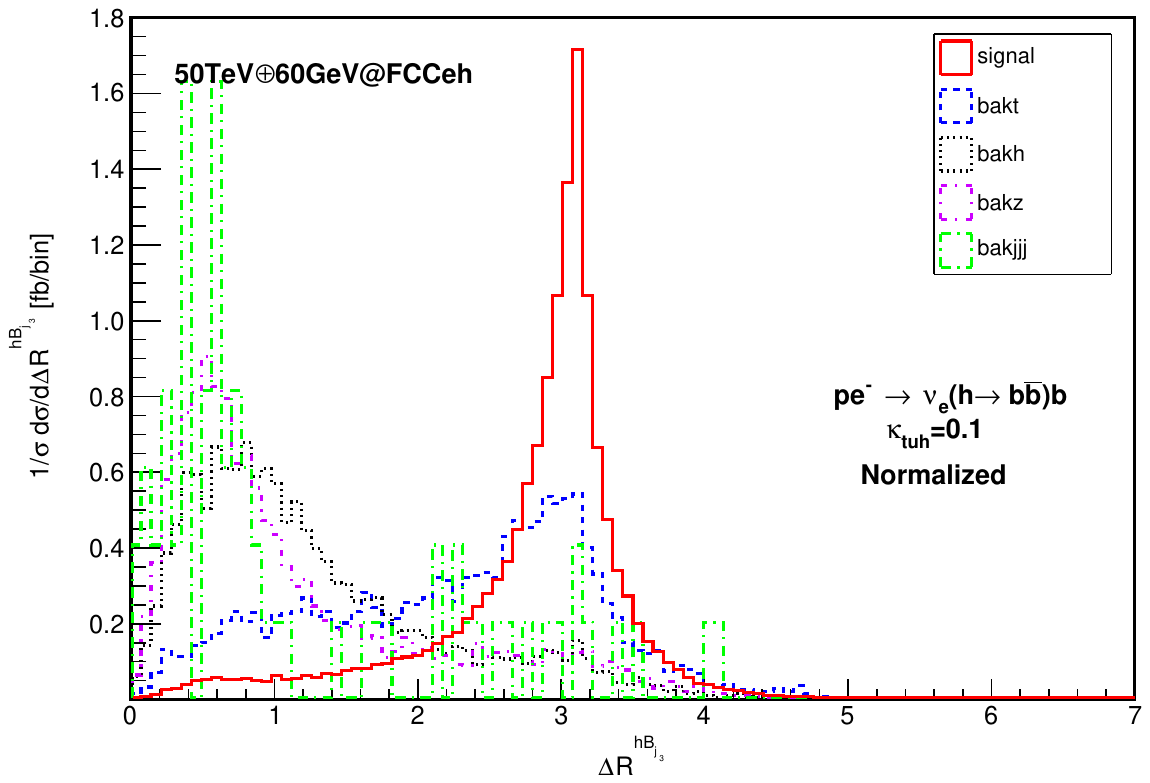}
\includegraphics[scale=0.27]{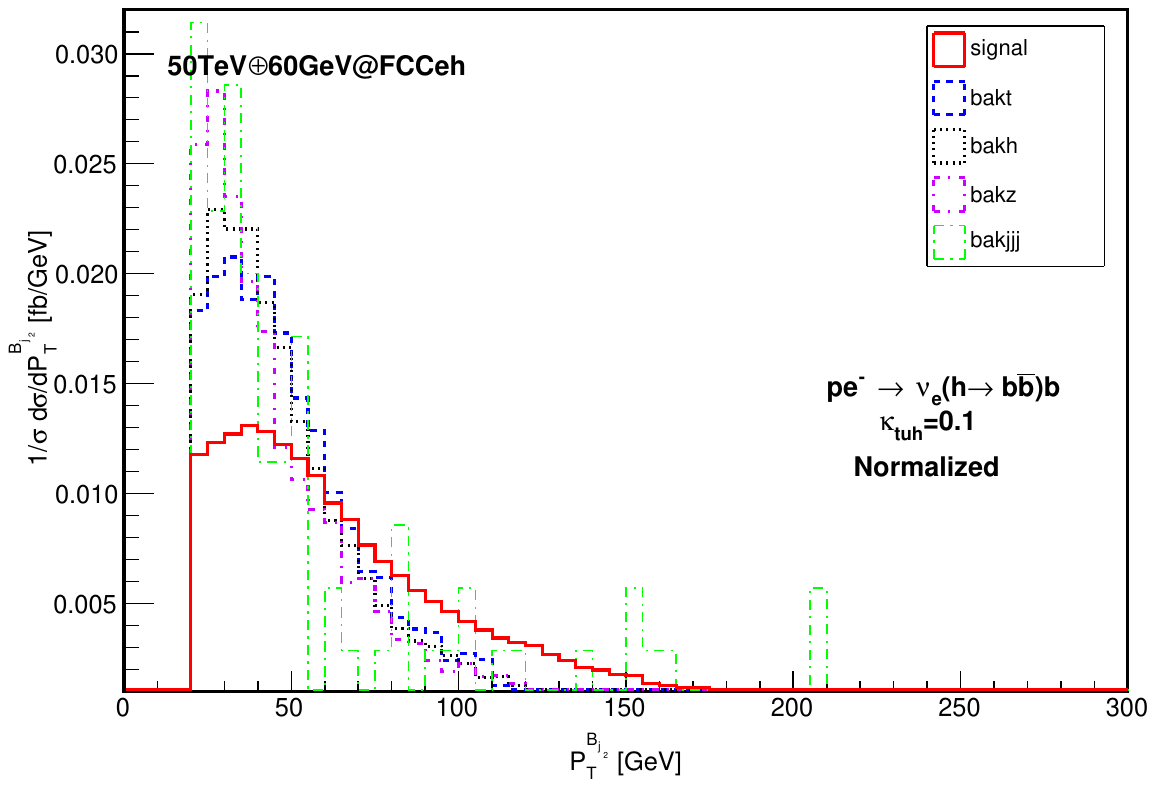}
\includegraphics[scale=0.27]{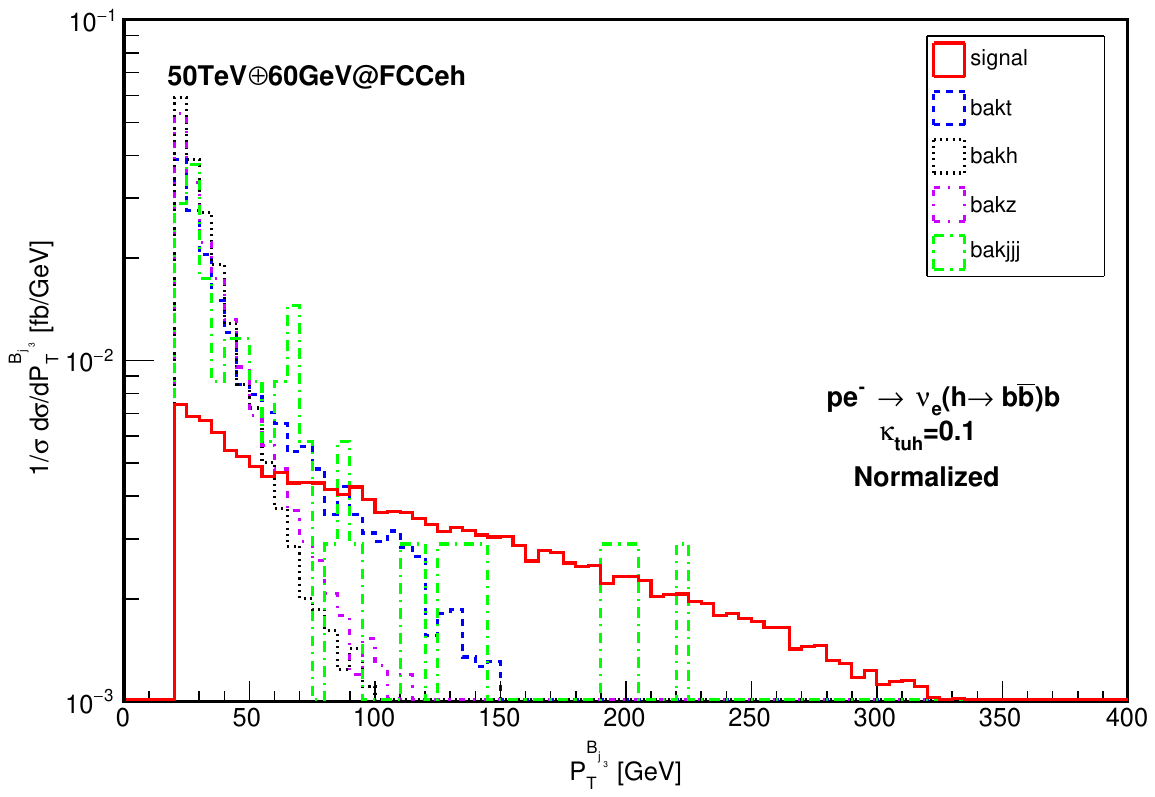}
\caption{\label{signalII_distributions}
Various kinematical distributions for signal.II
and backgrounds at the 50 TeV FCC-eh. The electron beam is 60 GeV.
Here $\rm \kappa_{tuh}=0.1$. Plots are unit normalized.}
\end{figure}

The analysis is similar for signal.II, where the selection is found directly to be
\begin{eqnarray}
\rm  E^{miss}_T\ +\ 0\ \ell\ +\ \geq 3\ jets,\ \ (with\ 3\ tagged\ Bjets).
\end{eqnarray}
We choose the distributions which show good potential to separate the signal over backgrounds
and plot them in Fig.\ref{signalII_distributions}.
The first one is to reconstruct the top system using three random jets.
We expect, and do find a similar peak for the signal as for bakz and bakh backgrounds.
However, there seems still overlap between the signal and bakt background.
The second distribution is the invariant mass of the three B-jets system, instead of calling it the top system.
In this case, the signal and bakt background are no longer a peak but a bump, with some long tails especially for the signal.
Choosing two of the B-jets, we can reconstruct the Higgs boson in the third distribution.
The forth and fifth ones are $\rm \Delta \Phi^{hB_{j_3}}$ and $\rm p_T^{B_{j_{2}}}$ distributions where $\rm B_{j_{1(2)}}$
is one of the B-jets that reconstruct the Higgs boson.
The most interesting distribution is the last one, $\rm p_T^{B_{j_{3}}}$,
the transverse momentum distributions for the B-jets which does not belong to anyone of the B-jets that reconstruct the Higgs.
There is a long tail for the signal in the large pt region,
while for all the other backgrounds, they seems prefer a forward B-jet (or a forward jet that faked as a B-jet),
which drops quickly in the low pt regions.

\subsection{The selections and discovery potential at the ep colliders}

\subsubsection{The comparison between the two signal channels}

We choose the optimized selections depending on the behavior of the distributions.
For signal.I, the optimized cuts are the mass windows of $\rm M_h$, $\rm M_{top}$($\rm M_{j_1j_2j_3}$), and the cuts on $\rm HT$.
The cross section and significance dependence on the cut flows are shown in Table.\ref{SBaftercuts_signal1}.
Notice when the colliding energy is different, the order of the optimized selections
(corresponding to the significance from small to large) and the values of cuts are not exactly the same.
Which one is the best cut in each step and what is the corresponding significance,
are determined in a somehow automatic way, relying on the machine computation.
Finally, we find that, with the integrated luminosity of 1$\rm ab^{-1}$ at the
7TeV$\oplus$60GeV$@$LHeC(50TeV$\oplus$60GeV$@$FCC-eh), the significance is 5.9(30.0) for signal.I($\rm \kappa_{tuh}=0.1$).
Here the significance is calculated by the following formula:
\begin{eqnarray}
\rm {\cal{SS}} = \sqrt{2[(n+b)\log(1+\frac{n}{b})-n]}
\end{eqnarray}
where n is the number of events and b is the number of backgrounds evaluated with the corresponding integrated luminosity.
We also find that among all the backgrounds, the bakt is indeed the most dangerous one,
and accounts for more than $70\%$ of the total backgrounds.
The question about how to suppress the single top background efficiently at the ep collider would be interesting to think of,
where Ref.\cite{tqh_LHC_th_Greljo} may give some ideas. We leave this into a deeper study in the future work.
\begin{table}[htbp]
\begin{center}
\begin{tabular}{c | c c c c c c c c c}
\hline\hline
\makecell {7TeV$\oplus$60GeV$@$LHeC \\ unpol. } & \makecell{ $\rm \sigma_{ini}$ \\  Basic cuts }           &  & \makecell { $\geq$3 jets with \\ 2 tagged Bjets}
& \makecell{ $\rm M_{j_1j_2j_3}\in$ \\ $[110, 180]$ }  &   \makecell{ $\rm M_{h}\in $ \\ $[105, 130]$ }    & \makecell{ HT $\in$ \\ $[60, 185]$  }   \\
\hline
signal.I[$\rm \kappa_{tqh}=0.1$]  &   7.96              &  &  1.05     & 0.87  &   0.48    & 0.4  \\
bakt       &   1321                                     &  &  60.9     & 33.82 &   6.4     & 3.33 \\
bakh       &   92.27                                    &  &  15.8     & 3.27  &   1.32    & 0.82 \\
bakz       &   70.73                                    &  &  10.0     & 2.88  &   0.08    & 0.03 \\
bakjjj     &   21730                                    &  &  14.7     & 6.87  &   0.70    & 0.22 \\
\hline
Total BG   &     -                                      &  & 101.4     & 46.84 &   8.5     & 4.4 \\
\hline
$\cal {SS}$[$\rm1 ab^{-1}$] & -                         &  & 3.28      &  4.0  &   5.19    & 5.9  \\
\hline\hline
\makecell {50TeV$\oplus$60GeV$@$FCC-eh \\ unpol. } &   \makecell{ $\rm \sigma_{ini}$ \\  Basic cuts }           &  & \makecell { $\geq$3 jets with \\ 2 tagged Bjets}
&  \makecell{ HT $\in$ \\ $[60, 175]$  }    &   \makecell{ $\rm M_{h}\in $ \\ $[90, 125]$ }    &  \makecell{ $\rm M_{j_1j_2j_3}\in$ \\ $[125, 170]$   }  \\
\hline
signal.I[$\rm \kappa_{tqh}=0.1$]  &   64.24             &  & 18.06      & 11.92  & 7.9       & 6.24 \\
bakt       &   10660                                    &  & 1296.45    & 328.2  & 74.2      & 34.24 \\
bakh       &   507.9                                    &  & 168.36     & 54.15  & 35.3      & 5.58 \\
bakz       &   357                                      &  & 104.88     & 25.97  & 1.33      & 0.32 \\
bakjjj     &   90070                                    &  & 203.20     & 41.79  & 1.98      & 1.08 \\
\hline
Total BG   &     -                                      &  & 1772.89    & 450.11 & 112.81    & 41.22 \\
\hline
$\cal {SS}$[$\rm 1\ ab^{-1}$] & -                       &  & 13.54      & 17.7   & 23.3      & 30.0 \\
\hline\hline
 \end{tabular}
 \end{center}
 \caption{\label{SBaftercuts_signal1}
Cross sections (in unit of fb) and significance depending on the cut flows for signal.I
$\rm \ e^- p \to \nu_e \bar{t} \to \nu_e h \bar{q} \to \nu_e b\bar{b} \bar{q}$ ($\rm \kappa_{tuh}=0.1$)
and backgrounds at the 7TeV$\oplus$60GeV$@$LHeC and 50TeV$\oplus$60GeV$@$FCC-eh.
$\rm {\cal {SS}}$ is evaluated with $\rm 1\ ab^{-1}$ integrated luminosity.
Polarization effects and systematic uncertainty are not considered yet.}
\end{table}

The optimized selections for signal.II include $\rm p_T^{B_{j_{(2,3)}}}$,
$\rm \Delta R^{hB_{j_{3}}}$ and mass windows of $\rm M_h$.
The cut flow dependence is shown in Table.\ref{SBaftercuts_signal2}.
Compare to signal.I, signal.II has one clear advantage, say,
the three tagged B-jets selection can reduce the backgrounds strongly.
However, its small production rate prominent it's disadvantage,
only 0.64(3.085) fb at the LHeC(FCC-eh) after the basic sample selections.
Considering $\rm 1\ ab^{-1}$ luminosity, the significance is calculated to be 4.02(16.7),
not small, showing good potential in the measurement of the anomalous tqh couplings.
Actually, soon we may find its discovery potential is already comparable to(at the LHeC)
or even better(at the FCC-eh) than signal.I.
\begin{table}[htbp]
\begin{center}
\begin{tabular}{c | c c c c c c c c c}
\hline\hline
\makecell {7TeV$\oplus$60GeV$@$LHeC \\ unpol. } &   \makecell{ $\rm \sigma_{ini}$ \\  Basic cuts }      &     & \makecell { 3 tagged \\ Bjets }
&   \makecell{ $\rm p_{T}^{B_{j_3}} \in $ \\ $[200, 480]$ }    &  \makecell{ $\rm M_{h}\in$ \\ $[100, 140]$   }
&   \makecell{  $\rm p_{T}^{B_{j_2}} \in $ \\ $[40, 140]$  }  \\
\hline
signal.II[$\rm \kappa_{tqh}=0.1$]  &  0.64              &  & 0.055       & 6.5$\times 10^{-3}$   &  5.28$\times 10^{-3}$    &   3.68$\times 10^{-3}$ \\
bakt       &        1320                                &  & 1.806       & 0                    &   0                       &  0 \\
bakh       &        92.27                               &  & 0.175       & 0.55$\times 10^{-3}$   & 0.554 $\times 10^{-3}$    &   0.185$\times 10^{-3}$ \\
bakz       &        70.73                               &  & 0.086       & 2.12$\times 10^{-3}$     & 0.283 $\times 10^{-3}$   & 0 \\
bakjjj     &        21730                               &  & 0.261       & 0                    &    0                        & 0  \\
\hline
Total BG   &     -                                      &  & 2.33     &  2.67$\times 10^{-3}$    & 0.837$\times 10^{-3} $   &   0.185$\times 10^{-3}$ \\
\hline
$\cal {SS}$[$\rm1 ab^{-1}$] & -                         &  & 1.14      &  3.1                  &   3.71                  &  4.02    \\
\hline\hline
\makecell {50TeV$\oplus$60GeV$@$FCC-eh \\ unpol. } &   \makecell{ $\rm \sigma_{ini}$ \\  Basic cuts }       &    &  \makecell { 3 tagged \\ Bjets }
&  \makecell{ $\rm p_{T}^{B_{j_3}} \in $ \\ $[265, 455]$ }    &   \makecell{ $\rm \Delta R^{hB_{j_3}}$ $\in$ \\ $[2.8, 3.5] $  }
&  \makecell{ $\rm M_{h}\in$ \\ $[95, 120]$   }   \\
\hline
signal.II[$\rm \kappa_{tqh}=0.1$]  &   3.085               &  &  0.54   & 0.083  & 0.071    &  0.044 \\
bakt       &   10660.0                                    &  &  101.1   & 0      & 0        &  0 \\
bakh       &   507.9                                    &  &  8.82      & 0.005  & 0.002   &   0.0007 \\
bakz       &   357.0                                    &  &  3.9       & 0.035  & 0.010    &  0 \\
bakjjj     &   90070.0                                    &  & 12.61    & 0      & 0        &  0 \\
\hline
Total BG   &     -                                      &  & 126.4      & 0.04  & 0.012  &  0.0007 \\
\hline
$\cal {SS}$[$\rm 1\ ab^{-1}$] & -                      &  &  1.51       & 10.5   &13.3   &  16.70  \\
\hline\hline
 \end{tabular}
 \end{center}
 \caption{\label{SBaftercuts_signal2}
Cross sections (in unit of fb) and significance depending on the cut flows for signal.II
$\rm \ e^- p \to \nu_e h b \to \nu_e b\bar{b} b$ ($\rm \kappa_{tuh}=0.1$)
and backgrounds at the 7TeV$\oplus$60GeV$@$LHeC and 50TeV$\oplus$60GeV$@$FCC-eh.
$\rm {\cal {SS}}$ is evaluated with $\rm 1\ ab^{-1}$ integrated luminosity.
Polarization effects and systematic uncertainty are not considered yet.}
\end{table}

\begin{figure}[hbtp]
\centering
\includegraphics[scale=0.35]{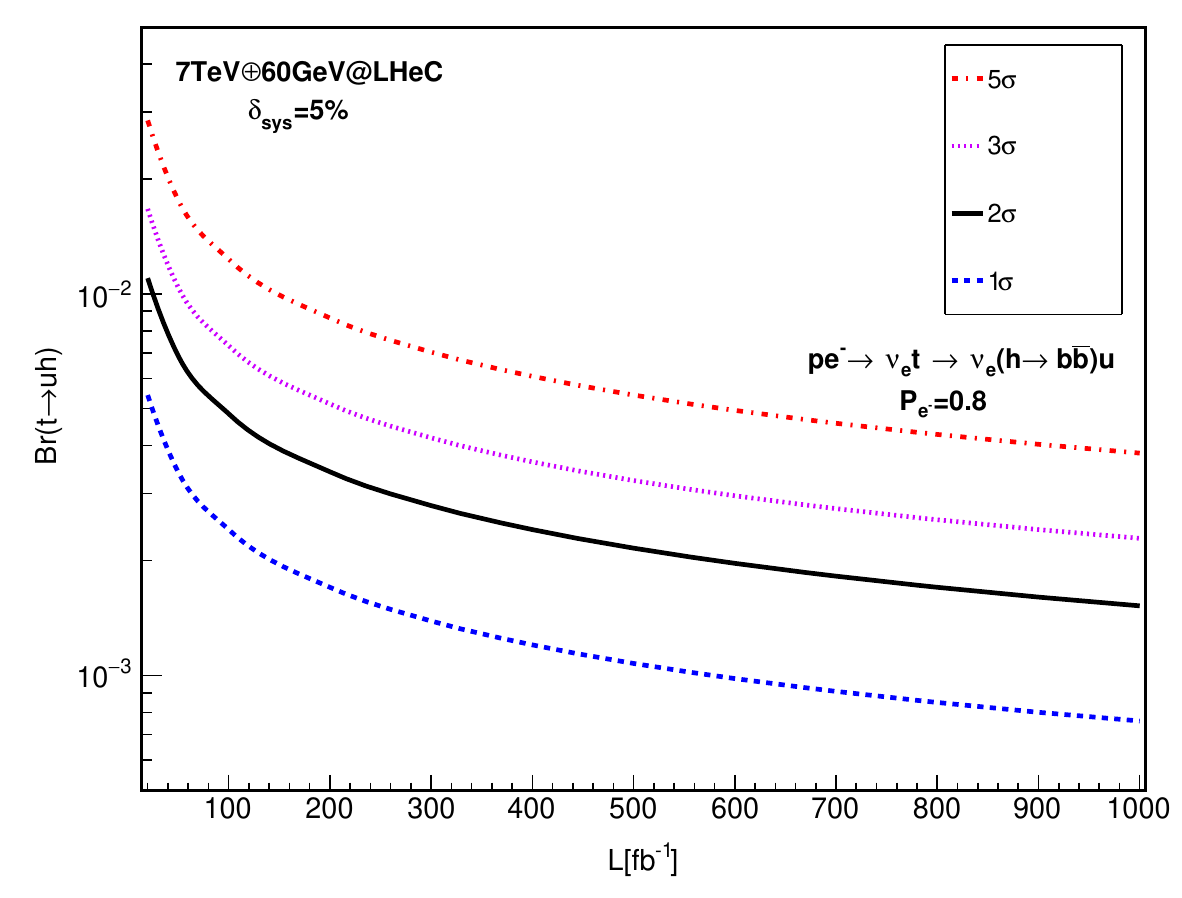}
\includegraphics[scale=0.35]{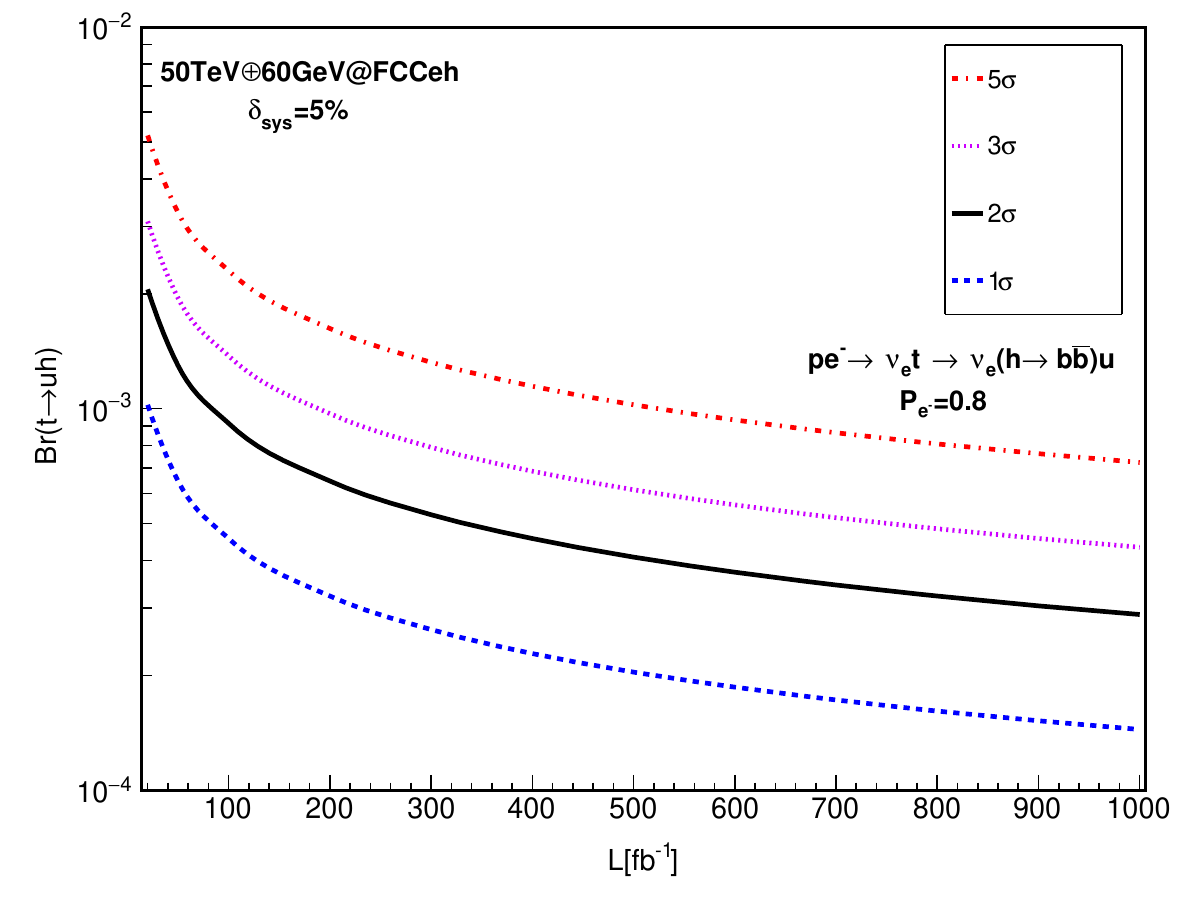}
\includegraphics[scale=0.35]{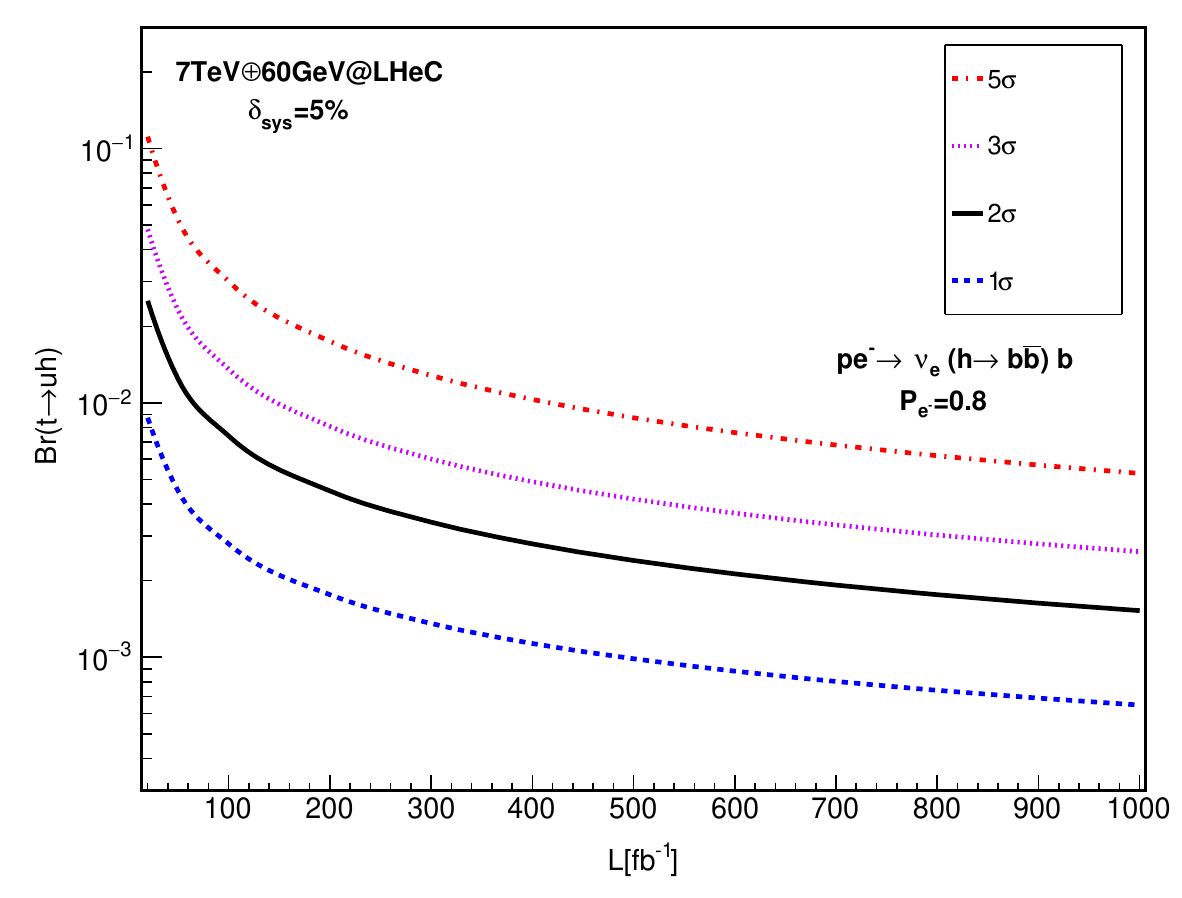}
\includegraphics[scale=0.35]{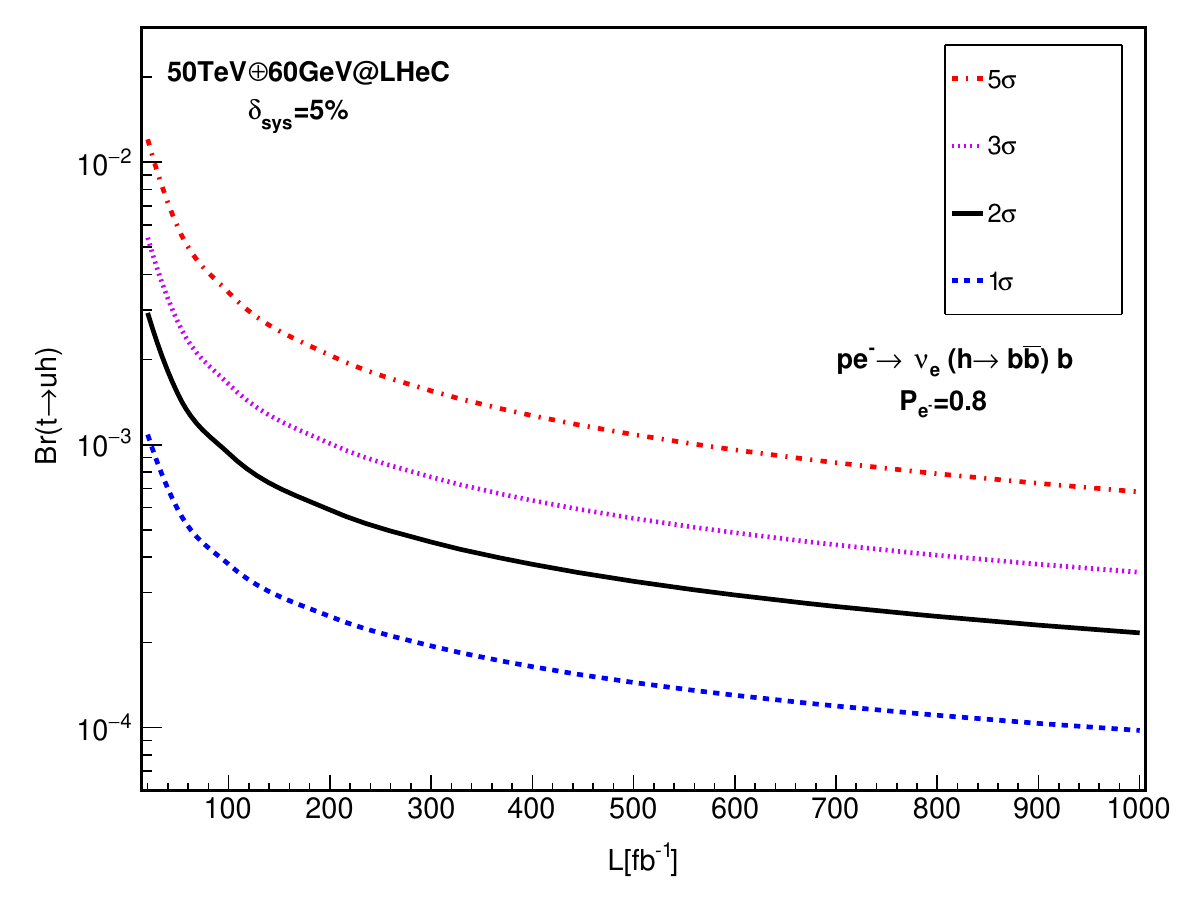}
\caption{\label{limit1}
The upper limit on $\rm Br(t\to uh)$ at 99.99, 99.73, 95.40, 68.27$\%$ C.L.
as a function of the integrated luminosity at the 7(50) TeV LHeC(FCC-eh) with 60 GeV electron beam
for signal.I and signa.II. The dashed blue, black solid, dotted violet and dash-dotted red curves present
1$\sigma$, 2$\sigma$, 3$\sigma$ and 5$\sigma$ significance, respectively. $80\%$ polarisation and
$5\%$ systematic uncertainty for backgrounds yields only are taken into account.}
\end{figure}

In Fig.\ref{limit1}, the upper limit on $\rm Br(t\to uh)$ at 99.99, 99.73, 95.40, 68.27$\%$ C.L.
as a function of the integrated luminosity at the 7(50) TeV LHeC(FCC-eh) with 60 GeV electron beam are plotted.
The dashed blue, solid black, dotted violet and dash-dotted red curves
present 1$\sigma$, 2$\sigma$, 3$\sigma$ and 5$\sigma$ significance, respectively.
The first two figures are for signal.I and the second two are for signal.II.
Our conclusion is that, for signal.I, at the high luminosity (up to 1$\rm ab^{-1}$) ep colliders
where the electrons have a polarisation of $80\%$ and electron energy is typical 60 GeV,
the 1$\sigma$, 2$\sigma$, 3$\sigma$ and 5$\sigma$ upper limit on $\rm Br(t\to uh)$
are $0.075\times 10^{-2}$($0.14\times 10^{-3}$), $0.15\times 10^{-2}$($0.29\times 10^{-3}$),
$0.22\times 10^{-2}$($0.43\times 10^{-3}$) and $0.38\times 10^{-2}$($0.72\times 10^{-3}$) at the LHeC(FCC-eh).
For signal.II, the boundaries are becoming
$0.064\times 10^{-2}$($0.097\times 10^{-3}$), $0.15\times 10^{-2}$($0.22\times 10^{-3}$),
$0.26\times 10^{-2}$($0.35\times 10^{-3}$) and $0.53\times 10^{-2}$($0.68\times 10^{-3}$)
at the LHeC(FCC-eh) respectively. We can see that signal.II can even have better potential than signal.I at the FCC-eh
due to its clean environment. Notice here we use $5\%$ systematic uncertainty for backgrounds yields only at both ep colliders.

\subsubsection{The comparison with the other limits}

Here we compare our discovery potential with the other studies.
Some references present limit on $\rm Br(t\to qh)$.
For example, Ref.\cite{tqh_ee_Hesari} probe the observability of the top-Higgs FCNC couplings through
the process $\rm e^-e^+\to t(\to \ell\nu\ell b)\bar{t}(\to qh)$.
It is shown that the branching ratio can be probed down to $1.12\times 10^{-3}$ at $95\%$ C.L.
at the center-of-mass energy of 500 GeV with the integrated luminosity of 3000 $\rm fb^{-1}$.
This limit can be further improved when the polarizations of both lepton beams are included\cite{tqh_ee_Monalisa}.
Ref.\cite{tqh_LHC_Whj_Liu}, present the study through the process $\rm pp \to W^-(\to \ell^- \bar{\nu}\ell )h(\to \gamma\gamma)j$,
and show that the branching ratios $\rm Br(t\to qh)$ can be probed to $0.16\%$ at $3\sigma$ level at 14 TeV LHC
with an integrated luminosity of 3000 $\rm fb^{-1}$.
Through some other channels, this limit can actually be pushed to even lower values.
As proposed in \cite{HLLHC_tqh_3000}, at the High-luminosity(HL)-LHC, the $95\%$ CL upper limit $\rm Br(t\to qh)$
can be estimated up to the order of $2\sim5\times 10^{-4}$
by a scaling with the luminosity, based on the studies in Ref.\cite{SM_tqH_multileptons}.

Some references present the limits on $\rm Br(t\to uh)$, which we can easily compare with.
As shown in Ref.\cite{tqh_LHC_tt_Wu},
through $\rm t\bar{t}\to W^{+}b + qh \to \ell^+\nu b + \gamma\gamma q $ channel at the LHC,
the branching ratios $\rm Br(t\to uh)$ can be respectively probed to $0.23\%$ at $3\sigma$ level at 14 TeV LHC with $\rm L=3000\ fb^{-1}$.
This limits can be improved in Ref\cite{tqh_LHC_th_Greljo} where the authors apply a development version of HEPTopTagger algorithm.
They found that, through multilepton searches ($\rm th\to \ell^+\nu b+\ell^+\ell^- X$),
vector boson plus Higgs search ($\rm th\to \ell^+\nu b + \tau^+\tau^-$)
and fully  hadronic  search ($\rm th\to jjb+b\bar{b}$), the limits are found to be $0.22\%$, $0.15\%$ and $0.36\%$
by using $\rm 100\ fb^{-1}$ of 13 TeV data.

Concerning our results, with $80\%$ electrons polarisation, 1$\rm ab^{-1}$ integrated luminosity,
and $\rm 5\%$ system uncertainty from background yields only,
the $3\sigma$ limits are $0.22\times 10^{-2}$ at the 7TeV$\oplus$60GeV$@$LHeC
and $3.5\times 10^{-4}$ at the 50TeV$\oplus$60GeV$@$FCC-eh.
Compare the limits we obtained with the others, on one hand, our limits are better than the limits form
the 8 TeV 20.3 (19.7) $\rm fb^{-1}$ data at the ATLAS (CMS), say,
$\rm Br(t\to uh) \leq 4.5(5.5)\times 10^{-3}$ \cite{tqh_rr_ATLAS},
on the other hand, comparable to or even better than some phenomenological studies at the other colliders.
In such case the ep colliders may play an important role of double checks
if the anomalous tqh couplings is really discovered at the LHC or (HL)-LHC.

\subsubsection{The sensitivity dependence on the electron beam energy change}

\begin{figure}[hbtp]
\centering
\includegraphics[scale=0.35]{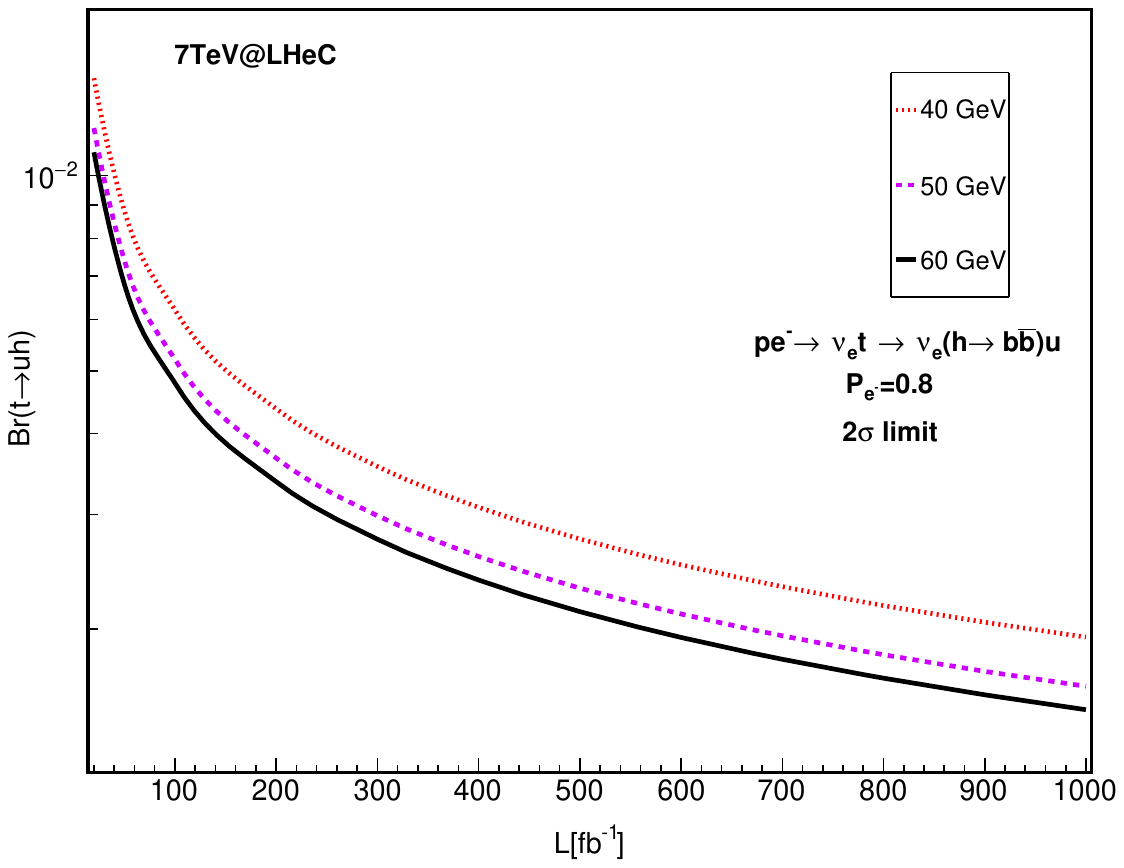}
\includegraphics[scale=0.35]{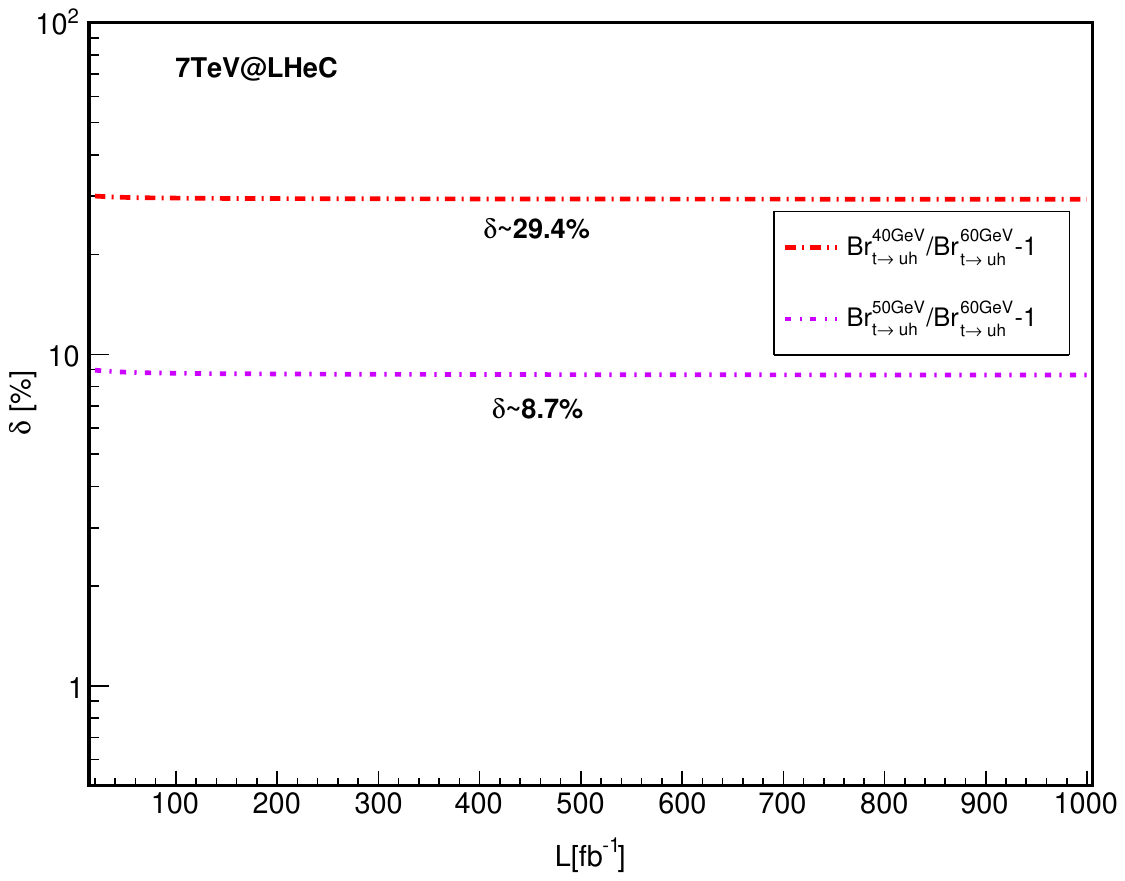}
\caption{\label{40GeV_50GeV_ratio_LHeC}
The left panel is the $2\sigma$ $\rm \kappa_{tuh}$ limit as a function of the luminosities.
The solid black, dashed violet and dotted red curves are for the 60, 50 and 40 GeV LHeC.
The right panel is the ratio defined as $\rm \delta^{E} = \frac{Br_{t\to uh}^{E} - Br_{t\to uh}^{60GeV} }{Br_{t\to uh}^{60GeV}}$,
where where E equal 50 or 40 GeV.}
\end{figure}

In the above analysis we explore the potentials
at the high luminosity (up to 1 $\rm ab^{-1}$) ep colliders
where the electrons have a polarisation of $80\%$. Electron energy is typical 60 GeV,
but lower energies are interesting due to the reason of cost.
Therefore we give an estimate on how our sensitivity (take signal.I as a example)
would change when we reduce the electron beam energy from 60 GeV to 50 GeV or even 40 GeV.
In Table.\ref{40GeV_50GeV_LHeC} we present the results at the 40 GeV and 50 GeV LHeC.
Compare to the 60 GeV LHeC, the significance is reduced from 5.9 to 5.37(4.52) for 50(40) GeV.
\begin{table}[htbp]
\begin{center}
\begin{tabular}{c c c c c c c c c c}
\hline\hline
\makecell {7TeV$\oplus$40GeV$@$LHeC \\ unpol. } &   \makecell{ $\rm \sigma_{ini}$ \\  Basic cuts }   &  & \makecell { $\geq$3 jets with \\ 2 tagged Bjets}
& \makecell{ $\rm M_{top}\in$ \\ $[110, 180]$ }  &  \makecell{ $\rm M_{h}\in $ \\ $[100, 130]$ }     & \makecell{ ht $\in$ \\ $[85, 190]$  } \\
\hline
signal.I[$\rm \kappa_{tqh}=0.1$]  &  4.52               &  & 0.55     & 0.46   &    0.30   & 0.24  \\
bakt       &        749.7                               &  & 28.0     & 16.8   &    3.95   & 1.92\\
bakh       &        57.68                               &  & 9.1      & 2.1    &    1.07   & 0.59 \\
bakz       &        45.84                               &  & 6.06     & 1.94   &    0.073  & 0.03 \\
bakjjj     &        15510                               &  & 9.3      & 4.6    &    0.47   & 0.09 \\
\hline
Total BG   &     -                                      &  & 52.5     & 25.42  &    5.55   & 2.64\\
\hline
$\cal {SS}$[$\rm1 ab^{-1}$] & -                         &  & 2.4      &2.9     &    4.0    & 4.52 \\
\hline\hline
\makecell {7TeV$\oplus$50GeV$@$LHeC \\ unpol. }  &   \makecell{ $\rm \sigma_{ini}$ \\  Basic cuts }  &  & \makecell { $\geq$3 jets with \\ 2 tagged Bjets}
& \makecell{ $\rm M_{top}\in$ \\ $[115, 180]$ }  &  \makecell{ $\rm M_{h}\in $ \\ $[105, 130]$ }        & \makecell{ ht $\in$ \\ $[75, 180]$  } \\
\hline
signal.I[$\rm \kappa_{tqh}=0.1$]  &  6.22               &  & 0.79      &  0.68   &    0.37    &  0.31  \\
bakt       &        1.032                               &  & 43.8      &  25.7   &    4.6     &  2.42 \\
bakh       &        75.25                               &  & 12.5      &  2.8    &    1.1     &  0.66 \\
bakz       &        58.54                               &  & 8.1       &  2.5    &    0.06    &  0.026 \\
bakjjj     &        18730                               &  & 10.5      &  5.4    &    0.34    & 0.075\\
\hline
Total BG   &     -                                      &  & 74.8      &  36.3   &    6.1     &  3.2\\
\hline
$\cal {SS}$[$\rm1 ab^{-1}$] & -                         &  & 2.88      &  3.54   &    4.74    & 5.37\\
\hline\hline
 \end{tabular}
 \end{center}
 \caption{\label{40GeV_50GeV_LHeC}
Cross sections (in unit of fb) and significance depending on the cut flows for signal.I
$\rm \ e^- p \to \nu_e \bar{t} \to \nu_e h \bar{q} \to \nu_e b\bar{b} \bar{q}$ ($\rm \kappa_{tuh}=0.1$)
and backgrounds at the 7TeV$\oplus$40GeV$@$LHeC and 7TeV$\oplus$50GeV$@$FCC-eh.
$\rm {\cal {SS}}$ is evaluated with $\rm 1\ ab^{-1}$ integrated luminosity.
Polarization effects and systematic uncertainty are not considered yet.}
\end{table}
A more straight comparison is presented in Fig.\ref{40GeV_50GeV_ratio_LHeC}.
The left panel is the $2\sigma$ $\rm Br(t\to uh)$ limit as a function of the luminosities.
The solid black, dashed violet and dotted red curves are for the results of 60, 50 and 40 GeV at the LHeC.
In the right panel, we define a ratio as
\begin{eqnarray}
\rm \delta^{E} = \frac{Br^{E}_{t\to uh} - Br_{t\to uh}^{60GeV} }{Br_{t\to uh}^{60GeV}}.
\end{eqnarray}
where E equal 50 or 40 GeV.
The dashed violet curve is for $\rm \delta^{50GeV}$ and the dotted red one is for $\rm \delta^{40 GeV}$.
We see that when the energy of electron beam reduce from 60 GeV to 50(40) GeV,
the discovery potential is reduced around 8.7(29.4) percent.
We also check that these numbers will not change no matter we are using the 1$\sigma$, 2$\sigma$, 3$\sigma$ or 5$\sigma$ limits.
So we conclude that the discovery potential reduce $8.7\%$($29.4\%$)
if the electron beam change from 60GeV to 50(40) GeV at the 7TeV LHeC.

\begin{figure}[hbtp]
\centering
\includegraphics[scale=0.35]{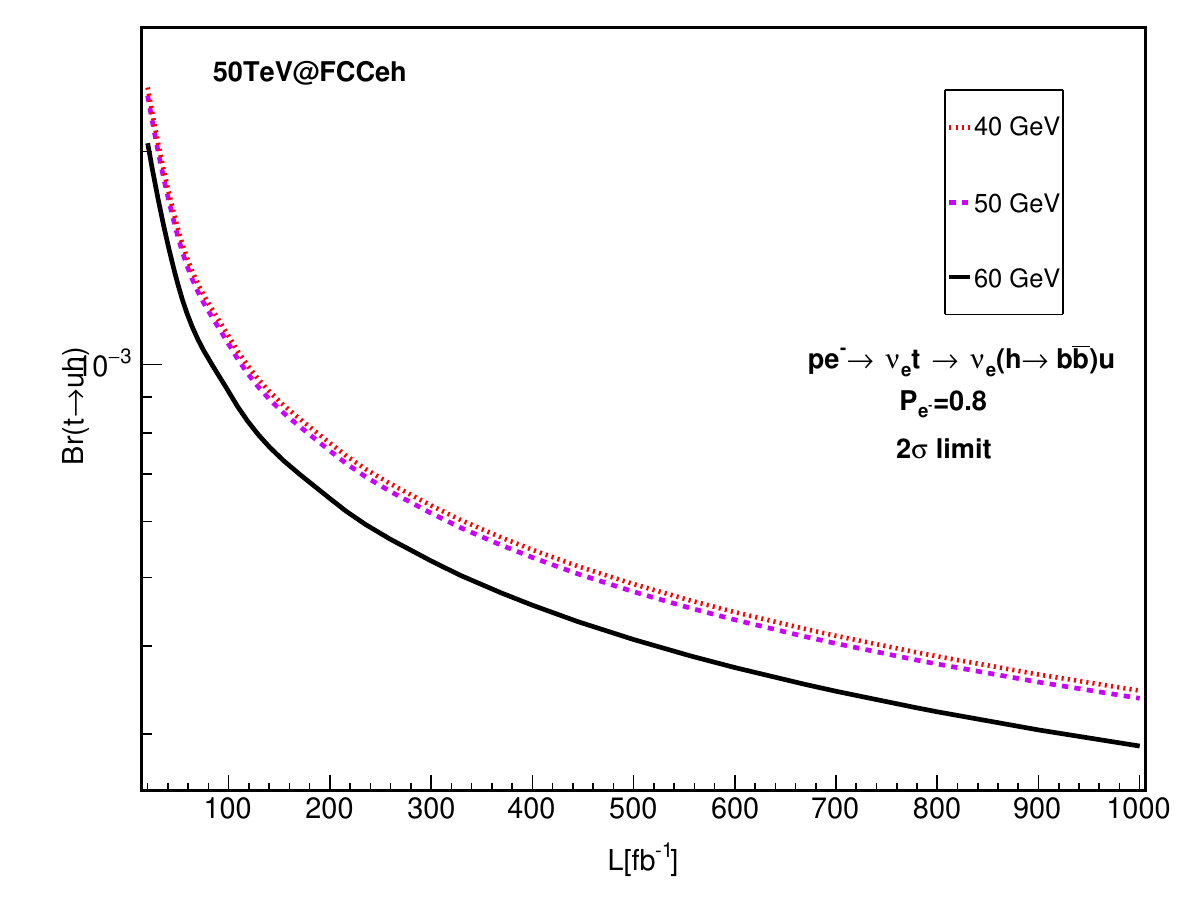}
\includegraphics[scale=0.35]{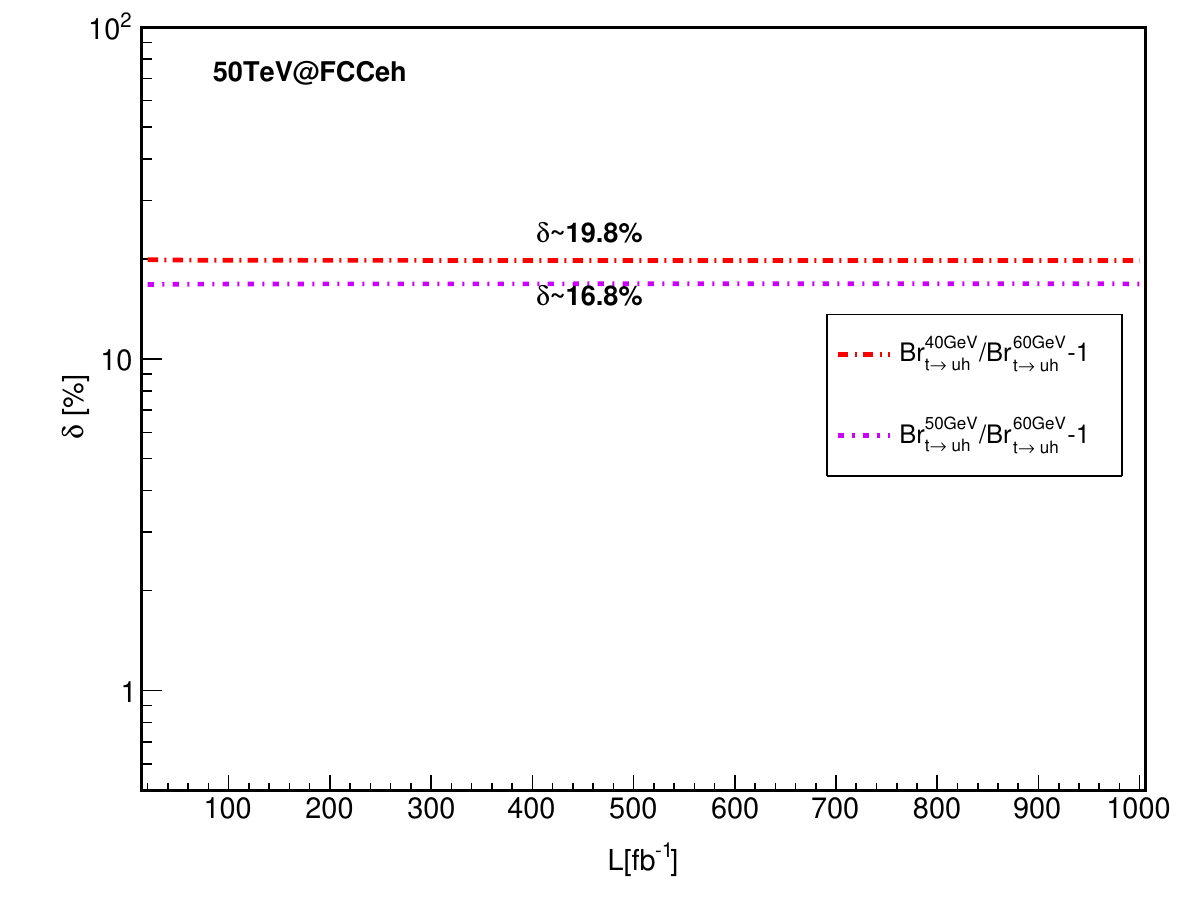}
\caption{\label{40GeV_50GeV_ratio_FCCeh}
The left panel is the $2\sigma$ $\rm \kappa_{tuh}$ limit as a function of the luminosities.
The solid black, dashed violet and dotted red curves are for the 60, 50 and 40 GeV FCC-eh.
The right panel is the ratio defined as $\rm \delta^{E} = \frac{Br_{t\to uh}^{E} - Br_{t\to uh}^{60GeV} }{Br_{t\to uh}^{60GeV}}$,
where where E equal 50 or 40 GeV.}
\end{figure}

The same comparison is done in Table.\ref{40GeV_50GeV_FCCeh} for the 40 GeV and 50 GeV FCC-eh.
Compare to the 60 GeV FCC-eh, the significance is reduced from 30.0 to 25.8(25.1) for 50(40) GeV FCC-eh(1 $\rm ab^{-1}$).
The similar ratio is plotted in Fig.\ref{40GeV_50GeV_ratio_FCCeh}.
It is found that when the energy of electron beam reduce from 60 GeV to 50(40) GeV,
the discovery potential is reduced about 16.8(19.8) percent correspondingly.
\vspace{1cm}
\begin{table}[htbp]
\begin{center}
\begin{tabular}{c c c c c c c c c c}
\hline\hline
\makecell {50TeV$\oplus$40GeV$@$FCC-eh \\ unpol. } &   \makecell{ $\rm \sigma_{ini}$ \\  Basic cuts }           &  & \makecell { $\geq$3 jets with \\ 2 tagged Bjets}
&  \makecell{ ht $\in$ \\ $[75, 165]$  }    &   \makecell{ $\rm M_{h}\in $ \\ $[90, 125]$ }    &  \makecell{ $\rm M_{top}\in$ \\ $[120, 170]$   } \\
\hline
signal.I[$\rm \kappa_{tqh}=0.1$]  &   44.57             &  &  11.9     & 8.2    &  5.41   & 4.56 \\
bakt       &   7393                                     &  &   762.9   & 207.8  &  45.9   & 25.0 \\
bakh       &   377.4                                    &  &   114.0   & 39.3   &  25.3   & 5.2 \\
bakz       &   267.8                                    &  &   71.9    & 19.4   &  0.9    & 0.26 \\
bakjjj     &   68370                                    &  &   127.6   & 32.7   &  2.3    & 0.96 \\
\hline
Total BG   &     -                                      &  &  1076.4    & 299.2 &  74.4   & 31.42 \\
\hline
$\cal {SS}$[$\rm 1\ ab^{-1}$] & -                      &  &   11.5    & 15.0  &  19.6   & 25.1  \\
\hline\hline
\makecell {50TeV$\oplus$50GeV$@$FCC-eh \\ unpol. } &   \makecell{ $\rm \sigma_{ini}$ \\  Basic cuts }           &  & \makecell { $\geq$3 jets with \\ 2 tagged Bjets}
&  \makecell{ ht $\in$ \\ $[80, 185]$  }    &   \makecell{ $\rm M_{h}\in $ \\ $[90, 125]$ }    &  \makecell{ $\rm M_{top}\in$ \\ $[125, 170]$   }  \\
\hline
signal[$\rm \kappa_{tqh}=0.1$]  &   54.67               &  &  15.18    & 10.4     & 6.8    & 6.0 \\
bakt       &   9074                                     &  &   1028.6  & 311.8    & 72.1   & 43.2\\
bakh       &   445.5                                    &  &   141.9   & 52.0     & 34.3   & 7.43\\
bakz       &   314.3                                    &  &   89.0    & 24.9     & 1.2    & 0.43 \\
bakjjj     &   79610                                    &  &   170.5   & 42.2     & 2.2    & 1.1\\
\hline
Total BG   &     -                                      &  &  1430.0   & 430.9    & 109.8  & 52.16\\
\hline
$\cal {SS}$[$\rm 1\ ab^{-1}$] & -                      &  &   12.7   &  15.8    &  20.4  & 25.8 \\
\hline\hline
 \end{tabular}
 \end{center}
 \caption{\label{40GeV_50GeV_FCCeh}
Cross sections (in unit of fb) and significance depending on the cut flows for signal.I
$\rm \ e^- p \to \nu_e \bar{t} \to \nu_e h \bar{q} \to \nu_e b\bar{b} \bar{q}$ ($\rm \kappa_{tuh}=0.1$)
and backgrounds at the 50TeV$\oplus$40GeV$@$FCC-eh and 50TeV$\oplus$50GeV$@$FCC-eh.
$\rm {\cal {SS}}$ is evaluated with $\rm 100\ fb^{-1}$ integrated luminosity.
Polarization effects and systematic uncertainty are not considered yet.}
\end{table}

\section{CONCLUSION}

In this paper we have investigated an updated analysis on searches for
the anomalous flavor changing neutral current(FCNC) Yukawa interactions
between the top quark, the Higgs boson, and either an up or charm quark ($\rm tqh, q=u, c$).
We probe the observability of the FCNC top-Higgs couplings through the process
$\rm e^- p\rightarrow \nu_e \bar{t} \rightarrow \nu_e h \bar{q}$ (signa.I)
and $\rm \ e^- p \to \nu_e h b$ (signal.II) at the ep colliders
where the Higgs boson decays to a $\rm b\bar{b}$ pair.
We perform the results from the cut-and-count based method.
Our results show that with $80\%$ electrons polarisation, 1$\rm ab^{-1}$ integrated luminosity,
and $\rm 5\%$ system uncertainty from background yields only,
the $3\sigma$ limits are $0.22\times 10^{-2}$ at the 7TeV$\oplus$60GeV$@$LHeC
and $3.5\times 10^{-4}$ at the 50TeV$\oplus$60GeV$@$FCC-eh.
These limits are, on one hand, better than the current limits for the experiments,
on the other hand, comparable to or even better than some phenomenological studies at the other colliders.
We also give an estimate on how our sensitivity (take signal.I as a example) would change
when we reduce the electron beam energy from 60 GeV to 50 GeV or even 40 GeV due to cost reason.
The conclusion is that the discovery potential reduce $8.7\%$($29.4\%$)
if the electron beam change from 60GeV to 50(40) GeV at the 7TeV LHeC,
and $16.8\%$($19.8\%$) at the 50 TeV FCC-eh.
In summary, we give a detailed overview on the searches potential for the anomalous top-Higgs couplings at the ep colliders
including the LHeC as well as the FCC-eh.

\section*{Acknowledgments} \hspace{5mm}
The author H. Sun would like to thank the comments and encourages
from the LHeC/FCC-eh (Top and Higgs and BSM) working Group.
This work is supported by the National Natural Science Foundation of China
(Grant No. 11675033), by the Fundamental Research Funds for the Central Universities
(Grant No. DUT15LK22).


\begin{thebibliography}{99}

\bibitem{SMHiggs_ATLAS}
G. Aad, et al., [ATLAS Collaboration], {\it Observation of a new particle in the search
for the Standard Model Higgs boson with the ATLAS detector at the LHC},
Phys. Lett. B 716, 1-29 (2012), {\tt [arXiv:1207.7214]}.

\bibitem{SMHiggs_CMS}
S. Chatrchyan, et al., [CMS Collaboration], {\it Observation of a new boson at a mass of
125 GeV with the CMS experiment at the LHC}, Phys. Lett. B 716, 30 (2012), {\tt [arXiv:1207.7235]}.

\bibitem{GIM_mechanism}
S. Glashow, J. Iliopoulos and L. Maiani, {\it Weak Interactions with Lepton-Hadron Symmetry}, Phys. Rev. D 2, 1285 (1970).

\bibitem{tqh_review}
Aguilar-Saavedra, {\it Top flavor-changing neutral interactions: Theoretical expectations and experimental detection},
Acta Phys. Polon. B 35, 2695 (2004), {\tt [hep-ph/0409342]}.

\bibitem{MSSM_tqh_1}
C. S. Li, R. J. Oakes and J. M. Yang, {\it Rare decay of the top quark in the minimal supersymmetric model},
Phys. Rev. D 49, 293 (1994). Erratum-ibid.D 56, 3156 (1997).

\bibitem{MSSM_tqh_2}
G. M. de Divitiis, R. Petronzio and L. Silvestrini, {\it Flavour-changing top decays in supersymmetric extensions of the standard model},
Nucl. Phys. B 504, 45 (1997), {\tt [hep-ph/9704244]}.

\bibitem{MSSM_tqh_3}
J. L. Lopez, D. V. Nanopoulos and R. Rangarajan, {\it New supersymmetric contributions to $t\to cV$},
Phys. Rev. D 56, 3100 (1997), {\tt [hep-ph/9702350]}.

\bibitem{MSSM_tqh_4}
Jin Min Yang, Bing-Lin Young and X. Zhang, {\it Flavor-changing Top Quark Decays In R-Parity Violating SUSY},
Phys. Rev. D 58, 055001 (1998), {\tt [arXiv:hep-ph/9705341]}.

\bibitem{MSSM_tqh_5}
J. Guasch and J. Sola, {\it FCNC top quark decays: A door to SUSY physics in high luminosity colliders?},
Nucl. Phys. B 562, 3 (1999), {\tt [hep-ph/9906268]}.

\bibitem{MSSM_tqh_6}
G. Eilam, A. Gemintern, T. Han, J.M. Yang and X. Zhang, {\it Top-quark rare decay $t\rightarrow c h$ in R-parity-violating SUSY},
Phys. Lett. B 510, 227-235 (2001), {\tt [arXiv:hep-ph/0102037]}.

\bibitem{MSSM_tqh_7}
D. Delepine and S. Khalil, {\it Top flavour violating decays in general supersymmetric models},
Phys. Lett. B 599, 62 (2004), {\tt [hep-ph/0406264]}.

\bibitem{MSSM_tqh_8}
J. J. Liu, C. S. Li, L. L. Yang and L. G. Jin, {\it $t\to cV$ via SUSY FCNC couplings in the unconstrained MSSM},
Phys. Lett. B 599, 92 (2004), {\tt [hep-ph/0406155]}.

\bibitem{MSSM_tqh_9}
J.J.Cao, G.Eilam, M.Frank, K.Hikasa, G.L.Liu, I.Turan and J. M. Yang, {\it SUSY-induced FCNC top-quark processes at the Large Hadron Collider},
Phys. Rev. D 75, 075021 (2007), {\tt [arXiv:hep-ph/0702264]}.

\bibitem{MSSM_tqh_10}
David Lopez-Val, Jaume Guasch and Joan Sola, {\it Single top-quark production by strong and
electroweak supersymmetric flavor-changing interactions at the LHC}, JHEP 0712, 054 (2007), {\tt [arXiv:0710.0587]}.

\bibitem{MSSM_tqh_11}
Junjie Cao, Zhaoxia Heng, Lei Wu and Jin Min Yang, {\it R-parity violating effects in top quark FCNC productions at LHC},
Phys. Rev. D 79, 054003 (2009), {\tt [arXiv:0812.1698]}.

\bibitem{MSSM_tqh_12}
A. Dedes, M. Paraskevas, J. Rosiek, K. Suxho and K. Tamvakis,
{\it Rare Top-quark Decays to Higgs boson in MSSM}, JHEP 1411, 137 (2014), {\tt [arXiv:1409.6546]}.

\bibitem{MSSM_tqh_13}
Junjie Cao, Chengcheng Han, Lei Wu, Jin Min Yang and Mengchao Zhang, {\it SUSY induced top quark FCNC decay $t\rightarrow cH$ after Run I of LHC},
Eur. Phys. J. C 74 (2014) 9, 3058, {\tt [arXiv:1404.1241]}.

\bibitem{MSSM_tqh_14}
Tie-Jun Gao, Tai-Fu Feng, Fei Sun, Hai-Bin Zhang and Shu-Min Zhao, {\it Top quark decay to a 125GeV Higgs in BLMSSM},
Chin. Phys. C 39 (2015) 7, 073101, {\tt [arXiv:1404.3289]}.

\bibitem{MSSM:DiazCruz:2001gf}
J.L. Diaz-Cruz, Hong-Jian He and C.-P. Yuan, {\it Soft Supersymmetry Breaking, Scalar Top-Charm Mixing and Higgs Signatures},
Phys. Lett. B 530, 179 (2002), {\tt [hep-ph/0103178]}.

\bibitem{2HDM_tqh_1}
B. Grzadkowski, J. F. Gunion and P. Krawczyk, {\it Neutral current flavor changing decays
for the Z boson and the top quark in two Higgs doublet models}, Phys. Lett. B 268, 106 (1991).

\bibitem{2HDM_tqh_2}
G. Eilam, J.L. Hewett and A. Soni, {\it Rare decays of the top quark in the standard and two Higgs doublet models},
Phys. Rev. D 44, 1473-1484 (1991), Erratum-ibid. D 59, 039901 (1999).

\bibitem{2HDM_tqh_3}
D. Atwood, L. Reina and A. Soni, {\it Phenomenology of two Higgs doublet models with flavor changing neutral currents},
Phys. Rev. D 55, 3156 (1997), {\tt [hep-ph/9609279]}.

\bibitem{2HDM_tqh_4}
S. Bejar, J. Guasch and J. Sola, {\it Loop induced flavor changing neutral decays of the top
quark in a general two-Higgs-doublet model}, Nucl. Phys. B 600, 21 (2001), {\tt [hep-ph/0011091]}.

\bibitem{2HDM_tqh_5}
Santi Bejar, Jaume Guasch and Joan Sola, {\it Higgs Boson Flavor-Changing Neutral Decays into Top Quark in a General Two-Higgs-Doublet Model},
Nucl. Phys. B 675, 270-288 (2003), {\tt [arXiv:hep-ph/0307144]}.

\bibitem{2HDM_tqh_6}
I. Baum, G. Eilam and S. Bar-Shalom, {\it Scalar FCNC and rare top decays in a two Higgs doublet model for the top},
Phys. Rev. D 77, 113008 (2008), {\tt [hep-ph/0802.2622]}.

\bibitem{2HDM_tqh_7}
Chung Kao, Hai-Yang Cheng, Wei-Shu Hou and Joshua Sayre, {\it Top Decays with Flavor Changing Neutral Higgs Interactions at the LHC},
Phys.Lett. B 716, 225-230 (2012), {\tt [arXiv:1112.1707]}.

\bibitem{2HDM_tqh_8}
K.-F. Chen, W.-S. Hou, C. Kao and M. Kohda, {\it When the Higgs meets the Top: Search for $t\to ch^0$ at the LHC},
Phys. Lett. B 725, 378 (2013), {\tt [arXiv:1304.8037]}.

\bibitem{2HDM_tqh_9}
D. Atwood, S. K. Gupta and A. Soni, {\it Constraining the flavor changing Higgs couplings to the top-quark at the LHC},
JHEP 1410, 57 (2014), {\tt [arXiv:1305.2427]}.

\bibitem{2HDM_tqh_10}
Kai-Feng Chen, Wei-Shu Hou, Chung Kao and Masaya Kohda,
{\it When the Higgs meets the Top: Search for $t\rightarrow ch^0$ at the LHC},
Phys. Lett. B 725, 378-381 (2013), {\tt [arXiv:1304.8037]}.

\bibitem{2HDM:He:2002fd}
Hong-Jian He, Shinya Kanemura and C.-P. Yuan, {\it Determining the Chirality of Yukawa Couplings via Single Charged Higgs Boson
Production in Polarized Photon Collision}, Phys. Rev. Lett. 89, 101803 (2002), {\tt [hep-ph/0203090]}.

\bibitem{2HDM:He:2002ak}
Hong-Jian He, Shinya Kanemura and C.-P. Yuan,
{\it Single Charged Higgs Boson Production in Polarized Photon Collision and the Probe of New Physics},
Phys. Rev. D 68, 075010 (2003), {\tt [hep-ph/0209376]}.

\bibitem{2HDM_tqH_Altunkaynak}
Baris Altunkaynak, Wei-Shu Hou, Chung Kao, Masaya Kohda, and Brent McCoy,
{\it Flavor Changing Heavy Higgs Interactions at the LHC}, Phys. Lett. B 751 (2015) 135-142, {\tt [arXiv:1506.00651]}.

\bibitem{2HDM_tqH:He:1998ie}
Hong-Jian He and C.-P. Yuan, {\it New Method for Detecting Charged (Pseudo-)Scalars at Colliders},
Phys. Rev. Lett. 83 (1999) 28, {\tt [hep-ph/9810367]}.

\bibitem{WED_tqH_1}
Aleksandr Azatov, Manuel Toharia and Lijun Zhu, {\it Higgs Mediated FCNC's in Warped Extra Dimensions},
Phys. Rev. D 80, 035016 (2009), {\tt [arXiv:0906.1990]}.

\bibitem{WED_tqH_2}
S. Casagrande, F. Goertz, U. Haisch, M. Neubert and T. Pfoh, {\it The Custodial Randall-Sundrum Model:
From Precision Tests to Higgs Physics}, JHEP 1009, 014 (2010), {\tt [arXiv:1005.4315]}.

\bibitem{ALRM_tqH}
R. Gaitan, O. Miranda and L. Cabral-Rosetti,
{\it  Rare top quark and Higgs boson decays in Alternative Left-Right Symmetric Models},
Phys. Rev. D72, 034018 (2005), {\tt [arXiv:hep-ph/0410268]};
{\it Rare top quark decays in extended models}, AIP Conf. Proc. 857, 179 (2006), {\tt [arXiv:hep-ph/0604170]}.

\bibitem{LHT_tqH_1}
Bingfang Yang, Ning Liu and Jinzhong Han,
{\it Top Quark FCNC Decay to 125GeV Higgs boson in the Littlest Higgs Model with T-parity},
Phys. Rev. D 89 (2014) 3, 034020, {\tt [arXiv:1308.4852]}.

\bibitem{QS_tqh_1}
F. del Aguila, J. A. Aguilar-Saavedra and R. Miquel, {\it Constraints on top couplings in models with exotic quarks},
Phys. Rev. Lett. 82, 1628 (1999).

\bibitem{QS_tqh_2}
J. A. Aguilar-Saavedra and B. M. Nobre, {\it Rare top decays $t\to c\gamma$, $t\to cg$ and CKM unitarity},
Phys. Lett. B 553, 251 (2003), {\tt [hep-ph/0210360]}.

\bibitem{QS_tqh_3}
J. Aguilar-Saavedra, {\it Effects of mixing with quark singlets},
Phys. Rev.D 67, 035003 (2003), {\tt [hep-ph/0210112]}. Erratum-ibid. D 69, 099901 (2004).

\bibitem{tqh_rr_ATLAS} 
G. Aad et al., [ATLAS Collaboration], {\it Search for top quark decays $t\rightarrow qH$
with $H\rightarrow \gamma\gamma$ using the ATLAS detector},
JHEP 06, 008 (2014), CERN-PH-EP-2014-036, {\tt [arXiv:1403.6293]}.

\bibitem{tqh_rr_ATLAS_2017}
[ATLAS Collaboration], {\it Search for top quark decays $t\rightarrow qH$
with $H\rightarrow \gamma\gamma$, in $\sqrt{13}=13\ TeV$ pp collisions using the ATLAS detector},
CERN-EP-2017-118, {\tt [arXiv:1707.01404]}.

\bibitem{tqh_rr_CMS} 
[CMS Collaboration], {\it Searches for heavy Higgs bosons in two-Higgs-doublet models and
for $t\rightarrow ch$ decay using multilepton and diphoton final states in pp collisions at 8 TeV},
Phys. Rev. D 90, 112013 (2014), CMS-HIG-13-025, CERN-PH-EP-2014-239, {\tt [arXiv:1410.2751]}.
[CMS Collaboration], {\it Combined multilepton and diphoton limit on t to cH}, CMS-PAS-HIG-13-034.

\bibitem{tqh_bb_ATLAS}
G. Aad et al., [ATLAS Collaboration], {\it Search for flavour-changing neutral current top quark
decays $t\to Hq$ in pp collisions at $\sqrt{s}$=8 TeV with the ATLAS detector},
JHEP 1512, 061 (2015), {\tt [arXiv:1509.06047]}.

\bibitem{tqh_bb_CMS}
CMS Collaboration, [CMS Collaboration], {\it Search for the Flavor-Changing Neutral Current
Decay $t\to qH$ Where the Higgs Decays to b$\bar{b}$ Pairs at $\sqrt{s}$=8 TeV}, CMS-PAS-TOP-14-020.

\bibitem{CMS_tqh_limit_comb}
V. Khachatryan et al., [CMS Collaboration], {\it Search for top quark decays via
Higgs-boson-mediated flavor-changing neutral currents in pp collisions at $\sqrt{s}$=8 TeV},
JHEP 02, 079 (2017), CMS-TOP-13-017, CERN-EP-2016-208, {\tt [arXiv:1610.04857]}.

\bibitem{Indirect_KK}
M. Bona et al., [UTfit Collaboration], {\it Model-independent constraints on $\Delta F$=2 operators
and the scale of new physics}, JHEP 0803, 049 (2008), {\tt [arXiv:0707.0636]}.

\bibitem{Indirect_BB}
G. Blankenburg, J. Ellis and G. Isidori, {\it Flavour-Changing Decays of a 125 GeV Higgs-like
Particle}, Phys. Lett. B 712, 386 (2012), {\tt [arXiv:1202.5704]}.

\bibitem{Indirect_DD}
J. I. Aranda, A. Cordero-Cid, F. Ramirez-Zavaleta, J. J. Toscano and E. S. Tututi,
{\it Higgs mediated flavor violating top quark decays $t\to u_i H$, $u_i \gamma$, $u_i\gamma\gamma$,
and the process $\gamma\gamma\to tc$ in effective theories},
Phys. Rev. D 81, 077701 (2010), {\tt [arXiv:0911.2304]}.

\bibitem{tqh_Ztocc}
F. Larios, R. Martinez and M. A. Perez, {\it Constraints on top quark FCNC from electroweak
precision measurements}, Phys. Rev. D 72, 057504 (2005), {\tt [hep-ph/0412222]}.

\bibitem{tqh_LHC_tt_Aguilar}
J. A. Aguilar-Saavedra and G. C. Branco, {\it Probing top flavor changing neutral scalar
couplings at the CERN LHC}, Phys. Lett. B 495, 347 (2000), {\tt [hep-ph/0004190]}.

\bibitem{tqh_LHC_tt_Atwood}
D. Atwood, S. K. Gupta and A. Soni, {\it Constraining the flavor changing Higgs couplings to
the top-quark at the LHC}, JHEP 1410, 57 (2014), {\tt [arXiv:1305.2427]}.

\bibitem{SM_tqH_multileptons}
Nathaniel Craig, Jared A. Evans, Richard Gray, Michael Park, Sunil Somalwar, Scott Thomas and Matthew Walker,
{\it Searching for $t\rightarrow ch$ with multileptons}, Phys. Rev. D 86 (2012) 075002, {\tt [arXiv:1207.6794]}.

\bibitem{tqh_LHC_tt_Kobakhidze}
A. Kobakhidze, L. Wu and J. Yue, {\it Anomalous Top-Higgs Couplings and Top Polarisation in
Single Top and Higgs Associated Production at the LHC}, JHEP 1410, 100 (2014), {\tt [arXiv:1406.1961]}.

\bibitem{tqh_LHC_tt_Wu}
L. Wu, {\it Enhancing thj Production from Top-Higgs FCNC Couplings}, JHEP 1502, 061 (2015), {\tt [arXiv:1407.6113]}.

\bibitem{tqh_LHC_th_Greljo}
A. Greljo, J. F. Kamenik and J. Kopp, {\it Disentangling Flavor Violation in the Top-Higgs
Sector at the LHC}, JHEP 1407, 046 (2014), {\tt [arXiv:1404.1278]}.

\bibitem{SM_tqH_ChargeRatio}
Sara Khatibi and Mojtaba Mohammadi Najafabadi, {\it Probing the Anomalous FCNC Interactions in Top-Higgs Final State
and Charge Ratio Approach}, Phys. Rev. D 89 (2014) 054011, {\tt [arXiv:1402.3073]}.

\bibitem{tqh_LHC_Whj_Liu}
Y. B. Liu and Z. J. Xiao, {\it Searches for top-Higgs FCNC couplings via Whj signal with $h \to \gamma\gamma$ at the LHC},
Phys. Rev. D 94, 054018 (2016), {\tt [arXiv:1605.01179}.

\bibitem{tqh_ee_Han}
T. Han, J. Jiang and M. Sher, {\it Search for $t\to ch$ at $e^+e^-$ linear colliders},
Phys. Lett. B 516, 337 (2001), {\tt [hep-ph/0106277]}.

\bibitem{tqh_ee_Behnke}
T. Behnke et al., {\it The International Linear Collider Technical Design Report - Volume 1: Executive Summary},
ILC-REPORT-2013-040, {\tt [arXiv:1306.6327]}.

\bibitem{tqh_ee_Aicheler}
M. Aicheler et al., {\it A Multi-TeV Linear Collider Based on CLIC Technology: CLIC Conceptual Design Report},
CERN-2012-007, {\tt doi:10.5170/CERN-2012-007}.

\bibitem{tqh_ee_Hesari}
H. Hesari, H. Khanpour and M. Mohammadi Najafabadi, {\it Direct and Indirect Searches for Top-Higgs FCNC Couplings},
Phys. Rev. D 92, no. 11, 113012 (2015), {\tt [arXiv:1508.07579]}.

\bibitem{tqh_ee_Monalisa}
Bla$\check{z}$enka Meli$\acute{c}$ and Monalisa Patra, {\it Exploring the Top-Higgs FCNC Couplings at Polarized Linear Colliders
with Top Spin Observables}, JHEP 01 (2017) 048, {\tt [arXiv:1610.02983]}.

\bibitem{tqh_LHeC_haosun1}
Wei Liu, Hao Sun, XiaoJuan Wang and Xuan Luo,
{\it Probing the anomalous FCNC top-Higgs Yukawa couplings at the Large Hadron Electron Collider},
Phys. Rev. D 92 (2015) 7, 074015, {\tt [arXiv:1507.03264]}.

\bibitem{tqh_LHeC_haosun2}
XiaoJuan Wang, Hao Sun and Xuan Luo, {\it Searches for the Anomalous FCNC Top-Higgs Couplings with Polarized Electron Beam at the LHeC},
Adv. High Energy Phys. 2017 (2017) 4693213, {\tt [arXiv:1703.02691]}.

\bibitem{DipoleM_tqH}
Martin Gorbahn and Ulrich Haisch, {\it Searching for $t\rightarrow c(u)h$ with dipole moments},
JHEP 1406 (2014) 033, {\tt [arXiv:1404.4873]}.

\bibitem{twb_NLO}
C. S. Li, R. J. Oakes, and T. C. Yuan,
{\it QCD corrections to $t\rightarrow W^+b$ }, Phys. Rev. D 43, 3759-3762 (1991).

\bibitem{decay_tqH}
Wei-Shu Hou, {\it Tree level $t \rightarrow ch$ or $h\rightarrow t{\bar c}$ decays},
Phys. Lett. B 296, 179-184 (1992).

\bibitem{FFL}
T. Hahn, {\it Generating Feynman diagrams and amplitudes with FeynArts 3},
Comput. Phys. Commun. 140 (2001) 418-431, [arXiv:hep-ph/0012260].
T. Hahn, {\it Automatic Loop Calculations with FeynArts, FormCalc, and LoopTools},
Nucl. Phys. Proc. Suppl. 89 (2000) 231-236, [arXiv:hep-ph/0005029].
S. Agrawal, T. Hahn, E. Mirabella, {\it FormCalc 7},
J. Phys. Conf. Ser. 368 (2012) 012054, [arXiv:1112.0124].
T. Hahn, M. Perez-Victoria,
{\it Automatized one loop calculations in four-dimensions and D-dimensions},
Comput. Phys. Commun. 118 (1999) 153-165, [arXiv:hep-ph/9807565].


\bibitem{tqh_NLO}
Cen Zhang and Fabio Maltoni, {\it Top-quark decay into Higgs boson and a light quark at next-to-leading order in QCD},
Phys. Rev. D88, 054005 (2013), {\tt [arXiv:1305.7386]};
Jure Drobnak, Svjetlana Fajfer and Jernej F. Kamenik,
{\it Flavor Changing Neutral Coupling Mediated Radiative Top Quark Decays at Next-to-Leading Order in QCD},
Phys. Rev. Lett. 104 (2010) 252001, {\tt [arXiv:1004.0620]};
Jia Jun Zhang, Chong Sheng Li, Jun Gao, Hao Zhang, Zhao Li, C.-P. Yuan and Tzu-Chiang Yuan,
{\it Next-to-leading order QCD corrections to the top quark decay via model-independent FCNC couplings},
Phys. Rev. Lett. 102 (2009) 072001, {\tt [arXiv:0810.3889]}.

\bibitem{LHeC_1}
M. Klein, {\it The Large Hadron Electron Collider Project}, Proceedings,
17th International Workshop on Deep-Inelastic Scattering and Related Subjects (DIS 2009),
Madrid, Spain, April 26-30, 2009, {\tt [arXiv:0908.2877]}.

\bibitem{LHeC_2}
J. L. Abelleira Fernandez, et al., [LHeC Study Group Collaboration],
{\it A Large Hadron Electron Collider at CERN: Report on the Physics and Design Concepts for Machine and Detector},
J. Phys. G 39 (2012) 075001, {\tt [arXiv:1206.2913]}.

\bibitem{LHeC_3}
O. Bruening and M. Klein, {\it The Large Hadron Electron Collider},
Mod. Phys. Lett. A 28 (2013) no.16, 1330011, {\tt [arXiv:1305.2090]}.

\bibitem{LHeC_4}
 M. Klein, {\it LHeC Detector Design},
 25th International Workshop on Deep Inelastic Scattering, 2017, Birmingham.
 {\tt https://indico.cern.ch/event/568360/contributions/2523637/}.

\bibitem{FCC-eh_1}
F. Zimmermann, M. Benedikt, D. Schulte and J. Wen-ninger, {\it Challenges for Highest Energy Circular Colliders},
IPAC-2014-MOXAA01, Proceedings, 5th International Particle Accelerator Conference (IPAC 2014), Dresden, Germany, June 15-20, 2014.

\bibitem{FCC-eh_2}
M. Klein, {\it Deep inelastic scattering at the energy frontier}, Annalen Phys. 528 (2016) 138-144.


\bibitem{LHeC_BSM_Lindner}
Manfred Lindner, Farinaldo S. Queiroz, Werner Rodejohann and Carlos E. Yaguna,
{\it Left-Right Symmetry and Lepton Number Violation at the Large Hadron Electron Collider},
JHEP 06(2016)140, {\tt [arXiv:1604.08596]}.

\bibitem{LHeC_BSM_Fischer}
Stefan Antusch, Eros Cazzato and Oliver Fischer,
{\it Sterile neutrino searches at future $e^-e^+$, pp, and $e^-p$ colliders},
Int.J.Mod.Phys. A32 (2017) no.14, 1750078, {\tt [arXiv:1612.02728]}.

\bibitem{LHeC_BSM_Subhadeep}
Subhadeep Mondal and Santosh Kumar Rai,
{\it Probing the Heavy Neutrinos of Inverse Seesaw Model at the LHeC},
Phys. Rev. D 94, 033008 (2016), {\tt [arXiv:1605.04508]}.

\bibitem{LHeC_BSM_ShouHua}
Yi-Lei Tang, Chen Zhang and Shou-hua Zhu,
{\it Invisible Higgs Decay at the LHeC}, Phys.Rev. D94 (2016) no.1, 011702, {\tt [arXiv:1508.01095]}.

\bibitem{LHeC_BSM_Kumar}
Baradhwaj Coleppa, Mukesh Kumar, Satendra Kumar and Bruce Mellado,
{\it Measuring CP nature of top-Higgs couplings at the future Large Hadron electron collider},
Phys.Lett. B770 (2017) 335-341, {\tt [arXiv:1702.03426]}.

\bibitem{LHeC_BSM_haosun}
Hao Sun, Xuan Luo, Wei Wei and Tong Liu,
{\it Searching for the doubly-charged Higgs bosons in the Georgi-Machacek model at the electron-proton colliders},
Phys. Rev. D 96, 095003 (2017), {\tt [arXiv:1710.06284]}.

\bibitem{LHeC_BSM_ohan}
H. Denizli, A. Senol, A. Yilmaz, I. T. Cakir, H. Karadeniz, O. Cakir,
{\it Top quark FCNC couplings at future circular hadron electron colliders}£¬
Phys. Rev. D 96, 015024 (2017), {\tt [arXiv:1701.06932]}.

\bibitem{seperate_tuh_tch}
[ATLAS Collaboration], {\it Performance and Calibration of the JetFitterCharm Algorithm
for c-Jet Identification}, ATL-PHYS-PUB-2015-001.

\bibitem{FeynRules2.0}
A. Alloul, N. D. Christensen, C. Degrande, C. Duhr and B.Fuks,
{\it FeynRules 2.0 - A complete toolbox for tree-level phenomenology},
Comput. Phys. Commun. 185, 2250-2300 (2014), {\tt [arXiv:1310.1921]}.

\bibitem{UFO}
C. Degrande, C. Duhr, B. Fuks, D. Grellscheid, O.Mattelaer and T. Reiter,
{\it UFO the universal FeynRules output},
Comput. Phys. Commun. 183, 1201-1214 (2012), {\tt [arXiv:1108.2040]}.

\bibitem{MadGraph5}
J. Alwall, R. Frederix, S. Frixione, V. Hirschi, F. Maltoni, O. Mattelaer, H.-S. Shao, T. Stelzer, P. Torrielli and M. Zaro,
{\it The automated computation of tree-level and next-to-leading order differential cross sections, and their
matching to parton shower simulations}, JHEP 1407, 079 (2014), {\tt [arXiv:1405.0301]}.

\bibitem{Pythia}
T. Sjostrand, S. Mrenna and P.Z. Skands, {\it PYTHIA 6.4 Physics and Manual},
JHEP 0605, 026 (2006), {\tt arXiv:hep-ph/0603175]}.

\bibitem{Delphes}
J. de Favereau, et al., [DELPHES 3 Collaboration],
{\it DELPHES 3, A modular framework for fast simulation of a generic collider experiment},
JHEP 1402, 057 (2014), {\tt [arXiv:1307.6346]}.

\bibitem{Fastjet}
M. Cacciari, G. P. Salam, and G. Soyez, {\it FastJet User Manual},
Eur. Phys. J. C 72, 1896 (2012), {\tt [arXiv:1111.6097]}.

\bibitem{antiKT}
M. Cacciari, G. P. Salam, and G. Soyez, {\it The Anti-k(t) jet clustering algorithm},
JHEP 0804, 063 (2008), {\tt [arXiv:0802.1189]}.

\bibitem{Ball:2012cx}
R.D. Ball, et al., {\it Parton distributions with LHC data},
Nucl. Phys. B867 (2013) 244-289, {\tt [arXiv:1207.1303]}.

\bibitem{Deans:2013mha}
C. S. Deans, {\it Progress in the NNPDF global analysis}, Proceedings,
48th Rencontres de Moriond on QCD and High Energy Interactions: La Thuile,
Italy, March 9-16, 2013, 2013, pp. 353-356, {\tt [arXiv:1304.2781]}.

\bibitem{HLLHC_tqh_3000}
K. Agashe et al. [Top Quark Working Group Collaboration], {\tt [arXiv:1311.2028]}.


\end{thebibliography}
\end{document}